\magnification1200

\rightline{KCL-MTH-12-06}

\vskip .5cm
\centerline {\bf Parameters, limits and higher derivative type II string corrections}

\vskip 1cm
\centerline{Finn Gubay and Peter West}
\centerline{Department of Mathematics}
\centerline{King's College, London WC2R 2LS, UK}

\vskip .5 cm
\noindent
String theory in $d$ dimensions has $n+1=11-d$ parameters that may be thought of as being inherited from the geometry of an $n+1$ torus which may be used to construct the theory using dimensional reduction from eleven dimensions.  We give the precise relationship between these parameters and the expectation values of the scalar fields that parameterise the $E_{n+1}$ coset of the $d$ dimensional theory.  This allows us to examine all possible limits of the automorphic forms which occur as the coefficient functions of the higher derivative corrections to the $d$ dimensional type II string effective action.
\vfill
\eject

\medskip
\noindent
{\bf {1. Introduction}}
\bigskip
The low-energy effective actions of the type IIA and IIB string theories are
the IIA [1-3] and IIB [4-6] supergravity theories.  Furthermore
eleven-dimensional supergravity [7] is the low-energy
effective action of one of the limits of M-theory. By virtue of the
large amount of supersymmetry they possess the type IIA and type IIB
supergravity theories contain all perturbative and non-perturbative
string effects at low-energy.  As a consequence their study has lead to many aspects
of what we now know about string theory not least of which was the
realisation that the underlying theory must contain branes on an equal
footing to strings. Unfortunately, we know very little about branes,
for example, we do not understand even their classical scattering and
essentially nothing is known about their quantum properties.
\par
The dimensional reduction of the IIA and IIB supergravity theories on an
$n$ torus, or  the eleven dimensional supergravity theory on
an $n+1$ torus, lead to the unique maximal supergravity theory in
$d = 10 - n$ dimensions which possesses a hidden
$E_{n+1}$ duality symmetry [8-11]. The IIB supergravity theory
possesses an $SL(2,R)$ symmetry [4]. The four-dimensional heterotic
supergravity theory possesses an analogous $SL(2,R)$ symmetry and taking
into account the fact that the brane charges are quantised [12,13] and
rotated by this symmetry it was proposed [14,15] that the four
dimensional heterotic string theory  was invariant under an $SL(2,Z)$
symmetry which included a transformation that mixed perturbative to
non-perturbative effects. This realisation was generalised to the
$E_{n+1}$ symmetry of type II theories in $d=10-n$ dimensions in [16].
\par
One might hope that this discrete $E_{n+1}$ group is a symmetry of
string theory in $d$ dimensions in which case it  must be a symmetry of
the effective action in $d$ dimensions. Since
the continuous
$E_{n+1}$ group is a symmetry of the lowest energy theory, that is, the
supergravity theory, it remains to show that it is a symmetry of the
higher derivative string corrections. The terms in the effective action
are polynomial in the Riemann tensor, the field strengths of the gauge
fields and expressions containing the derivatives of the scalars.  The
coefficients of these terms are functions of the scalar fields and not
their space-time derivatives. In the supergravity theory the terms are
only bilinear in space-time derivatives and their coefficients are
numerical factors. Indeed, the scalar  fields in the
supergravity theory are contained in a
non-linear realisation of
$E_{n+1}$ with the local subgroup  being the maximal compact subgroup,
denoted $I_c(E_{n+1})$; put another way the scalars belong to the coset
$E_{n+1}$/$I_c(E_{n+1})$. As such the scalar fields in the supergravity
theories appear through their derivatives which are contained in  Cartan
forms of
$E_{n+1}$ and so the way the scalar fields can occur is highly
constrained by the non-linear realisation of the continuous group.
In the higher derivative terms in the effective action the
derivatives of the scalars are also contained in the $E_{n+1}$ Cartan
forms but the coefficients are functions of the naked scalars, although
they must transform under the discrete $E_{n+1}$ symmetry which can be
read off from the terms that they multiply.
\par
The higher derivative corrections to string theory have been
most studied in the context of the effective action of IIB string theory
which should possess an
$SL(2,Z)$ symmetry. For the terms with no more than twelve space-time
derivatives and constructed from the Riemann curvature alone, it has
been
proposed that the coefficients are certain Eisenstein automorphic forms.
Although these are not holomorphic functions of the two scalars they are
eigenvalues of the Laplacian acting on the coset space $SL(2,R)$/
$SO(2)$. Such objects are highly constrained and this places very strong
constraints on the theory [17-24]; indeed one finds restrictions of
string scattering, such as non-renormalisation theorems for the
perturbative corrections. Quite a number of these effects have been
checked against known string results and have been found to be true;
this
provides strong evidence that the $SL(2,R)$ symmetry of the IIB
supergravity theory [4],  when discretised to $SL(2,Z)$,  really is a
symmetry of ten dimensional IIB string theory.
\par
One of the simplest ways to see that the maximal supergravity theory in
$d=10-n$ dimensions might have  an $E_{n+1}$ symmetry is to compute the
dependence on the diagonal components of the metric in the dimensional
reduction on the $n$-torus of, for example, the ten dimensional IIB supergravity theory.
These fields $\vec \phi$ occur in the resulting action in the form
$e^{{1\over \sqrt 2}\vec \phi\cdot \vec w}$ and one finds that the
vectors $\vec w$ that occur are the roots of $E_{n+1}$ [25-27]. 
If one repeats such a calculation for the dimensional reduction of the higher 
derivative terms of the ten dimensional IIB effective action to $d$ dimensions 
then one does not find the roots of $E_{n+1}$. However, one does find that the
corresponding vectors are certain weights of $E_{n+1}$ which can be
accounted for if the coefficients in the effective action in $d$
dimensions are automorphic forms of $E_{n+1}$ constructed from the representations 
of $E_{n+1}$ that contain the weights that appear [27].  The dimensional reduction of the eleven dimensional M-theory effective action on an $n+1$ torus leads to a similar result [26].  The appearance of the weights of $E_{n+1}$ upon dimensional reduction of arbitrary higher derivative terms in the type IIA/B and M-theory effective actions to $d=10-n$ dimensions puts constraints on the automorphic forms that appear as the coefficient functions of the corresponding higher derivative terms in $d$ dimensions.  In particular, the $E_{n+1}$ coefficient functions should contain the $E_{n+1}$ weights found when a given higher derivative term in type IIA/B string theory and M-theory are dimensionally reduced to $d$ dimensions, these weights suggest the coefficient functions contain automorphic forms constructed from particular representations of $E_{n+1}$ [28,29].  This provided evidence that the higher
derivative string corrections do possess a discrete $E_{n+1}$ symmetry
and that the coefficients are automorphic forms constructed from
scalars that are contained in coset group elements of
$E_{n+1}$/$I_c(E_{n+1})$ acting on the representations found. 
\par
Recently the higher derivative corrections of type II
string theories in less than ten dimensions have been systematically
studied  and specific automorphic forms, constructed from particular representations of $E_{n+1}$, have been proposed for the
coefficients of certain higher derivative terms  which have fewer than
twelve space-time derivatives [30-33]. This has lead to
predictions for string scattering, especially for perturbative results,
that have been checked against the known string theory results.  This
work provides further evidence for a discrete $E_{n+1}$
symmetry of string theory [30-33].  These papers have in
particular generalised the previous results on non-renormalisation
theorems [34-36].   Earlier results were found in eight dimensions
[37,38] and more recently in [39], discussions can also be found in [40-42] and references
therein. 
\par
Given the effective action for type IIA, or IIB, string theory in ten
dimensions one can dimensionally reduce it on an $n$ torus to obtain
terms in the effective action in $d=10-n$ dimensions. The scalars in the
lower dimensional theory that arise from the diagonal components of the metric
in the compactified directions parameterise, in a one to one way, the
volume of the $n$-torus and all the sub-tori that are used in the dimensional
reduction.   As explained above, these scalar fields together with the
other scalar fields are encoded in the higher derivative effective
action as a non-linear realisation of $E_{n+1}$ with local subgroup
$I_c(E_{n+1})$; the diagonal components of the metric, together with
the ten dimensional dilaton, appear as the part of this group element
constructed from the Cartan subalgebra of  $E_{n+1}$.
\par
One can also carry out the dimensional reduction from the eleven
dimensional effective action of M-theory to $d=10-n$ dimensions on an
$n+1$ torus and then the $n+1$ diagonal components of the metric
parameterise the volumes of the $n+1$ dimensional torus and all the
possible sub-tori. In the dimensionally reduced theory these $n+1$
scalar fields are encoded in a group element that belongs to the Cartan
subalgebra of $E_{n+1}$. As is well known, the volume of the sub-tori
which sits in the eleventh dimension, essentially the radius of the
circle in the eleventh direction, becomes the dilaton in the ten
dimensional IIA string theory and it is related to the string coupling
$g_s$ in this theory.
\par
Unlike the other scalar fields, the scalar fields in the $d$
dimensional theory that are associated with the Cartan
subalgebra in the theory in $d$ dimensions appear in exponentials. The expectation
values of these fields are parameters that define the theory in $d$ dimensions. Indeed, the expectation value of one of the scalars $\phi_d$ in the
$d$-dimensional effective action is related to the string coupling in
that dimension by  $g_d=e^{\phi_d}$. The perturbative corrections can be
found by taking the limit $g_d\to 0$, or equivalently  $\phi_d\to
-\infty$ in the effective action. To find these one must find the
expression for the automorphic form in this limit
and one can then check if one indeed finds a perturbation theory that
is consistent with string theory; the simplest check being that it
contains
$g_d^{-2+2n}$ for
$n=0,1,2,\ldots$. This test was indeed carried out for the coefficient functions of the eight and 12 derivative terms constructed purely from the Riemann curvature with the
proposed automorphic forms [30-33].
\par
The scalar field which is related to the string coupling is just
one of $n+1$ scalar fields and we can also consider limits of their
expectation values which correspond to various limits as certain sub-tori become
large or small.
Computing this limit for the automorphic form we can find
the effect on the effective action in $d$ dimensions.
In particular if we let the radius of the circle
in the $d+1$th direction become large then we are
decompactifying the theory to $d+1$ dimensions and the
automorphic form of $E_{n+1}$ becomes an automorphic form of $E_{n}$ with
coefficients that are powers of the exponential of the scalar
involved in the limit. This was carried out in [31-33] for the eight, twelve and fourteen derivative terms constructed purely from the Riemann curvature in various dimensions.  However, the precise connection to the
parameters of the theory and the possible limits one can take in general
have not been previously discussed.
\par
In this paper we will consider the effective action in $d$ dimensions
and find the precise relationship between the scalar fields associated
with the Cartan subalgebra of $E_{n+1}$ and the parameters of the theory in ten dimensions, that is
the string coupling and the tori sub-volumes. This will allow us to find a
one to one correspondence between the nodes of the $E_{n+1}$ Dynkin diagram
and the parameters of the $d$ dimensional string theory. 
\par
As mentioned earlier, upon dimensional reduction on a torus to $d<10$ dimensions type IIA/B string theory and M-theory are equivalent and possess a hidden $E_{n+1}$ symmetry.  The $d=10-n$ dimensional maximal supergravity theory may be viewed as arising from the dimensional reduction of type IIA/B string theory on an $n$ torus where the natural parameters to consider are the $d$ dimensional string coupling $g_{d}$ and the volume of the $n$ torus and the volume of its $1,2,...,n-1$ subtori.  Alternatively, dimensional reduction of M-theory on an $m=n+1$ torus leads to a $d=10-n$ dimensional supergravity theory with an $E_{n+1}$ symmetry where the natural parameters of the $d$ dimensional theory are the volume of the $m$ torus and the volume of its $1,2,...,m-1$ subtori.
\par
For example, the volume $V_{m(M)}$ of the $m=n+1$ torus, upon which M-theory is compactified, is associated with the exceptional node in the $E_{m}$ Dynkin diagram below.  To be precise $V_{m(M)}=e^{3 \phi_{m}}$, where $\phi_{m}$ is the field associated with Cartan subalgebra element $m$ in Chevalley basis, which in our notation corresponds to the exceptional node in the $E_{m}$ Dynkin Diagram in figure 1.
The remaining nodes of the $E_{m}$ Dynkin diagram are associated 
with a specific combination of the $m$ parameters $V_{m(M)}$ and the $m-1$ tori sub-volumes $V_{j}$, $j \leq m-1$ as given below each node in figure 1.
$$
\matrix {
 & & & & & & V_{m(M)}^{{1 \over 3}} & &  &&\cr
 & & & & & & \bullet & & && \cr
 & & & & & & | & & &&\cr
  \bullet & - &
\ldots &-&\bullet&-&\bullet&-&\bullet&-&\bullet \cr
 V_{1} & &  & & V_{m-4}  &  & V_{m-3} &  & V_{m-2}V_{m(M)}^{-1} && V_{m-1}V_{m(M)}^{-{2 \over 3}} }
$$
\bigskip
\centerline{Figure 1. The $E_{m}$ Dynkin diagram labelled by the $d$
dimensional M-theory parameters}
\bigskip
The powers of the parameters labelling node $i$ are such that they equal $e^{\phi_{i}}$, where $\phi_{i}$ is the field associated with Cartan subalgebra element $i$.  The correspondence between the fields $\phi_{i}$ and the parameters is derived in section 2.
\par
From the type IIB point of view, the nodes of the $E_{n+1}$ Dynkin diagram are associated with the $n+1$ parameters that include the $d$ dimensional string coupling $g_{d}$, the volume $V_{n(B)}$ of the $n$ torus, upon which the type IIB theory is compactified and the $n-1$ tori sub-volumes $V_{j}$, $j \leq n-1$.  The sub-volumes of the torus $V_{j}$, $j \leq n-2$ are related to the Chevalley fields $\phi_{i}$ by $V_{j}=e^{\phi_{j}}$ and are identified with $n-2$ nodes of the $E_{n+1}$ Dynkin diagram, as shown in figure 2.  The specific combination of the $n+1$ parameters associated with each node is given in figure 2.
$$
\matrix {
 & & & & & & V_{n-1} V_{n(B)}^{-{
1 \over 2}} & &  &&\cr  & & & & & & \bullet & & && \cr
 & & & & & & | & & &&\cr
  \bullet & - &
\ldots &-&\bullet&-&\bullet&-&\bullet&-&\bullet \cr
 V_{1} & &  & & V_{n-3}  &  &
V_{n-2} &  & V_{n (B)}^{{1 \over 2}} &&
g_{d}^{-{4 \over 8-n}} }
$$
\bigskip
\centerline{Figure 2. The $E_{n+1}$ Dynkin diagram labelled by the $d
$ dimensional type IIB parameters}
\bigskip
\par
The nodes of the $E_{n+1}$ Dynkin diagram are similarly associated with the $n+1$ parameters, of the type IIA theory dimensionally reduced on an $n$ torus.  The $E_{n+1}$ Dynkin diagram labelled by these parameters is given in figure 3.
$$
\matrix {
 & & & & & & V_{n (A)}^{{1 \over 2}}g_{d}^{{2 \over 8-n}} & &  &&\cr
 & & & & & & \bullet & & && \cr
 & & & & & & | & & &&\cr
  \bullet & - &
\ldots &-&\bullet&-&\bullet&-&\bullet&-&\bullet \cr
 V_{1} & &  & & V_{n-3}  &  & V_{n-2} &  & V_{n-1} V_{n(A)}^{-{1 \over 2}}g_{d}^{-{2 \over 8-n}} &&
g_{d}^{-{4 \over 8-n}} }
$$
\bigskip
\centerline{Figure 3. The $E_{n+1}$ Dynkin diagram labelled by the $d
$ dimensional type IIA parameters}
\bigskip
\par
These result allows us, in section 4,  to discuss in general all
possible limits of the automorphic form and the effective action in
these
limits.
\par
To carry out these calculations we will use the $E_{11}$ formulation
of the maximal supergravity theories. We will not rely on the conjecture
that $E_{11}$ is a symmetry of the underlying theory of strings and
branes [43] but results [43-48] that relate the non-linear
realisations of $E_{11}$ at low levels to the maximal supergravity theories. One could
also find these results by explicitly carrying out the dimensional
reduction to find the $d$ dimensional theory, identifying the way
the $E_{n+1}$ symmetry acts on the fields and then relating the
parameters to the fields associated with the Cartan subalgebra of $E_{n+1}$. This is
a lengthy process, however, using the $E_{11}$ approach it is simple
to identify the $E_{n+1}$ symmetry and in particular the part of the
group element that contains the Cartan subalgebra fields. The $E_{11}$
non-linear realisation also permits one to move between type IIA, type IIB
and M-theory in a simple way.

\bigskip
\noindent
{\bf {2. Group interpretation of parameters in string theory and
M-theory} }
\bigskip
Dimensional reduction of type IIA/B supergravity on an $n$
torus, or equivalently eleven dimensional supergravity on an $n+1$
torus, leads to scalars belonging to a non-linear realisation of an
$E_{n+1}$ symmetry.  The coefficient functions of higher derivative
terms
in the effective action of type IIA/B string theory in $d=10-n$
dimensions   transform as $E_{n+1}$  automorphic forms which  are
functions of these  scalars.
In this section we will use the $E_{11}$ formulation of
type IIA/B and eleven dimensional supergravity theories to derive the
relationship between  these scalar   fields  and the physical parameters
of these theories.  In particular, we will find the relationship
between  the expectation values of the fields
$\dot{\varphi}_{i}$,
$i=d+1,...,11$, parameterising the Cartan subalgebra part of the
$E_{n+1}$ group element and  the
physical parameters of type IIA/B and eleven dimensional supergravity
when dimensionally reduced on an $n$ torus.
This will lead us  to a correspondence between the physical parameters
and the nodes of the $E_{n+1}$ Dynkin diagram.
\medskip
\noindent
{ \bf{2.1 Review of parameters in string theory and M-theory} }
\medskip
The low-energy effective actions of type  IIA/B string theory in ten
dimensions are the type IIA/B supergravity theories.  As a consequence,
the parameters of type IIA/B string theory are related to those of
corresponding supergravity theory in $d=10$ dimensions.  The type IIA/B
supergravity theories have two parameters, the Newtonian coupling
constant $\kappa_{10}$ and $e^{<\phi>}$, where $<\phi>$ is the
expectation value of the type IIA/B dilaton $\phi$.  Note that the type
IIA dilaton and the Newtonian coupling are not equal to the
type
IIB dilaton and Newtonian coupling. Thus on each equation one should add
the subscripts A or B to indicate what theory is being discussed.
However,
the generic discussion applies to each theory in the same way and  in
order to not over complicate the equations we will suppress the
labels A
and B, but they are to be understood to be present.
\par
The $d$ dimensional Newtonian coupling constant  $\kappa_{d}$ by
definition  appears in the action, multiplying the $d$ dimensional
Einstein-Hilbert term in Einstein frame, in the form
$$
{1 \over 2\kappa_{d}^{2}}\int d^{d} x \sqrt{-g} R.
\eqno(2.1.1)
$$
The $d$ dimensional Planck length $l_{d}$ is
defined by
$$
l_{d}^{d-2} = 2\kappa_{d}^{2},
\eqno(2.1.2)
$$
where the form of the relation is determined by dimensional analysis.
\par
The parameters of type IIA/B supergravity in $d=10-n$ dimensions may be
found by dimensional reduction of these theories on an $n$ torus.  To be
general we consider a theory in $D$ dimensions and reduce it on an $n$
torus to $d=D-n$ using  the  ansatz
$$
d\hat{s}^{2}=e^{2\alpha \tilde{\rho}}ds^{2}
+e^{2\beta \left( \tilde{\rho} + <\rho>
\right)}G_{ij}\left(dx^{i} +A^{i}_{\mu}dx^{\mu} \right)\left(dx^{j}
+A^{j}_{\mu}dx^{\mu} \right),
\eqno(2.1.3)
$$
where,
$\tilde{\rho}=\rho- <\rho>$, $det(G)=1$ and the constants $\alpha$ and
$\beta$ are given by
$$
\alpha =\sqrt{{n \over 2\left( d-2 \right) \left( D-2 \right)} }, \ \
\ \ \
\beta =-{\left(d-2\right) \alpha \over n}.
\eqno(2.1.4)
$$
This ansatz is designed so that the dimensionally reduced theory in $d$ dimensions 
has an  Einstein term that involves just the graviton, but  no scalar fields such as 
$\tilde \rho$.  
Equation (2.1.3) is a generalisation of the ansatz used in our previous papers [26-30] 
that carefully
takes into account the expectation values of the fields which  will turn
out to appear  in the definitions of the physical parameters of the
various theories.  Although many of the results in this section are well
known, we are not aware of a very careful discussion and as this is
central to our paper we have tried to present this material here in a
diligent way.
\par
Compactifying a $D$ dimensional theory to $d=D-n$ dimensions,  via this
ansatz, gives the $d$ dimensional theory in Einstein frame.  The
coordinate reparameterisation invariant length of a torus cycle in the
$i$ direction is
$$
\int_{0}^{ l_{D}}  <\hat{e}_{i}^{\ i}> dx^{i}= 2 \pi r_{i},
\eqno(2.1.5)
$$
where we have parameterised the circle by the $D$
dimensional Planck length $l_{D}$ and $r_{i}$ is the physical radius in
the
$i$ direction.  
Note that the vielbein of the internal metric $\hat{e}_{i}^{\ i}
$ is
taken to be independent of the compactified coordinates, therefore the
integration is trivial and one finds
$$
< \hat{e}_{i}{}^{i} > = {2 \pi r_{i} \over l_{D}}.
\eqno(2.1.6)
$$
This allows one to express the volume of an $n$ torus which is given by $\sqrt{<{det(\hat{G})}>}$, where the components of $\hat{G}$ are 
$\hat{G}_{ij}=e^{2\beta \left( \tilde{\rho} + <\rho> \right)} G_{ij}$, as
$$
\sqrt{<{det(\hat{G})}>}=< \hat{e}_{D}^{\ D} >< \hat{e}_{D-1}^{\ D-1} >...
< \hat{e}_{d+1}^{\ d+1} >=(2 \pi )^{n}{r_{D}r_{D-1}...r_{d+1} \over \left( l_{D} \right)^{n}},
\eqno(2.1.7)
$$
where our chosen vielbein frame has lower triangular components for $\hat{e}$ set to zero. Since $det(G)=1$, the expression  in equation (2.1.7) is given by
$$
\sqrt{<{det(\hat{G})}>} =e^{n \beta <\rho> }.
\eqno(2.1.8)
$$
We note that we have taken the vielbein in Einstein frame. It will turn out to be simpler to take a rescaled definition of the volume $V_{n}$ of the $n$ torus by defining
$$
V_{n} = (2 \pi )^{n} {r_{D}r_{D-1}...r_{d+1} \over \left(l_{d}\right)^{n}} = (2 \pi )^{n} \left( {l_{D} \over l_{d}}  \right)^{n} {r_{D}r_{D-1}...r_{d+1} \over \left(l_{D}\right)^{n}}
.
\eqno(2.1.9)
$$
where the factor of $(2 \pi)^{n}$ is chosen to simplify later expressions.  
\par
We will also be interested in subtori of the $n$ torus.  The dimensionless volume of the $j$ subtorus, where $j<n$, is defined by
$$
V_{j}=(2 \pi )^{j} { r_{d+1} r_{d+2}...r_{d+j} \over \left( l_{d} \right)^{j}}.
\eqno(2.1.10)
$$

\par
The dimensional reduction of $D$ dimensional supergravity on a circle
of radius $r_{D}$ gives
$$
{1 \over 2\kappa_{D}^{2}}\int d^{D} x \sqrt{-\hat{g}} \hat{R} + \dots
=
{1\over 2 \kappa_{D}^{2}} \int \hat{e}_{D}{}^{D} dx^{D} \int d^{D-1} x
\sqrt{-g} R +\dots
$$
$$
= 2 \pi r_{D} {1 \over 2\kappa_{D}^{2}}\int d^{D-1} x \sqrt{-g} R +
\dots
\eqno(2.1.11)
$$
where ... denotes all terms in the action beyond the  Einstein-Hilbert
term and fields possessing a $\hat{}$ are $D$ dimensional fields while
fields without are $D-1$ dimensional fields. Comparing the coupling of
the $D-1$ dimensional theory on the the right hand side of equation
(2.1.11)
with the expected coupling in the action of the $D-1$ dimensional theory
given by
$$
{1 \over 2 \kappa_{D-1}^{2}}\int d^{D-1} x \sqrt{-g} R + ...
\eqno(2.1.12)
$$
one finds that the relationship between the $D$ dimensional  Newtonian
coupling constant and the $D-1$ dimensional Newtonian coupling constant
is
$$
\left( \kappa_{D} \right)^{2}= 2 \pi r_{D} \left( \kappa_{D-1} \right)
^{2}.
\eqno(2.1.13)
$$
One may rewrite equation (2.1.13) in terms of the Planck length rather
than
the Newtonian coupling constant, this leads to
$$
\left( l_{D} \right)^{D-2}= 2 \pi r_{D} \left( l_{D-1} \right)^{D-3}.
\eqno(2.1.14)
$$
We may write the ratio of the $D$ dimensional Planck length $l_{D}$ to the $d$ dimensional Planck length $l_{d}$ by iterating equation (2.1.14) $n$ times and dividing by $\left(l_{D} \right)^{n}$, this gives
$$
\left( { l_{D} \over l_{d} } \right)^{d-2}=(2 \pi)^{n} {r_{D} r_{D-1}...r_{d+1} \over (l_{D})^{n} }.
\eqno(2.1.15)
$$
Substituting equation (2.1.15) back into (2.1.9) and using equations (2.1.7) and (2.1.8) to express the result in terms of $<\rho>$ one finds that the volume $V_{n}$ of the $n$ torus as a function of the field $\rho$ is given by
$$
V_{n}= e^{\left(D-2 \over D-2-n \right)n \beta <\rho>}.
\eqno(2.1.16)
$$
\par
A generic feature of supergravity theories compactified on an $n$
torus to $d=D-n$ dimensions is the scaling of the $d$ dimensional
gauge fields by factors of the dimensionless volume $V_{n}$ of the $n
$ torus.  For example, the ten dimensional Einstein-Hilbert term
dimensionally reduced on an $n$ torus with the ansatz
(2.1.3) gives the $d=10-n$ dimensional action
$$
{1 \over 2 \kappa_{10}^{2}} \int d^{10}x \sqrt{-g} \hat{R}
$$
$$= {1 \over (2 \pi)^{7} l_{10}^{8} } l_{10}^{n} e^{n \beta
<\rho>}  \int d^{d}x \sqrt{-g} \left( R - {1 \over 4} e^{2
(\beta\left( \tilde{\rho} + <\rho> \right) - \alpha \tilde{\rho}
) } G_{ij}F^{i}_{\mu \nu}F^{j \mu \nu} - S_{m i}^{j}S^{\mu i}_{j} \right.
$$
$$ 
\left. -
\gamma^{2} \partial_{\mu}\rho \partial^{\mu} \rho \right.
\left.  - 2\left( n \beta + \left(d-1\right) \alpha  \right) \nabla^
{2} \rho \right)
$$
$$
={1 \over (2 \pi)^{7} l_{10}^{8-n}}  V_{n}^{8-n \over 8}  \int d^{d}x \sqrt{-g}
\left( R - {1 \over 4} V_{n}^{{2 \over n}} e^{2 \left(\beta -
\alpha   \right) \tilde{\rho}  } G_{ij}F^{i}_{\mu \nu}F^{j \mu \nu} -
S_{m i}^{j}S^{\mu i}_{j} - \gamma^{2} \partial_{\mu}\rho \partial^
{\mu} \rho \right.
$$
$$
\left.  - 2\left( n \beta + \left(d-1\right) \alpha  \right) \nabla^
{2} \rho \right).
\eqno(2.1.17)
$$
So we see that in this  case each field strength $F^{i}_{\mu \nu}$
constructed from the gauge field $A^{i}_{\nu}$ appears with a factor of
$V_{n}^{{1 \over n}}$.  In general one finds that each gauge
field $A_{i_{1}...i_{k}}$ carrying $k$ contracted downstairs compact
indices is scaled by a factor of $V_{n}^{-{k \over n}}$.
Similarly,
each gauge field carrying $k$ contracted upstairs compact indices is
scaled by a factor of $V_{n}^{{k \over n}}$. 
\par
The type IIA supergravity is the low energy effective theory for 
IIA  string theory.  As mentioned earlier, the parameters
of the type IIA  supergravity theory in ten dimensions are the
Newtonian coupling constant $\kappa_{10(A)}$ and $e^{<\phi_A>}$.  The type
IIA string theory in ten dimensions also has two parameters, the
string length $l_{s(A)}$ and the string coupling constant $g_{s(A)}$, where
the
string length may be defined in terms of $\alpha'_{(A)}$ as
$l_{s(A)}=\sqrt{\alpha'_{A}}$.   Similar statements hold for the IIB theory and 
for simplicity we have omitted the A or B labels on the parameters. 
The parameters of interest to us in taking
various limits of type IIA and IIB string theory and M-theory
compactified to $d=10-n$ dimensions are the $d$ dimensional Planck
length
$l_{d}$, the string  coupling $g_{d}$ in $d=10-n$ dimensions, the
physical radii
$r_{i}$, $i=d+1,...,D$, of the $n$ torus and the volume of the torus
upon
which type IIA/B string theory are compactified on, denoted $V_{n(A)}$
and $V_{n(B)}$ respectively, along with the volume $V_{m(M)}$ of the torus
upon which M-theory is compactified upon.
\par
The relationship between the type IIA and IIB  supergravity coupling
$\kappa_{10}$ and the string length $l_{s}$ is found by comparing the
supergravity action and the effective action derived from type IIA and
IIB string theory, this yields [49,50]
$$
g_{s}=e^{<\phi>}, \quad {\rm and }\quad
2\kappa_{10}^{2}=(2\pi)^{7}l_{s}^{8}g_{s}^{2}.
\eqno(2.1.18)
$$
Each of the symbols in this equation should carry the label
A or B denoting the string theory to which it belongs, but here, and in much of what follows below, we have
suppressed these for simplicity.
The ten dimensional Planck length $l_{10}$ in type IIA or type IIB
string theory is related to the ten dimensional Newtonian coupling constant
through equation (2.1.2) so one has
$$
l_{10}^{8}=(2\pi)^{7}l_{s}^{8} g_{s}^{2}.
\eqno(2.1.19)
$$
\par 
Our aim is to derive an expression
for the effective coupling
$g_{d}$ of type IIA/B string theory compactified on a torus to $d=10-n$
dimensions.  The  string coupling $g_{d}$ is the $d$ dimensional
analogue of the ten dimensional type IIA/B string coupling $g_{s}$.  The
ten dimensional type IIA/B supergravity action in Einstein frame takes
the schematic form
$$
S  = {1 \over 2 \kappa_{10}^{2}} \int d^{10} x \sqrt{-g} R + ...
\eqno(2.1.20)
$$
Rewriting the action in terms of the ten dimensional string parameters
$l_{s}$ and $g_{s}$, using  equation (2.1.19), the low-energy effective
action of type IIA/B string theory in ten dimensions takes the form
$$
S  = {1\over (2\pi)^{7} {l_{s}}^8 g_{s}^{2}}\int d^{10}  x \sqrt{-
g} R
+ ...
\eqno(2.1.21)
$$
Dimensionally reducing the low-energy effective action of equation
(2.1.21)
on an
$n$ torus via the ansatz (2.1.3), using equations (2.1.7) and (2.1.8),  one finds
$$
S = {1\over (2\pi)^{7} l_{s}^8 e^{2 <\phi>} } (l_{10})^{n} e^{n
\beta <\rho>} \int d^{d}
x \sqrt{-g}R+...
$$
$$
= {1\over (2\pi)^{7} } {l_{s}^{n} e^{{n \over 4}<\phi>} e^{n \beta <
\rho>}  \over l_{s}^{8} e^{  2 <\phi> }} \int d^{d}
x \sqrt{-g}R+...
$$
$$
= {1\over (2\pi)^{7} l_{s}^{8-n}} {1 \over[ e^{  \left(8-n \over 8
\right) <\phi> - \left(n \beta <\rho> \over 2  \right) }]^2} \int d^{d}
x \sqrt{-g}R+...
\eqno(2.1.22)
$$
We now define $g_{d}$ as the $d=10-n$ dimensional analogue of $g_{s}$
in equation (2.1.21).  The compactified $d=10-n$ action takes the form
$$
S = {1\over (2\pi)^{7} {l_{s}}^{8-n} g_{d}^{2}}\int
d^{d} x \sqrt{-g} R+... \,
\eqno(2.1.23)
$$
Comparing the expected $d=10-n$ dimensional action in  equation (2.1.20)
with the dimensionally reduced action in equation (2.1.21) one observes
that the $d$ dimensional effective coupling $g_{d}$ is given by
$$
g_{d}=e^{{8-n
\over 8}<\phi>-{n \beta <\rho> \over 2}}=V_{n}^{-\left( {8-n \over 16 } \right) }g_{s}^{\left( {8-n \over 8 } \right)},
\eqno(2.1.24)
$$
where $g_{s}$ is the ten dimensional type IIA/B string coupling  
and $V_{n}$ is the volume of the $n$ torus of equation (2.1.16).  Defining
the
shifted $d$ dimensional dilaton $\phi_{d}$ by $\phi_{d}= \left( {8-n \over
8} \right)\phi-{n \beta  \over 2}\rho$, one may write $g_{d}=e^{<\phi_{d}>}$.
Note that in $d$ dimensions, the components of the Einstein frame
metric $g_{\mu \nu}$ are related to
the components of the string frame metric $g_{(S) \mu \nu }$ by $g_
{\mu \nu} = e^{-{4 \over
d-2}\tilde{\phi}_{d}}g_{(S) \mu \nu}$, where $\tilde{\phi}_{d}=\phi_
{d} -
<\phi_{d}>$, that is it does not involve the expectation value of $\rho$. 

\par
Our expressions for the volume $V_{n}$, coupling constant $g_{d}$ and
the ten dimensional Planck length $l_{10}$ then allows us to write
$$
l_{d}^{d-2} =g_{d}^{2}l_{s}^{d-2}.
\eqno(2.1.25)
$$
In principle one has different type IIA and type IIB string
couplings in
$d$ dimensions, $g_{s(A)}$ and $g_{s(B)}$.  However, there is a unique
maximal supergravity theory in $d<10$ dimensions and the type IIA and
type IIB string theories in
$d<10$ dimensions are related by T-duality.  Moreover, T-duality
provides
a map between the ten dimensional type IIA dilaton $\phi$ and the moduli
of the torus compactifying the type IIA theory with their type IIB
counterparts, as a result one may show that the string coupling $g_{d}$
takes
the same form for both the type IIA and type IIB theories.  Similarly
the
$d$ dimensional Planck length, or equivalently the $d$ dimensional
Newtonian coupling, takes the same form for both the compactified type
IIA and type IIB theories.  We will revisit the correspondence between
the physical fields of type IIA and type IIB supergravity in $d<10$
dimensions in section three. The same conclusion applies to the Newtonian 
coupling constant $\kappa_d$ and so the Planck length in $d$ dimensions $l_d$. 
\par
Eleven dimensional supergravity is conjectured to be the low-energy  effective action of
M-theory. The IIA supergravity theory was derived [43,51] from the eleven dimensional supergravity theory using dimensional reduction on a circle using a special case of the torus ansatz (2.1.3) and is
given by
$$
ds^{2}=e^{-{2 \over 3} \tilde{\phi} }g_{(S)  \mu \nu}dx^{\mu}dx^{\nu} +
e^{{4 \over 3} \left( \tilde{\phi} + <\phi> \right)}\left(dx^{11}+A_
{\mu}
dx^{\mu} \right)^{2},
\eqno(2.1.26)
$$
where the subscript $S$ indicates that the  ansatz is written in
terms of
the string frame ten dimensional metric and $\tilde{\phi}=\phi - <
\phi>$.
 From the ansatz, we may identify $e^{{2 \over 3} \left( \tilde{\phi} +
<\phi> \right)}$ as the vielbein on the circle $\hat{e}_{11}^{\ 11}$ and
so using equations (2.1.6) and (2.1.18) one finds
$$
r_{11}=g_{s(A)}^{{2 \over 3}}l_{11}.
\eqno(2.1.27)
$$
It then follows that upon compactification of  the eleventh
dimension, we
have
$$
2\kappa_{11}^{2}=2\pi r_{11} \kappa_{10}^{2}=2\pi r_{11} l_{s}^{8}g_{s
(A)}^{2},
\eqno(2.1.28)
$$
where $\kappa_{11}$ is the $d=11$ supergravity coupling constant.
Using our expression for the eleven dimensional Planck length $l_{11}
$ in terms of the IIA string coupling $g_{s(A)}$ and the radius of
the eleventh dimension $r_{11}$, we find, from the above expression,
$$
r_{11}=g_{s}l_{s}.
\eqno(2.1.29)
$$
We may use equation (2.1.15) to write the $d=11-m$
dimensional Planck length $l_{d}$ as a function of the volume $V_{m(M)}$
of the torus used to dimensionally reduce M-theory and the eleven
dimensional Planck length $l_{11}$, this yields
$$
l_{d}= l_{11} V_{m(M)}^{-{1 \over 9}}.
\eqno(2.1.30)
$$
\par
Reinstating the labels denoting type IIA and type IIB quantities we
have, in summary, that the Planck length in ten and eleven
dimensions are related to the string length and IIA/IIB coupling by
$$
l_{11} = g_{s(A)}^{{1 \over 3}}l_{s},
\eqno(2.1.31)
$$
$$
l_{10 (A)} = g_{s(A)}^{{1 \over 4}}l_{s},
\eqno(2.1.32)
$$
$$
l_{10 (B)} = g_{s(B)}^{{1 \over 4}} l_{s},
\eqno(2.1.33)
$$
$$
r_{11}  = g_{s(A)} l_{s},
\eqno(2.1.34)
$$
\par
By dimensional analysis, an arbitrary higher
derivative term in the $d$ dimensional Einstein frame effective
action of type IIA/B string theory and M-theory compactified on a
torus to $d=10-n$ dimensions contains a factor of $l_{d}^{p-d}$,
where $p$ is
the number of derivatives in the term. To examine the various limits
in the parameters of type IIA/B string and M-theory compactified to
$d=10-n$ dimensions we will rewrite the $d$ dimensional Planck length
$l_{d}$ in terms of these parameters.
\par
The $d$ dimensional Planck length $l_{d}$ is related to the $d+1$
dimensional quantities $l_{d+1}$ and the radius $r_{d+1}$ on which a
$d+1$ dimensional theory is compactified on, relevant to the
decompactification of a single dimension limit, by
$$
l_{d}=r_{d+1}^{-{1 \over d-2}}\left( l_{d+1} \right)^{{d-1 \over d-2}},
\eqno(2.1.35)
$$
where we have made use of equation (2.1.14).
The $d$ dimensional Planck length $l_{d}$ is related to the volume of
the $n$ torus $V_{n(A)}$ upon which $D=10$ type IIA string theory is
compactified by
$$
l_{d}=l_{10 (A) }V_{n(A)}^{-{1 \over 8}},
\eqno(2.1.36)
$$
this may be derived by iterating equation (2.1.14) $n$ times.
Similarly, the $d$ dimensional Planck length $l_{d}$ is related to
the volume of the $n$ torus $V_{n(B)}$ upon which $D=10$ type IIB
string theory is compactified by
$$
l_{d}=l_{10 (B) }V_{n(B)}^{-{1 \over 8}}.
\eqno(2.1.37)
$$
While the  $d$ dimensional Planck length $l_{d}$ is related to the
volume of the $m$ torus $V_{m(M)}$ upon which M-theory is compactified by
$$
l_{d}=l_{11}V_{m(M)}^{-{1 \over 9}},
\eqno(2.1.38)
$$
again, this may be derived by iterating equation (2.1.14) $m$ times.  
The $d$ dimensional Planck length $l_{d}$ is related to the volume of a $j$ dimensional subtorus $V_{j}$ by 
$$
l_{d}=l_{d+j}V_{j}^{-{1 \over d+j-2}}.
\eqno(2.1.39)
$$
Finally, the transition from $d$ dimensional Einstein frame to $d$
dimensional string frame, relevant to the $d$ dimensional
perturbative limit, is given by
$$
g_{\mu \nu} = e^{-{4 \over d-2}\tilde{\phi}_{d}}g_{(S) \mu \nu},
\eqno(2.1.40)
$$
where $g_{\mu \nu}$ are the components of the $d$ dimensional
Einstein frame metric and $g_{(S) \mu \nu}$
are the components of the $d$ dimensional string frame metric.

\medskip
\noindent
{ {\bf 2.2. Parameters and fields in M-theory}}
\medskip
Upon dimensional reduction on
an $n$ torus to $d=10-n$ dimensions the type IIA and Type IIB
supergravity theories lead to the same theory which  possess an
$E_{n+1}$ symmetry. The scalars that appear  belong to a non-linear
realisation  of this
$E_{n+1}$ symmetry. As such the scalar fields appear as parameters of an
$E_{n+1}$ group element, strictly a coset element. Indeed,   the
diagonal components of the metric and the
dilaton, in the case of the IIA and IIB theories,   parameterise
the part of this group element that is in the Cartan subgroup  of
$E_{n+1}$.  In the previous section we described the
connection between the expectation values of certain scalar
fields
in type IIA/B supergravity compactified on an
$n$ torus, or equivalently eleven dimensional supergravity on an $m=n+1$
torus, and the parameters of the compactified $d$ dimensional maximal
supergravity theory, or equivalently string theory in $d$ dimensions.
By certain scalar fields we mean the scalar fields just
mentioned, namely, the diagonal components of the metric and the
dilaton, in the case of the IIA and IIB theories.
In this section we will derive the connection between these   scalar
fields  and the nodes of the
$E_{n+1}$ Dynkin diagram.  As a result, we will find that the parameters
in type IIA/B supergravity compactified on an
$n$ torus, or equivalently eleven dimensional supergravity on an $n+1$
torus, are associated with specific nodes of the $E_{n+1}$ Dynkin
diagram.
\par
The eleven dimensional, IIA and IIB supergravity theories, as well
as the maximal type II supergravity theories in lower dimensions, can be
formulated as non-linear realisations which at low levels are
non-linear realisations of the Kac-Moody algebra $E_{11}$ [43-48].  The
different theories arise by taking the different decompositions of
$E_{11}$ into the subalgebras that arise from deleting the different
nodes in the $E_{11}$ Dynkin diagram.  The
fields of these theories appear as the parameters of an  $E_{11}$ group
element, or more precisely a coset element,  and it turns out that the
fields are in one to one correspondence with the generators of the
Borel subalgebra of
$E_{11}$. As such
$E_{11}$ encodes the fields of each of these theories and, as there is
only one $E_{11}$ algebra, it provides us with a way of relating the
fields in the different supergravity theories to each other [52].
As we now explain this connection is particularly simple for the subcase
of interest to us in this paper.
\par
The $E_{11}$ algebra is formulated in terms of generators that are the
multiple commutators of the so called Chevalley generators. There are
three  Chevalley generators associated with every node of the $E_{11}$
Dynkin diagram and they obey conditions that are encoded in the Cartan
matrix of $E_{11}$. In this paper,  we will not need a detailed
knowledge
of this construction, as  we will be interested in the $E_{11}$ group
element restricted to the Cartan subgroup. As such it will suffice to
know that the Chevalley generators contain the Cartan subalgebra of
$E_{11}$ and we will  denote these generators by
$H_{\hat a}$, $\hat a=1,2,\ldots ,11$; the generator $H_{\hat a}$ being
associated with node $a$ in the
$E_{11}$ Dynkin diagram. Thus  the
non-linear realisation of $E_{11}$  restricted to  the Cartan subgroup
is of the form $e^{\sum_{\hat a}\phi_{\hat a} H_{\hat a}}$ where
$\phi_{\hat a}$ will be called the Chevalley fields.  It is important 
to realise that the generators $H_{\hat a}$ are essentially uniquely
specified by the above construction and so as a consequence are the
fields $\phi_{\hat a}$.  We will denote the Cartan subalgebra part of the same group element in Weyl basis as $e^{\vec{\phi}. \vec{H}}$, where $\vec{H}$ are the Cartan subalgebra generators in Weyl basis and $\vec{\phi}$ are the corresponding fields.
\par
The maximal supergravity theories in $d$ dimensions are
described in terms of the fields that have been used to describe the
propagating degree of freedom since long ago, such as the graviton and
gauge fields, these include the fields of interest to us, that is the
diagonal components of the metric and the dilaton, and we will refer to these as
the physical fields.  However, these are not the same as the fields that arise in the $E_{11}$
non-linear realisation when we formulate the algebra using the Chevalley
generators and in particular the Chevalley fields $\phi_{\hat a}$ that
arise when we write the Cartan subalgebra in
terms of the generators $H_{\hat a}$. However since the $E_{11}$
non-linear realisation leads to the supergravity theories there is a one
to one relationship between the Chevalley fields  $\phi_{\hat a}$
and the physical fields of interest to us. We note that in the different
supergravity theories one finds different sets of physical fields but
only one set of Chevalley fields.  In this section we
will find the relationship between the physical fields consisting of the
diagonal components of the metric and dilaton, where present, in the
eleven dimensional, type IIA and type IIB theories in ten dimensions
and their reduction on
tori and the Chevalley fields. As such we will be able to connect the
physical parameters of the various theories with the nodes of the Dynkin
diagram.
\par
One could also derive the results of this section in a more conventional
way by dimensionally reducing the eleven and ten dimensional theories to
$d$ dimensions on tori, identifying the fields that belong to the
$E_{n+1}$ coset and finding the relations between the fields in the
resulting $d$ dimensional theories. However, this requires the use of
dualisations to find all the scalars and is generally rather
complicated.
The $E_{11}$ approach has the advantage that it is rather technically
simple and that the presence  of the $E_{n+1}$ symmetry is very
transparent. We also do not rely on the $E_{11}$ conjecture although
there is now good evidence for this.  See [47,53] for a short review. 
\par
In the rest of this paper we are really working with the expectation
values of the fields, but in order to not clutter the equations we will
not explicitly show the  the expectation
value,  but it is to be understood to be present.
\par
The $E_{11}$ Kac Moody algebra is encoded in the Dynkin
diagram given in figure 4.
$$
\matrix {
& & & & & & & & 11 & &  &&\cr
& & & & & & & & \bullet & & && \cr
& & & & & & & & | & & &&\cr
\bullet &  - &  \ldots & -
&\bullet&-&\bullet&-&\bullet&-&\bullet&-&\bullet
\cr 1& & & & 6 & & 7  &  & 8 &  & 9&&10 }
$$
\bigskip
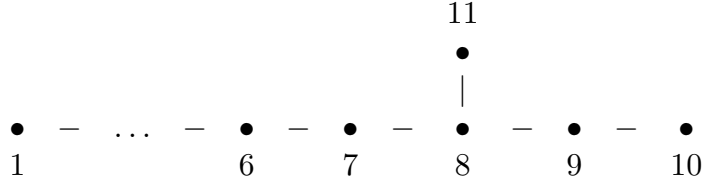
\centerline {Figure 4. The $E_{11}$ Dynkin diagram with eleven
dimensional supergravity labeling }
\bigskip
The eleven dimensional supergravity theory emerges from the $E_{11}$
non-linear realisation if we decompose the
$E_{11}$ algebra in terms of the algebra that results from
deleting the exceptional node labelled eleven, namely the algebra $GL
(11)$.
This subalgebra has the generators
$K^{\hat a}{}_{\hat b}$, $
\hat a, \hat b=1,\ldots ,11$ and it includes all the Cartan subalgebra
generators of $E_{11}$. In particular the relation to the Chevalley
generators $H_{\hat a},\ \hat{a}=1,\ldots ,11  $ being [45]
$$
H_{\hat a} = K^{\hat a}{}_{\hat a}- K^{\hat a+1}{}_{\hat a+1},
 \ \hat{a}=1,\ldots ,10,
$$
$$
H_{11} = - {1 \over 3}\left(K^{1}{}_{1}+...
+ K^{8}{}_{8} \right) + {2 \over 3} \left( K^{9}{}_{9} + K^{10}{}_{10} +
K^{11}{}_{11} \right).
\eqno(2.2.1)
$$
The first ten generators being the Cartan subalgebra generators of
$SL(11)$.
\par
The contribution of the $GL(11)$ subgroup to the $E_{11}$ group
element in the
non-linear realisation is of the form
$$
g=e^{h_{\hat a}{}^{\hat  b}K^{\hat a}{}_{\hat  b}},
\eqno(2.2.2)
$$
where we have added the space-time translation
generators $P^{\hat a}$. The non-linear realisation of $GL(11)$ and the space-time
translations  is known to give  rise to eleven dimensional gravity
and as a result, when dealing with the eleven dimensional theory, the line in
the above Dynkin diagram that is from nodes one to ten inclusive is
known
as the gravity line.  In the non-linear realisation of $GL(11)$ together
with the translations one finds that the Cartan form contains the
object
$$
g^{-1} dx^{\hat a} P_{\hat a } g=   dx^{\hat \mu} \hat e_{\hat
\mu}{}^{\hat a} (\det e_{\hat
\mu}{}^{\hat a} )^{-{1\over 2}}P_{\hat a}
\eqno(2.2.3)
$$
and it turns out that $e_\mu{}^a=(e^h)_a{}^b$ is the
eleven-dimensional vielbein. Thus the fields $h_a{}^a$ are related
to the gravity fields and can be thought of as physical fields. In
equation (2.2.3) we have used the relation
$[K^{\hat
a}{}_{\hat b}, P_{\hat c} ]= -\delta _{\hat a}^{\hat b} K^{\hat
a}{}_{\hat b}+{1\over 2} \delta _{\hat a}^{\hat b} K^{\hat c}{}_{\hat
c}$ [54]. This commutator between the lowest level $E_{11}$ generators,
that is  the $K^{\hat a}{}_{\hat b}$,  and the  lowest level
generators of
the fundamental representation associated with node one, that is
$P_{a}$, follows from the fact that this is a highest weight
representation.
\par
We are interested in the diagonal components of the metric, or
equivalently  the fields $h_{\hat a}{}^{\hat a}$. As already
mentioned  the generators
$K^{\hat a}{}_{\hat a}$ span the Cartan subalgebra of $E_{11}$ and   the
group element of $E_{11}$ restricted to the Cartan subalgebra can also
be written in the form
$ e^{\sum_{\hat{a}}  h_{\hat a}{}^{\hat a} K^{\hat a}{}_{\hat a}}$.
However, the
generators $H_{\hat a}$ also span the Cartan subalgebra and so we can,
as discussed above,  also write the  the group element of $E_{11}$
restricted to the Cartan subalgebra in the form $ e^{\phi_{\hat
a}{}H_{\hat a}}$. These are just two different ways of parameterising
the
group element and we may equate them to find
$$
e^{\sum_{\hat{a}}  \hat{\phi}_{\hat a}\hat H_{\hat a}}= e^{\sum_{\hat{a}} h_{\hat a}{}^{\hat a} K^{\hat
a}{}_{\hat a}}.
\eqno(2.2.4)
$$
Comparing coefficients of $K^{\hat
a}{}_{\hat a}$ using equations (2.2.1) we find the
following relations between the physical fields and the Chevalley fields
$$
\phi_{i} = h^{1}{}_{1} + h^{2}{}_{2} ... +  h^{i}{}_{i}
- {i \over 2} \sum_{j=1}^{11} h^{j}{}_{j}, \quad {\rm  for}  \quad   1
\leq i
\leq 8,
$$
$$
\phi_{9} = h^{1}{}_{1} + h^{2}{}_{2} ... +  h^{9}{}_{9} - 3 \sum_{j=1}
^{11} h^{j}{}_{j},
\quad
\phi_{10} = h^{1}{}_{1} + h^{2}{}_{2} ... +  h^{10}{}_{10} - 2 \sum_
{j=1}^{11}
h^{j}{}_{j},
$$
$$
\phi_{11} = - {3 \over 2} \sum_{j=1}^{11} h^{j}{}_{j}.
\eqno(2.2.5)
$$
It is possible to find the  the correspondence between   the physical
fields and Chevalley fields beyond those associated with the Cartan
subalgebra. In the next section we carry out the identification of the
physical and Chevalley fields, when restricted to the Cartan subalgebra,
in the dimensionally reduced theory.
\medskip
{\bf {2.2.1. Dimensionally reduced M-theory} }
\medskip
The maximal supergravity theory in  $d$ dimensional theory can be
found
by   dimensional reduction of eleven dimensional
supergravity on an
$m=n+1$ torus. This theory also appears in the  non-linear
realisation of
$E_{11}$  if we decompose  $E_{11}$ into the subalgebra that
arises when we delete node
$d$ of the Dynkin diagram, as shown in figure 5; that is the
subalgebra $GL(d)\otimes E_{n+1}$.
$$
\matrix {
&&&&&&& & & & & & & & 11 & &  &&\cr
&&&&&&& & & & & & & & \bullet & & && \cr
&&&&&&& & & & & & & & | & & &&\cr
\bullet & - & \ldots & - & \bullet & - & \times &  - &  \bullet & - &
\ldots &-&\bullet&-&\bullet&-&\bullet&-&\bullet \cr
1 &&&&d-1&&d& & d+1& &  & & 7  &  & 8 &  & 9&&10 }
$$
\bigskip
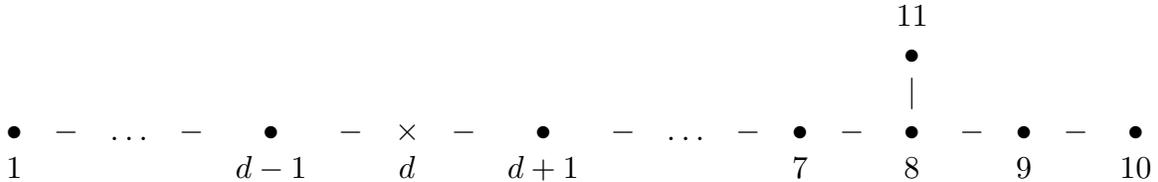
\centerline{Figure 5. The $E_{11}$ Dynkin diagram appropriate to
maximal supergravity in $d<10$ dimensions}
\bigskip
The $GL(d)$ algebra is responsible for
$d$ dimensional gravity and restricted to the Cartan subalgebra of
$E_{11}$ these are given by $K^{a}{}_{ a}$, $a=1,...,d$. The embedding
of the
$E_{m}$ algebra in $E_{11}$ is fixed by requiring that it commutes
with the $GL(d)$ subalgebra of the gravity line and the space-time
translations
$P_a$ in the $d$ dimensions.  Looking at the Dynkin diagram of figure 5 we
see that the $E_m$ algebra contains an $GL(m)$ algebra corresponding to
nodes $d+1$ to $10$ and the generators of this subalgebra are given by
$$
\dot{K}^{i}{}_{j}\equiv K^{i}{}_{j}-{1 \over
9-m}\delta^{i}_{j}\sum_{a=1}^{d}K^{a}{}_{a},\  i,j=d+1,...,11.
\eqno(2.2.6)$$
One
may verify that these generators obey the condition
$[P_{a},\dot{K}^{i}{}_{j}]=0$, for all
$a=1,...,d$ and $i,j=d+1,...,11$. In deriving this equation we
have again used the equation $[K^{\hat
a}{}_{\hat b}, P_{\hat c} ]= -\delta _{\hat a}^{\hat b} K^{\hat
a}{}_{\hat b}+{1\over 2} \delta _{\hat a}^{\hat b} K^{\hat c}{}_{\hat
c}$ [54].   The
Chevalley generators of $E_{m}$ belonging to its  Cartan subalgebra
element  are given by
$$
T_{i} = \dot{K}^{i+1}{}_{i+1} - \dot{K}^{i+2}{}_{i+2} \quad
i=d+1,...,10,
$$
$$
T_{11}  = -{1\over 3}\left( \dot{K}^{d+1}{}_{d+1} + ... +
\dot{K}^{8}{}_{8}
\right) + {2\over 3} \left( \dot{K}^{9}{}_{9} + \dot{K}^{10}{}_{10} +
\dot{K}^{11}{}_{11}  \right).
\eqno(2.2.7)$$
Substituting the expressions of equation (2.2.6)
into equation (2.2.7) one finds that
$$H_{i}=T_{i}
\eqno(2.2.8)$$
for
$i=d+1,...,11$, that is the Chevalley generators $T_i$  in $E_{m}$ are
equal to the Chevalley generators of $E_{11}$ associated with nodes $d+1$
to $10$.
\par
The Cartan subalgebra of $E_{11}$ consists of the generators $K^a{}_a$
and
the generators $T_i,\ i=d+1,...,11$ and so we can write the group
element
of $E_{11}$ restricted to the Cartan subalgebra in the form
$$
k_{M}= e^{\sum_{a=1}^{d} \dot h^{a}_{\ a} K^{a}_{\ a}+ e_{3}\rho
\sum_{a=1}^{d}K^{a}{}_{a}}
e^{\sum_{a=d+1}^{11} \dot{\varphi}_{a} T_{a}}.
\eqno(2.2.9)
$$
where $\dot{\varphi}_{a}$ are   $E_{m}$ Chevalley fields and $e_3$ is a
constant.
\par
Examining the ansatz of equation (2.1.3)  we see that the gravity
fields in
$d$ dimensions are scaled  by powers of $e^{\rho}$ and so to arrive at
the
correct gravity fields in $d$ dimensions using the non-linear
realisation
we must incorporate this shift into the way the group element is
written. As such we consider rewriting the group element of equation
(2.2.2) in the form
$$
g=e^{\sum_{a=1}^{d}  h^{a}{}_{a} K^{a}{}_{a} +e_{1}\rho
\sum_{a=1}^{d}K^{a}{}_{a}}e^{ \sum_{i=d+1}^{11} h^{i}{}_{i} K^{i}{}_{i} +e_{2}\rho
\sum_{i=d+1}^{11}K^{i}{}_{i}},
\eqno(2.2.10)
$$
where $e_1$ and $e_2$ are constants. Proceeding in an analogous way  to
the steps leading to equation (2.2.3) we find that for the new
parameterisation of equation (2.2.10) the
vielbein is given by
$$
(dx^{\mu}\hat{e}_{\mu}{}^{ a}P_{a}+
dx^{i}\hat{e}_{i}{}^{ j}P_{j}) (\det \hat{e}_{\mu}{}^{ a})^{-{1\over 2}}
= g^{-1}\left(dx^{a}P_{a}+dx^{i}P_{i} \right)g
$$
$$
= \sum_{a=1}^{d} dx^{\mu}
e_\mu ^{a}( \det {e}_{\mu}{}^{ a})^{-{1\over 2}}
( \det {e}_{i}{}^{ j})^{-{1\over 2}}
e^{{\rho e_1}}e^{-{\rho\over
2}(de_1+m e_2)}P_{a}
$$
$$
+
\sum_{i=d+1}^{11}dx^{i }  e_i{}^{j}( \det {e}_{\mu}{}^{ a})^{-{1\over
2}}
( \det {e}_{i}{}^{ j})^{-{1\over 2}}e^{{\rho e_2}} e^{-{\rho\over
2}(m e_2+de_1)}P_{j},
\eqno(2.2.11)$$
where we have defined
$$
e_{i}^{\ j}= e^{h^{i}{}_{i}}\delta _i^j, \ i=d+1,\ldots ,11\quad {\rm
and
}
\quad e_{\mu}{}^{ a}=e^{h_{a}{}^{ a}}\delta_\mu^a,
\eqno(2.2.12)
$$
consistent with the fact that we have a diagonal metric.
The condition that the
internal metric satisfies
$\det(G)=\det e_i{}^j=1$ simplifies the above equation and
corresponds  to
the constraint
$\sum_{i=d+1}^{11}h^{i}{}_{i} =0$.
Taking the resulting metric to be of the form
$$
\hat{e}_{\mu}^{\ a}=e^{\alpha \rho}e_{\mu}^{\ a}\quad {\rm  and}\quad
\hat{e}_{j}^{\ i}=e^{\beta \rho}e_{j}^{\ i}
\eqno(2.2.13)$$
and substituting into the first line of equation (2.2.11) we find the
last line of the same equation but with  $\alpha$ and $\beta$
replaced by
$e_1$ and
$e_2$.
\par
We can now consider the same maneuver but with the group element of
equation (2.2.9) and we find that $k_M^{-1}
\left(dx^{a}P_{a}+dx^{i}P_{i} \right) k_M$
leads to a $P_a$ term with a factor $e^{-{\rho\over
2}(d-2)e_3}$. Thus to find the
same result as the group element of equation (2.2.10) as given in
equation (2.2.11) we must choose $(d-2)e_3=(d-2)e_1+m e_2$.
\par
Let us first consider the ansatz of equation (2.1.3) for the part of
$\rho$ that has zero expectation value, that is $\tilde \rho$. As the
above discussion makes clear, we will find the ansatz of equation
(2.1.3) provided we take
$e_{1}=
\alpha (D=11)$,
$e_{2}=
\beta (D=11)$  We note that if $(d-2)e_1+m e_2=0$ then $\hat e_\mu{}^a
( \det \hat {e}_{\mu}{}^{ a})^{-{1\over 2}}$ contains  no $\rho$,  which
also  implies that the product of two inverse vielbeins times the
determinant of the vielbein contains no $\rho$. However, this is just
the
condition that the dimensional reduction leads to an Einstein term with
no $\rho$ and so is in Einstein frame. We recognise this as the same
condition as that of equation (2.1.4). From the viewpoint of the group
element of equation (2.2.9)  we will recover the ansatz of
equation
(2.1.3) for the  part
$\tilde\rho$ if  we set  $e_3=0$.
\par
Let us now apply the above discussion to the part of the ansatz of
equation (2.1.3) that involves the expectation value of $\rho$, that is
$<\rho >$; this is the case of interest to us in this paper. In this case
$e_2=\beta$ has the same value as above, as it multiplies $\tilde \rho+
<\rho>$ in the compactified part, but $e_1=\alpha=0$. We note that in
this case we do find that the Einstein term contains $<\rho>$.
 From the viewpoint of the group element of equation (2.2.9),  we
now must take
$(d-2)e_3= me_2$. We note that the value of the coefficient $e_3$ only
affects the uncompactified part.
\par
We now proceed to compare the $E_{11}$ group element, restricted to the
Cartan subalgebra, when written in terms of the physical fields, that is
in equation (2.2.10),  and the Chevalley fields, that is equation
(2.2.9).  For
$d<9$ dimensions and only  considering  the part of the group element
that is associated with the
$n$ torus to the group element in equation (2.2.10) we find that
$$
e^{ \left( h^{d+1}{}_{d+1}+e_{2}\rho \right)
K^{d+1}{}_{d+1} }...e^{   \left( h^{9}{}_{9}+e_{2}\rho \right)
K^{9}{}_{9} }e^{ \left( h^{10}{}_{10}+e_{2}\rho \right) K^{10}{}_{10}
}e^{ \left( h^{11}{}_{11}+e_{2}\rho \right) K^{11}{}_{11}}
=e^{\phi_i T_i}=
$$
$$
=  e^{\dot{\varphi}_{d+1} \left(K^{d+1}{}_{d+1} -K^{d+2}{}_{d+2}
\right)  }...e^{ \dot{\varphi}_{10} \left( K^{10}{}_{10} - K^{11}{}_
{11}  \right)  } e^{\dot{\varphi}_{11} \left( {2 \over 3} \sum_{a=d+1}^
{11} K^{a}{}_{a} - \sum_{a=d+1}^{8} K^{a}{}_{a}  \right) }.
\eqno(2.2.14)
$$
where we have used equations (2.2.7).
Equating the coefficients of the generators gives the relations between
the physical fields in the dimensionally reduced theory and the $E_{n
+1}$
Chevalley fields
$$
h^{d+1}{}_{d+1}+e_{2} \rho
= \dot{\varphi}_{d+1} - {1 \over 3} \dot{\varphi}_{11} , \quad
h^{d+2}{}_{d+2}+e_{2} \rho = -\dot{\varphi}_{d+1} + \dot{\varphi}_{d
+2} - {1 \over 3} \dot{\varphi}_{11},
\ldots
$$
$$
h^{8}{}_{8}+e_{2} \rho = -\dot{\varphi}_{7} + \dot{\varphi}_{8}
- {1 \over 3} \dot{\varphi}_{11}, \quad
h^{9}{}_{9}+e_{2} \rho = -\dot{\varphi}_{8} + \dot{\varphi}_{9}
+ {2 \over 3}\dot{\varphi}_{11},
$$
$$
h^{10}{}_{10}+e_{2} \rho = - \dot{\varphi}_{9} + \dot{\varphi}_{10}
+ {2 \over 3} \dot{\varphi}_{11}, \quad
h^{11}{}_{11}+e_{2} \rho = - \dot{\varphi}_{10}
+ {2 \over 3} \dot{\varphi}_{11}.
\eqno(2.2.15)
$$
Solving these equations for the $E_{m}$ fields in terms of the
scalars $h^{d+1}{}_{d+1},...,h^{11}{}_{11}$ and $\rho$ found upon
dimensional reduction we find
$$
\dot{\varphi}_{i}= h^{d+1}_{\ \ d+1} + h^{d+2}_{\ \ d+2}+ ...
+ h^{i}{}_{ i}+ \left( m-11+i \right){9 \over 9-m} e_{2} \rho, \quad
d+1\leq i < 8.
$$
$$
\dot{\varphi}_{9}=h^{d+1}{}_{d+1} + h^{d+2}{}_{d+2} +...+h^{9}{}_{9}
+ {6\left(m- 3\right) \over 9-m}e_{2}
\rho,
$$
$$
\dot{\varphi}_{10}=h^{d+1}{}_{d+1} + h^{d+2}{}_{d+2} +...+h^{10}{}_
{10} + {3\left(m- 3\right) \over 9-m} e_{2} \rho,
\quad
\dot{\varphi}_{11}= {3m \over 9-m}e_{2} \rho.
\eqno(2.2.16)
$$
\par
The reader may also like to find the analogous equations for the
uncompactified part. For example, the coefficient of $K^1{}_1$
implies the
equation
$h^1{}_1=h^1{}_1+e_3\rho -{1\over 3}\dot\varphi_{11}$. Using the
relation $(d-2)e_3= me_2$ and the value of $\dot{\varphi}_{11}$ given in
equation (2.2.16) we find that this equation is automatically satisfied.


\medskip
\noindent
{\bf {2.2.2 M-theory Parameters}}
\medskip
Determining the fields $\dot{\varphi}_{i}$  in terms of the scalars
$h^{d+1}{}_{d+1},...,h^{11}{}_{11}$ and $\rho$, in equations
(2.2.16) allows one to express the M-theory parameters given in
section 2.1, namely the volume of the $m$ torus $V_{m(M)}$
and the ratio of the radius $r_{d+1}$ of a compact dimension to the $d$
dimensional Planck length, in terms of the $E_{m}$ fields
$\dot{\varphi}_{i}$.   For the volume of the M-theory torus $V_{m(M)}$ one
finds from equations (2.1.16) and (2.2.16),
$$
V_{m(M)}=(2 \pi)^{m} {r_{11}r_{10}r_{9}...r_{d+1} \over l_{d}^{m}}=e^{\left(9 \over
9-m \right)m\beta \rho}
=e^{3\dot{\varphi}_{11}}.
\eqno(2.2.17)
$$
We therefore find that the
volume $V_{m(M)}$  is closely related to the expectation value (not
explicitly shown) of the field
$\dot \varphi_{11}$ which is itself  associated with the node eleven of
the Dynkin diagram in figure 5.
\par
Equations (2.1.7) and (2.1.29) may be used to show
that the ratio of the radius of the circle in the
$d+1$ direction
$r_{d+1}$ to the
$d$ dimensional Planck length $l_{d}$ is
$$
{r_{d+1} \over l_{d}}={l_{11} \over l_{d}}{r_{d+1} \over l_{11}} =e^{{9
\over 9-m} \beta \rho + h^{d+1}{}_{d+1}}=e^{\dot{\varphi}_{d+1}}.
\eqno(2.2.18)
$$
\par
Thus ${r_{d+1} \over l_{d}}$ is closely related to the expectation
value of the field
$\dot \varphi_{d+1}$ which is itself  associated with the node $d+1$  of
the Dynkin diagram in figure 5.
\par
Furthermore, nodes $d+1$ to $8$ are associated with subtori of
the $m$ torus.  The dimensionless volume of the $j$ torus $V_{j}$
contained within the $m$ torus $V_{m(M)}$ is defined in equation (2.1.10) and may be
written
$$
V_{j}=(2 \pi)^{j} \left({l_{11} \over l_{d} }\right)^{j} {r_{d+1} r_{d+2} ... r_{d
+j} \over \left( l_{11} \right)^{j}}=e^{j\left({ 9 \over 9-m}\right)
\beta \rho  + h^{d+1}{}_{d+1} + h^{d+2}{}_{d+2} +...+ h^{d+j}{}_{d+j}}
=e^{\dot{\varphi}_{d+j}},
\eqno(2.2.19)
$$
where we have made use of equations (2.1.15), (2.1.16), (2.1.6) and equations
(2.2.16) to express the $d$ dimensional M-theory physical fields in
terms of the $E_{n+1}$ Chevalley field.
Similarly, the volume of the $m-2$ subtorus is given by
$$
V_{m-2}=(2 \pi)^{m-2} \left({l_{11} \over l_{d} }\right)^{m-2} {r_{d+1} r_{d+2} ...
r_{9} \over \left( l_{11} \right)^{m-2}}=e^{\left(m-2\right)\left({ 9
\over 9-m}\right) \beta \rho  + h^{d+1}{}_{d+1} + h^{d+2}{}_{d+2} +...
+ h^{9}{}_{9}}
$$
$$
=e^{\dot{\varphi}_{9}+ \dot{\varphi}_{11}},
\eqno(2.2.20)
$$
and the volume of the $m-1$ subtorus is 
$$
V_{m-1}=(2 \pi)^{m-1} \left({l_{11} \over l_{d} }\right)^{m-1} {r_{d+1} r_{d+2} ...
r_{10} \over \left( l_{11} \right)^{m-1}}=e^{\left(m-1\right)\left({ 9
\over 9-m}\right) \beta \rho  + h^{d+1}{}_{d+1} + h^{d+2}{}_{d+2} +...
+ h^{10}{}_{10}}
$$
$$
=e^{\dot{\varphi}_{10}+ 2\dot{\varphi}_{11}}.
\eqno(2.2.21)
$$
It follows that each node of the $E_{m}$ Dynkin diagram is associated
with a specific combination of the $m$ parameters $V_{i}$, $i=1,...,m$, as given below each node in figure 6.
$$
\matrix {
 & & & & & & V_{m(M)}^{{1 \over 3}} & &  &&\cr
 & & & & & & \bullet & & && \cr
 & & & & & & | & & &&\cr
  \bullet & - &
\ldots &-&\bullet&-&\bullet&-&\bullet&-&\bullet \cr
 {r_{d+1} \over l_{d}} & &  & & V_{m-4}  &  & V_{m-3} &  & V_{m-2}V_{m(M)}^{-{1 \over 3}} && V_{m-1}V_{m(M)}^{-{2 \over 3}} }
$$
\bigskip
\centerline{Figure 6. The $E_{m}$ Dynkin diagram labelled by the $d$
dimensional M-theory parameters}
\bigskip

\medskip
\medskip
{\bf {2.3 Parameters and fields in type IIA supergravity} }
\medskip
Let us now consider the ten dimensional  IIA supergravity theory
which can
be obtained
  from the supergravity theory in eleven dimensions by dimensional
reduction on a circle. In this process, the diagonal components of the
eleven dimensional metric result in the diagonal components of the ten
dimensional metric and a scalar
$\phi$, which is the dilaton of the IIA theory.
In terms of the $E_{11}$ non-linear realisation  we obtain the
IIA theory by decomposing $E_{11}$ into the algebra that results
from deleting nodes ten and eleven of the Dynkin diagram below (see
figure
7), that is   the subalgebra
$GL(10)\otimes GL(1)$ algebra. The
$GL (10)$ algebra leads to ten
dimensional gravity and the $GL(1)$ factor leads to the IIA dilaton.
$$
\matrix {
& & & & & & & & 11 & & 10 &&\cr
& & & & & & & & \bullet & & \bullet && \cr
& & & & & & & & | & & | &&\cr
\bullet &  - &  \ldots & - &\bullet&-&\bullet&-&\bullet&-&\bullet&& \cr
 1& & & & 6 & & 7  &  & 8 &  & 9&& }
$$
\bigskip
\centerline{Figure 7.
The $E_{11}$ Dynkin diagram appropriate to type IIA supergravity}

\bigskip
The gravity line is now the horizontal line of the Dynkin diagram of
figure 7.
\par
Let us denote the
generators of $GL(10)$  by
$\tilde{K}^a{}_b$, $a, b=1,\ldots , 10$ and let $\tilde R$ be the $GL
(1)$
generator. These contain the generators of the Cartan subalgebra of
$E_{11}$.  The group element in the Cartan subalgebra of
$E_{11}$ can therefore be written in the form
$$
g=e^{\sum_{a} \tilde h^{a}_{\ a} \tilde K^{a}_{\ a}}e^{\tilde{\sigma} \tilde R}.
\eqno(2.3.1)
$$
The tilde distinguishes the fields and generators from those used in
eleven dimensions. However, in terms of the Chevalley generators in the
Cartan subalgebra of $E_{11}$, the group element has the  form
$g=e^{ \sum_{\hat{a}}  \phi_{\hat a} H_{\hat a}}$. This is the same form as in
eleven dimensions as the $E_{11}$ algebra has essentially   a unique set
of generators
$H_{\hat a}$.
\par
The derivation of the IIA supergravity theory as a non-linear
realisation leads to the following relation between the Cartan
sub-algebra generators
$H_{\hat a}$ of the $E_{11}$ algebra and those in the $GL(10)\otimes
GL(1)$ algebra [45]
$$
H_{a}= \tilde K^{a}_{\ a}- \tilde K^{a+1}_{\ a+1} , \quad
a=1,...,9,
$$
$$
H_{10}=-{1 \over 8}\left( \tilde K^{1}{}_{
1}+...+ \tilde K^{9}{}_{ 9} \right) + {7
\over 8} \tilde K^{10}{}_{ 10} - { 3 \over 2 }\tilde R,
$$
$$
H_{11}=-{1 \over
4}\left( \tilde K^{1}{}_{ 1} +...+ \tilde K^{8}{}_{ 8}  \right) + {3
\over 4}\left( \tilde K^{9}_{\
\ 9}+ \tilde K^{10}{}_{ 10}  \right) + \tilde R.
\eqno(2.3.2)
$$
Equating the group element $g$ in the Cartan subalgebra written in terms
of the two different set of generators we find that
$$
g = e^{\sum_{a=1}^{10}\tilde h^{a}_{\ a} \tilde K^{a}_{\ a}}
e^{\sigma \tilde R} =e^{\sum_{a=1}^{11}\phi_{\hat a} H_{\hat a}}
=e^{\phi_{1} \left(  \tilde K^{1}{}_{ 1}- \tilde K^{2}{}_{ 2}
\right)}
...e^{\phi_{9}\left(  \tilde K^{9}{}_{ 9} -  \tilde K^{10}{}_{ 10}
\right)}
$$
$$
\times e^{\phi_{10}\left( - {1 \over 8}\left(  \tilde K^{1}{}_{ 1} +...+
\tilde K^{9}{}_{ 9} \right) + {7 \over 8}  \tilde K^{10}{}_{ 10} - {3
\over
2}\tilde R\right)}
e^{\phi_{11}\left( - {1 \over 4}\left(  \tilde K^{1}{}_{ 1} +...+
\tilde K^{8}{}_{ 8} \right) + {3 \over 4} \left(  \tilde K^{9}{}_{ 9}
+  \tilde K^{10}{}_{
10} \right) + \tilde R \right)}.
\eqno(2.3.3)
$$
using equations (2.3.2). Comparing the coefficients of the generators
$\tilde R$ and
$\tilde K^{a}{}_{a}$  we find the physical fields are related to the
the $E_{11}$ Chevalley  fields $\phi_{\hat a}$  by
$$
\tilde{\sigma} =-{3 \over 2}\phi_{10} + \phi_{11},\quad
\tilde h^{1}{}_{ 1} = \phi_{1} - {1 \over 8}\phi_{10}
- {1 \over 4} \phi_{11} ,
$$
$$
\tilde h^{i}{}_{ i} = -\phi_{i-1} + \phi_{i} - {1 \over 8}\phi_{10}
- {1 \over 4} \phi_{11},\quad  {\rm  for} \quad  2 \leq i<9 ,
$$
$$
\tilde h^{9}{}_{ 9} = - \phi_{8} + \phi_{9} - { 1 \over 8  }
\phi_{10}  + {3 \over 4 } \phi_{11} ,\quad
\tilde h^{10}{}_{ 10} = - \phi_{9} + \phi_{10} + { 7 \over 8  }
\phi_{10}  + {3 \over 4 } \phi_{11}.
\eqno(2.3.4)
$$
\medskip
{\bf {2.3.1. Dimensionally reduced type IIA}}
\medskip
The theory in $d$ dimensions that results from dimensionally reducing
the IIA theory on an $n=10-d$ torus  is found in the
$E_{11}$ non-linear realisation by decomposing $E_{11}$  with respect to
the algebra that remains when one   deletes node
$d$ of the Dynkin diagram, after deletion of nodes 10 and 11, that is
the
subalgebra $GL(d)\otimes GL(n) \otimes GL(1)$.
$$
\matrix {
&&&&&&& & & & & & & & 11 & & 10 &&\cr
&&&&&&& & & & & & & & \otimes & & \otimes && \cr
&&&&&&& & & & & & & & | & & | &&\cr
\bullet & - & \ldots & - & \bullet & - & \otimes &  - &  \bullet & -
& \ldots &-&\bullet&-&\bullet&-&\bullet& & \cr
1 &&&&d-1&&d& & d+1& &  & & 7  &  & 8 &  & 9& & }
$$
\bigskip
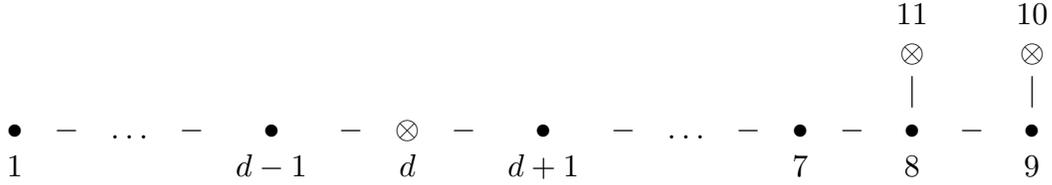
\centerline{Figure 8. The $E_{11}$ Dynkin diagram appropriate to
type IIA supergravity in $d=10-n$ dimensions}
\bigskip
The  $SL(d)$ subalgebra has the generators
$\tilde K^{a}_{\ \ a}$, $a=1,...,d$ and it is responsible in the
non-linear realisation for gravity in
$d$ dimensions. The gravity line consists of  nodes 1 to $d-1$
inclusive.   The $GL(1)$ factor has generator
$\tilde R$ while the   $GL(n)$ factor,  associated with
nodes
$d+1$ to
$9$,  has the generators
$$
\dot{\tilde K}{}^{i}{}_{j}=\tilde{K}^{i}{}_{j}-{1\over
8-n}\delta^{i}_{j}\sum_{a=1}^{d}\tilde{K}^{a}{}_{a}, \quad i,j=d
+1,...,10.
\eqno(2.3.5)$$
One can verify that these generators commute with those of $GL(d)$ and the
space-time translations in $d$ dimensions.
\par
It is evident from the Dynkin diagram of figure 8 that the theory
possess an $E_{n+1}$ symmetry arising from the nodes $d+1$ to 11. This
contains the subalgebra $ GL(n)\otimes GL(1)$. The Chevalley generators
in the Cartan subalgebra of $E_{n+1}$ are given by
$$
\tilde{T}_{i}=\dot{\tilde K}{}^{i}{}_{i}-\dot{\tilde K}{}^{i+1}{}_{i
+1} ,
\quad a=d+1,...,9,
$$
$$
\tilde{T}_{10}=-{1 \over 8}\left(
\dot{\tilde K}{}^{d+1}{}_{d+1}+...+\dot{\tilde K}{}^{9}{}_{9} \right)
+ {7
\over 8}\dot{\tilde K}{}^{10}{}_{10} - {3 \over 2} \tilde{R},
$$
$$
\tilde{T}_{11} =-{1 \over 4}\left( \dot{\tilde K}{}^{d+1}{}_{d+1}
+...+\dot{\tilde K}{}^{8}{}_{8}  \right) + {3 \over 4}\left(
\dot{\tilde K}{}^{9}{}_{9}+\dot{\tilde K}{}^{10}{}_{10}  \right) +
\tilde{R}.
\eqno(2.3.6)
$$
Substituting equations (2.3.5)  into equations
(2.3.6) one finds ${H}_{i}=\tilde{T}_{i}$ for $i=d+1,...,11$ where
$H_i$ are Chevalley generators of $E_{11}$.
\par
The $E_{11}$ group element restricted to the Cartan subgroup can
therefore  be written in the form
$$
k_{IIA}= e^{\sum_{a=1}^{d} \tilde{h}^{a}{}_{a} \tilde{K}^{a}{}_{a}+
\tilde
e_3 \tilde \rho \sum_{a=1}^{d}  \tilde{K}^{a}{}_{a}}
e^{\sum_{i=d+1}^{11} \dot{\varphi}_{i} \tilde{T}_{i}},
\eqno(2.3.7)
$$
where $\dot{\varphi}_{a}$ are the $E_{n+1}$  Chevalley fields that
parameterise the $E_{n+1}$ Cartan subalgebra.
\par
As for the reduction of  eleven dimensional supergravity, we must also
find the gravity fields in the theory in $d$ dimensions   corresponding
to our  compactification ansatz of equation  (2.1.3).
Following similar arguments as in that case we find the ansatz is
encoded in the non-linear realisation if we take the
$E_{11}$ group element, restricted to the Cartan subalgebra to be
given by
$$
g=e^{ \sum_{a=1}^{d}  \tilde{h}^{a}{}_{a} \tilde{K}^{a}{}_{a} +\tilde e_{1}\tilde{\rho}
\sum_ {a=1}^{d}\tilde{K}^{a}{}_{a}}e^{ \sum_{i=d+1}^{10} \tilde{h}^{i}{}_{i}
\tilde{K}^{i}{}_ {i} +\tilde e_{2}\tilde{\rho}
\sum_{i=d+1}^{10}\tilde{K}^{i}{}_{i}}e^{\tilde {R} \hat{\tilde
{\sigma}}}.
\eqno(2.3.8)
$$
\par
We are interested in the part of the ansatz of equation (2.13)
that involves  $<\tilde \rho>$ and so, following the discussion around
equations (2.2.11) and (2.2.13),  we take $\tilde e_1=0$,
$\tilde e_{2}=\beta (D=10)$, while $\tilde e_3= {n\tilde e_2\over 8-n}$.
\par
We now have two different way of expressing the $E_{11}$ group
element; the formulation of equation (2.3.7) and that of equation
(2.3.8). Equating these two and keeping only those parts associated with
the compactified directions we find that
$$
e^{ \left( \tilde{h}^{d+1}{}_{d+1}
+\tilde{e}_{2}\tilde{\rho} \right) \tilde{K}^{d+1}{}_{d+1} }...e^{
\left( \tilde{h}^{9}{}_{9}+\tilde{e}_{2}\tilde{\rho} \right)
\tilde{K}^{9}{}_{9} }e^{ \left(
\tilde{h}^{10}{}_{10}+\tilde{e}_{2}\tilde{\rho} \right)
\tilde{K}^{10}{}_{10} }e^{\tilde{\sigma} \tilde{R}}
$$
$$
=e^{{\tilde  T}_i \dot{\varphi}_i}
=  e^{\dot{\varphi}_{d+1} \left(\tilde{K}^{d+1}{}_{d+1}
-\tilde{K}^{d+2}{}_{d+2} \right)  }...e^{\dot{\varphi}_{8} \left(
\tilde{K}^{8}{}_{8}-\tilde{K}^{9}{}_{9}  \right) } e^{ \dot{\varphi}_{9}
\left( \tilde{K}^{9}{}_{9} - \tilde{K}^{10}{}_{10} \right)  }
$$
$$
e^{\dot{\varphi}_{10} \left( -{1\over 8} \left( \tilde{K}^{d+1}{}_{d+1}
+...+
\tilde{K}^{9}{}_{9} \right) + {7 \over 8} \tilde{K}^{10}{}_{10} -{3\over 2}
\tilde{R} \right)}  e^{\dot{\varphi}_{11} \left( -{1\over 4} \left(
\tilde{K}^{d+1}{}_{d+1} +...+ \tilde{K}^{8}{}_{8} \right) + {3\over 4}
\left(  \tilde{K}^{9}{}_{9} +
\tilde{K}^{10}{}_{10} \right) + \tilde{R} \right)}.
\eqno(2.3.9)
$$
Equating the coefficients of the generators we find the physical
fields in the $d$ dimensional type IIA theory in terms of the $E_{n+1}
$ fields
$$
\tilde{h}^{d+1}{}_{d+1}+\tilde{e}_{2} \tilde{\rho} = \dot{\varphi}_{d
+1} -
{1\over 8} \dot{\varphi}_{10} - {1\over 4} \dot{\varphi}_{11},
$$
$$
\tilde{h}^{d+2}{}_{d+2}+\tilde{e}_{2} \tilde{\rho} = -\dot{\varphi}_{d
+1} + \dot{\varphi}_{d+2} -
{1\over 8} \dot{\varphi}_{10} - {1\over 4} \dot{\varphi}_{11}, \ldots
$$
$$
\tilde{h}^{8}{}_{8}+\tilde{e}_{2} \tilde{\rho} = -\dot{\varphi}_{7} +
\dot{\varphi}_{8} -
{1\over 8} \dot{\varphi}_{10} - {1\over 4} \dot{\varphi}_{11}, \quad
\tilde{h}^{9}{}_{9}+\tilde{e}_{2} \tilde{\rho} = -\dot{\varphi}_{8} +
\dot{\varphi}_{9} -
{1\over 8} \dot{\varphi}_{10} + {3\over 4} \dot{\varphi}_{11},
$$
$$
\tilde{h}^{10}{}_{10}+\tilde{e}_{2} \tilde{\rho} = -\dot{\varphi}_{9} +
{7\over 8}\dot{\varphi}_{10} + {3\over 4} \dot{\varphi}_{11}, \quad
\tilde{\sigma} = -{3\over 2} \dot{\varphi}_{10} + \dot{\varphi}_{11}.
\eqno(2.3.10)
$$
Solving these equations  for the IIA $E_{n+1}$ Chevalley fields
$\dot{\varphi}_{i}$ in terms of the physical fields
$\tilde{h}^{d+1}{}_{d+1},
\tilde{h}^{d+2}{}_{d+2}, ... ,\tilde{h}^{10}{}_{10}, \tilde{\sigma},
\tilde{\rho} $ of the theory in $d$ dimensions  we find
$$
\dot{\varphi}_{i}= \tilde{h}^{d+1}{}_{d+1}
+ \tilde{h}^{d+2}{}_{d+2}+ ... + \tilde{h}^{i}{}_{i} + \left( n-10+i
\right){8 \over 8-n} \tilde{e}_{2} \tilde{\rho}, \ \ for \ i=d+1,...,8.
$$
$$
\dot{\varphi}_{9}=  \tilde{h}^{d+1}{}_{d+1} + \tilde{h}^{d+2}{}_{d+2}
+ ... + \tilde{h}^{9}{}_{9} + {5n-8\over
8-n} \tilde{e}_{2} \tilde{\rho} -{1\over 4} \tilde{\sigma},
$$
$$
\dot{\varphi}_{10}=-{1\over 2} \tilde{\sigma} + {2\over 8-n}n
\tilde{e}_{2}
\tilde{\rho},
$$
$$
\dot{\varphi}_{11}={1\over 4} \tilde{\sigma} + {3\over 8-n}n \tilde{e}
_{2}
\tilde{\rho}.
\eqno(2.3.11)
$$

\medskip
{\bf {2.3.2. Type IIA parameters}}
\medskip
Finally we can express the parameters
$V_{n(A)}$, $g_{d}$, ${r_{d+1}\over l_d}$ and the volume $V_{j}$ of the $j$ dimensional subtorus of $V_{n(A)}$ in terms of the expectation
values, not explicitly shown, of the $E_{n+1}$   Chevalley fields
$\dot{\varphi}_{i}$.  For the volume of
the IIA torus $V_{n(A)}$ one finds from equations (2.1.16) and (2.3.11) 
$$
V_{n(A)}=(2 \pi)^{n} {r_{10}r_{9}...r_{d+1} \over l_{d}^{n}}
=e^{\left(8 \over 8-n  \right)n \beta \tilde{\rho}}=e^{\left( \dot
{\varphi}_{10} + 2
\dot{\varphi}_{11} \right)}.
\eqno(2.3.12)
$$
Thus the IIA volume is related to the expectation value of the
Chevalley fields
$\dot{\varphi}_{10}$ and $\dot{\varphi}_{11}$ which are associated with
nodes ten and eleven of the $E_{11}$ Dynkin diagram.
\par
The effective coupling in
$d$ dimensions is given in equation (2.2.21) and may be written
$$
g_{d}=e^{{8-n \over 8} \sigma
- {n \beta \tilde{\rho} \over 2}}=e^{-2\left({8-n \over 8}
\right)\dot{\varphi}_{10}}.
\eqno(2.3.13)
$$
the string coupling in $d$ dimension is associated with the
expectation value of the Chevalley field $\dot{\varphi}_{10}$ and so
with
node ten of the $E_{11}$ Dynkin diagram.
\par
Equations (2.1.15), (2.1.16) and (2.1.6) may
be used to show that the ratio of the radius of the circle in the $d+1$
direction
$r_{d+1}$ to the
$d$ dimensional Planck length $l_{d}$ is given by
$$
{r_{d+1} \over l_{d}}={l_{10} \over l_{d}}{r_{d+1} \over l_{10}}
=e^{{8 \over 8-n} \beta \tilde{\rho} +
\tilde{h}^{d+1}{}_{d+1}}=e^{\dot{\varphi}_{d+1}}.
\eqno(2.3.14)
$$
Thus this last parameter is related to the expectation value
of the Chevalley field $ \dot{\varphi}_{d+1}$ and so with node $d+1$
of the $E_{11}$ Dynkin diagram.
As in the M-theory case, nodes $d+1$ to $8$ are associated with
subtori of the $n+1$ torus.  The dimensionless volume of the $j$
torus $V_{j}$ contained within the $n$ torus is defined in equation
(2.1.10) and may be written
$$
V_{j}=(2\pi)^{j} \left({l_{10} \over l_{d} }\right)^{j} {r_{d+1} r_{d+2} ... r_{d
+j} \over \left( l_{10} \right)^{j}}=e^{j\left({ 8 \over 8-n}\right)
\tilde{\beta} \tilde{\rho} + \tilde{h}^{d+1}{}_{d+1} + \tilde{h}^{d+2}
{}_{d+2} +...+ \tilde{h}^{d+j}{}_{d+j}}=e^{\dot{\varphi}_{d+j}},
\eqno(2.3.15)
$$
where we have made use of equations (2.1.15), (2.1.16), (2.1.6) and equations
(2.3.11) to express the $d$ dimensional type IIA physical fields in
terms of the $E_{n+1}$ Chevalley field.
Similarly, the volume of the $n-1$ subtorus is given by
$$
V_{n-1}=(2 \pi)^{n-1} \left({l_{10} \over l_{d} }\right)^{n-1} {r_{d+1} r_{d+2} ...
r_{9} \over \left( l_{10} \right)^{n-1}}=e^{\left(n-1\right)\left({ 8
\over 8-n}\right) \tilde{\beta} \tilde{\rho} + \tilde{h}^{d+1}{}_{d
+1} + \tilde{h}^{d+2}{}_{d+2} +...+ \tilde{h}^{9}{}_{9}}
$$
$$
=e^{\dot
{\varphi}_{9}+ \dot{\varphi}_{11}}.
\eqno(2.3.16)
$$
It follows that each node of the $E_{n+1}$ Dynkin diagram is
associated with a specific combination of the $n+1$ parameters $V_{i}$, $i=1,...,n$, and $g_{d}$ as given below each node in
figure 9.
$$
\matrix {
 & & & & & & V_{n (A)}^{{1 \over 2}}g_{d}^{{2 \over 8-n}} & &  &&\cr
 & & & & & & \bullet & & && \cr
 & & & & & & | & & &&\cr
  \bullet & - &
\ldots &-&\bullet&-&\bullet&-&\bullet&-&\bullet \cr
 {r_{d+1} \over l_{d}}& &  & & V_{n-3}  &  & V_{n-2} &  & V_{n-1} V_{n(A)}^{-{1 \over 2}}g_{d}^{-{2 \over 8-n}} &&
g_{d}^{-{4 \over 8-n}} }
$$
\bigskip
\centerline{Figure 9. The $E_{n+1}$ Dynkin diagram labelled by the $d
$ dimensional type IIA parameters}
\bigskip

\medskip
{\bf {2.4 Parameters and fields in type IIB supergravity} }
\medskip
The $E_{11}$ formulation of type IIB supergravity emerges from  the
non-linear realisation of $E_{11}$ after decomposing the $E_{11}$ algebra
in terms of the algebra that results from deleting the node labelled
nine
in the $E_{11}$ Dynkin diagram in figure 10, namely the subalgebra
$GL(10)\otimes SL(2)$.
$$
\matrix {
& & & & & & & & \bullet &10 &  \cr
& & & & & & & & | & &  \cr
& & & & & & & & \otimes &9 &  \cr
& & & & & & & & | & & \cr
\bullet &  - &  \ldots & -
&\bullet&-&\bullet&-&\bullet&-&\bullet&
\cr 1& & & & 6 & & 7  &  & 8 &  & 11 }
$$
\bigskip
\centerline{Figure 10. The $E_{11}$ Dynkin diagram appropriate to type
IIB
supergravity in $10$ dimensions}
\bigskip
The $GL(10)$ factor gives rise to ten dimensional gravity and the
the gravity line in the type IIB
theory that consists of nodes one to eight in addition to node 11. The
$SL(2)$ factor arises from node ten and it is the $SL(2)$ symmetry of
the IIB theory.  The
$GL(10)$ subalgebra is generated by
$\hat{K}^{a}{}_{b}$, $
 a,  b=1,\ldots ,10$.  Together with the generator  $\hat{R}$ of the
$SL(2)$ symmetry, the generators $\hat{K}^{a}{}_{a}$, $a=1,...,10$
provide a basis for the Cartan subalgebra generators of $E_{11}$.  The
relation between the Cartan subalgebra generators of the $E_{11}$
algebra in Chevalley basis, the $H_{\hat a}$, and the above
generators is
given by [44]
$$
{H}_{a}=\hat{K}^{a}{}_{a}-\hat{K}^{a+1}{}_{a+1} , \  \ a=1,...,8,
$$
$$
{H}_{9}=\hat{K}^{9}{}_{9} + \hat{K}^{10}{}_{10} - {1 \over 4}
\sum_{a=1}^{10}\hat{K}^{a}{}_{a} +
\hat{R},
$$
$$
{H}_{10}=-2\hat{R},
$$
$$
{H}_{11}=\hat{K}^{9}{}_{9}-\hat{K}^{10}{}_{10}.
\eqno(2.4.1)
$$
\par
The $E_{11}$ group  element restricted to the
Cartan subalgebra takes the usual form, that is  $e^{ \sum_{a} \hat{\phi}_{a}
{H}_{a}}$, where $\hat{\phi}$ are the fields associated with the $E_
{11}$
Chevalley generators ${H}_{a}$ in the Cartan subalgebra which are the
same
no matter what theory we are considering. However, we can also express
this group element in terms of the above generators as
$$
g=e^{\sum_{a} h_{ a}{}^{  a}K^{ a}{}_{  a}}e^{\hat{R} \hat{\sigma}},
\eqno(2.4.2)
$$
where  $\hat{\sigma}$ is the type IIB dilaton. Equating these two group
elements we find that
$$
e^{ \sum_{a} \hat{\phi}_{ a} \hat{H}_{ a}}= e^{  \sum_{a} h_{ a}{}^{ a} K^{
a}{}_{ a}}e^{\hat{\sigma} \hat{R}}.
\eqno(2.4.3)
$$
Comparing the coefficients of the generator $R$ we find
$$
\hat{\phi}_{10}=-{1 \over 2} \hat{\sigma},
\eqno(2.4.4)
$$
one may similarly compare the coefficients of  the
$GL(10)$ generators to find an expression for the rest of the $E_{11}$
Chevalley fields in terms of the physical fields $\hat{h}^{a}{}_{a}$,
$a=1,...,10$.
\medskip
\noindent
{\bf {2.4.1. Dimensionally reduced type IIB}}
\medskip
The theory obtained from the dimensionally reduced type IIB theory on
an $n=10-d$ torus appears if one decomposes $E_{11}$ into the sub-algebra
that arises when one deletes node
$d$ of the Dynkin diagram of figure 11, that is the algebra
$GL(d)\otimes SL(2)\otimes GL(9-d) $.
 $$
\matrix {
&&&&&&& & & & & & & & \bullet & 10&  \cr
&&&&&&& & & & & & & & | & &  \cr
&&&&&&& & & & & & & & \otimes &9 & \cr
&&&&&&& & & & & & & & | & & \cr
\bullet & - & \ldots & - & \bullet & - & \otimes &  - &
\bullet & - & \ldots &-&\bullet&-&\bullet&-&\bullet \cr
1 &&&&d-1&&d& & d+1& &  & & 7  &  & 8 &  & 11}
$$
\bigskip
\centerline{Figure 11. The $E_{11}$ Dynkin diagram
appropriate to type IIB supergravity in $d=10-n$ dimensions}
\bigskip
The $GL(d)$ algebra has the generators $\hat K^{a}{}_{b}$, $a,b=1,...,d$
and  gives rise to ten dimensional
gravity. The
$GL(9-d)$ algebra  arises from nodes $d+1$ to eight inclusive and node
nine.  the generators of $GL(9-d)$ are given by
$$
\dot{\hat K}{}^{i}{}_{j}=\hat{K}^{i}{}_{j}-{1 \over
8-n}\delta^{i}_{j}\sum_{a=1}^{d}\hat{K}^{a}{}_{a}, \quad
i,j=d+1,...,10,
\eqno(2.4.5)$$
and one can verify that they commute with those of $GL(d)$, $SL(2)$ and the
space-time translations in $d$ dimensions.
\par
As is obvious from the Dynkin diagram of figure 11 the theory in $d$
dimensions has an $E_{n+1}$ symmetry which contains the $ GL(9-d)\otimes
SL(2)$ symmetry. The generators in the Cartan subalgebra of $E_{n+1}$
are
given by
$$
\hat{T}_{i}=\dot{\hat K}{}^{i+1}{}_{i+1} - \dot{\hat K}^{i+2}{}_{i+2}
\quad {\rm for}\quad   i=d+1,...,8,
$$
$$
\hat{T}_{9}=-{1\over 4}\left( \sum_{i=d+1}^{8} \dot{\hat K}{}^{i}{}_{i}
\right) + {3
\over 4} \left( \dot{\hat K}{}^{9}{}_{9} + \dot{\hat K}{}^{10}{}_{10}
\right) +
\hat{R},
$$
$$
\hat{T}_{10}=-2\hat{R}, \quad
\hat{T}_{11} = \dot{\hat K}{}^{9}{}_{9} - \dot{\hat K}{}^{10}{}_{10}.
\eqno(2.4.6)
$$
Substituting the equation (2.4.5)  into equations
(2.4.6) one finds ${H}_{i}=\hat{T}_{i}$ for $i=d+1,...,11$. This
equality between the Cartan subalgebra elements of $E_{11}$ and those of
$E_{n+1}$ is to be expected as the
$E_{n+1}$ algebra is the same algebra regardless as whether the
theory in
$d$ dimensions  is found from dimensional reduction of the eleven,
IIA or
IIB supergravity theories. This fact is particularly obvious once one
looks at the corresponding Dynkin diagrams of figures 5, 8 and 11 that
specify
the subalgebras used to find the non-linearly realised theory.
\par
As a result the  group element of $E_{11}$, viewed from the IIB
perspective and  restricted to the Cartan subalgebra, $k_{IIB}$ can  be
written in the form
$$
k_{IIB}= e^{\sum_{a=1}^{d} \hat{h}^{a}{}_{a} \hat{K}^{a}{}_{a}
+\hat e_3\hat \rho  \sum_{a=1}^{d}  \hat{K}^{a}{}_{a}}
e^{\sum_{i=d+1}^{11} \dot{\varphi}_{i} \hat{T}_{i}},
\eqno(2.4.7)
$$
where $\hat e_3$ is a constant and $\dot{\varphi}_{a}$ are  the $E_{n
+1}$
Chevalley fields that parameterise the $E_{n+1}$ Cartan subalgebra.
\par
As for the case of the dimensional reduction of the eleven dimensional and type IIA supergravity
theories the gravity fields in eleven dimensions and ten dimensions, respectively, are related to those in
$d$ dimensions by factors of exponentials of the scalar fields as given
in the ansatz of equation (2.1.3).  Following similar arguments as for
these previous cases we  take the group element to be given by
$$
e^{  \sum_{a=1}^{d} \hat{h}^{a}{}_{a} \hat{K}^{a}{}_{a}
+e_{1}\hat{\rho} \sum_{a=1}^{d}\hat{K}^{a}{}_{a}}e^{ \sum_{i=d+1}^{10} \hat{h}^{i}{}_{i}
\hat{K}^{i}{}_{i} +e_{2}\hat{\rho}
\sum_{i=d+1}^{10}\hat{K}^{i}{}_{i}}e^{\hat{R} \hat{\sigma}}.
\eqno(2.4.8)
$$
Following the analogous discussions for the M-theory and type IIA cases
we take
$\hat e_1= \alpha(D=10)$, $\hat e_2=\beta (D=10)$ and $\hat e_3={n\hat e_2\over 8-
n}$ .
\par
We can now equate the two ways of writing the $E_{11}$ group element,
restricted to the Cartan subalgebra and for $d<9$ dimensions;  that of
equations (2.4.7) and (2.4.8). Keeping only parts not involving the
$SL(d)$ generators we find the
relations
$$
e^{ \left( \hat{h}^{d+1}{}_{d+1}+e_{2}\hat{\rho} \right) \hat{K}^{d+1}
{}_{d+1} }...e^{   \left( \hat{h}^{9}{}_{9}+\hat{e}_{2} \hat{\rho}
\right) \hat{K}^{9}{}_{9} }e^{ \left( h^{10}{}_{10}+\hat{e}_{2} \hat
{\rho} \right) \hat{K}^{10}{}_{10} }e^{\hat{\sigma} \hat{R}}
=e^{\dot \varphi _iT_i}$$
$$
=  e^{\dot{\varphi}_{d+1} \left(\hat{K}^{d+1}{}_{d+1} -\hat{K}^{d+2}{}
_{d+2} \right)  }...e^{\dot{\varphi}_{9}\left( - {1 \over 4} \sum_{i=d
+1}^{8} \hat{K}^{i}{}_{i} + {3 \over 4} \left( \hat{K}^{9}{}_{9} +
\hat{K}^{10}{}_{10} \right)  \right)   }e^{ - 2 \dot{\varphi}_{10}
\hat{R}   } e^{\dot{\varphi}_{11} \left( \hat{K}^{9}{}_{9} - \hat{K}^
{10}{}_{10} \right) }.
\eqno(2.4.9)
$$
Using the relations of equation (2.4.6) and equating the
coefficients  of
the generators gives the relations between the $d$ dimensional physical
fields and the $E_{n+1}$ Chevalley fields
$$
\hat{h}^{d+1}{}_{d+1}+\hat{e}_{2} \hat{\rho}
= \dot{\varphi}_{d+1} -{1 \over 4} \dot{\varphi}_{9} , \quad
\hat{h}^{d+2}{}_{d+2}+\hat{e}_{2} \hat{\rho}  = -\dot{\varphi}_{d+1} +
\dot{\varphi}_{d+2} - {1 \over 4} \dot{\varphi}_{9}, \ldots
$$
$$
h^{8}{}_{8}+\hat{e}_{2} \hat{\rho}  = -\dot{\varphi}_{7} +
\dot{\varphi}_{8} - {1 \over 4} \dot{\varphi}_{9}, \quad
h^{9}{}_{9}+\hat{e}_{2} \hat{\rho} = -\dot{\varphi}_{8}
+ {3 \over 4} \dot{\varphi}_{9} + \dot{\varphi}_{11},
$$
$$
h^{10}{}_{10}+\hat{e}_{2} \hat{\rho} = {3 \over 4}\dot{\varphi}_{9} -
\dot{\varphi}_{11},
\hat{\sigma} = -2 \dot{\varphi}_{10}.
\eqno(2.4.10)
$$
Solving these equations for the $E_{n+1}$ fields in terms of the
scalars $\hat{\sigma}$, $\hat{h}^{d+1}{}_{d+1},...,\hat{h}^{10}{}_{10}
$ and $\hat{\rho}$ found upon dimensional reduction yields
$$
\dot{\varphi}_{i}= \hat{h}^{d+1}{}_{d+1} + \hat{h}^{d+2}{}_{d+2}
+ ... + \hat{h}^{i}{}_{i} + \left( n-10+i \right){8 \over 8-n}
\hat{e}_{2} \hat{\rho}, \ \ d+1\leq i < 8.
$$
$$
\dot{\varphi}_{9}= {4n \over 8-n}\hat{e}_{2} \hat{\rho}, \quad
\dot{\varphi}_{10}=-{1 \over 2} \hat{\sigma} + {2 \over 8-n}n \hat{e}_
{2}
\hat{\rho}, \quad
\dot{\varphi}_{11}=\sum_{a=d+1}^{9} h^{a}{}_{a} + {4\left(n-2\right)
\over 8-n} \hat{\rho}.
\eqno(2.4.11)
$$
\medskip
{\bf {2.4.2. Type IIB parameters}}
\medskip
The Volume of the torus in
the type IIB theory $V_{n(B)}$, the $d$ dimensional effective coupling
$g_{d}$, the ratio of the radius $r_{d+1}$ to the Planck length
$l_{d}$ and the volume $V_{j}$ of the $j$ dimensional subtorus of $V_{n(B)}$ are expressible in terms of the $E_{n+1}$ Chevalley fields
$\dot{\varphi}_{i}$.   For the volume of
the IIB torus $V_{n(B)}$ one finds from equations (2.1.16) and (2.4.11),
$$
V_{n(B)}=(2 \pi)^{n} {r_{10}r_{9}...r_{d+1} \over l_{d}^{n}}=e^{\left( 8 \over 8-
n \right)n\beta \hat
{\rho}}=e^{2 \dot{\varphi}_{9}}.
\eqno(2.4.12)
$$
Thus the volume on which the IIB theory is compactified is related to
the
expectation value of the Chevalley field $\dot{\varphi}_{9}$ which is
associated with node nine of the $E_{11}$ Dynkin diagram.
\par
The effective coupling in $d$ dimensions, given  in equation (2.1.24),
may be written
$$
g_{d}=e^{{8-n \over 8} \dot{\varphi} - {n \beta \hat{\rho} \over 2}}
=e^{-2\left({8-n \over 8} \right)\dot{\varphi}_{10}}.
\eqno(2.4.13)
$$
thus the string coupling constant of the IIB string theory in $d$
dimensions is related to the
expectation value of the Chevalley field $\dot{\varphi}_{9}$ which is
associated with node ten of the Dynkin diagram.
\par
Equations (2.1.15),(2.1.16) and (2.1.6) may be used to rewrite the ratio of the
radius of the circle in the $d+1$ direction $r_{d+1}$ to the $d$ dimensional
Planck length $l_{d}$ as
$$
{r_{d+1} \over l_{d}}={l_{10} \over l_{d}} {r_{d+1} \over l_{10}}=e^
{{8 \over 8-n} \beta \hat{\rho} + \hat{h}^{d+1}{}_{d+1}}=e^{\dot
{\varphi}_{d+1}}.
\eqno(2.4.14)
$$
Thus the above ratio is related to the expectation value of the
Chevalley
fields $\dot{\varphi}_{d+1}$ which is
associated with node $d+1$  of the Dynkin diagram.
\par
Again, nodes $d+1$ to $8$ are associated with subtori of the $n+1
$ torus.  The dimensionless volume of the $j$ torus $V_{j}$ contained
within the $n$ torus is defined in equation (2.1.10) and may be written
$$
V_{j}=(2 \pi)^{j} \left({l_{10} \over l_{d} }\right)^{j} {r_{d+1} r_{d+2} ... r_{d
+j} \over \left( l_{10} \right)^{j}}=e^{j\left({ 8 \over 8-n}\right)
\hat{\beta} \hat{\rho} + \hat{h}^{d+1}{}_{d+1} + \hat{h}^{d+2}{}_{d
+2} +...+ \hat{h}^{d+j}{}_{d+j}}=e^{\dot{\varphi}_{d+j}},
\eqno(2.4.15)
$$
where we have made use of equations (2.1.15), (2.1.16), (2.1.6) and equations
(2.4.11) to express the $d$ dimensional type IIB physical fields in
terms of the $E_{n+1}$ Chevalley field.
Similarly, the volume of the $n-1$ subtorus is given by
$$
V_{n-1}= ( 2 \pi)^{n-1}  \left({l_{10} \over l_{d} }\right)^{n-1} {r_{d+1} r_{d+2} ...
r_{9} \over \left( l_{10} \right)^{n-1}}=e^{\left(n-1\right)\left({ 8
\over 8-n}\right) \hat{\beta} \hat{\rho}  + \hat{h}^{d+1}{}_{d+1} +
\hat{h}^{d+2}{}_{d+2} +...+ \hat{h}^{9}{}_{9}}=e^{\dot{\varphi}_{9}+
\dot{\varphi}_{11}}.
\eqno(2.4.16)
$$
It follows that each node of the $E_{n+1}$ Dynkin diagram is
associated with a specific combination of the $n+1$ parameters $V_{i}$, $i=1,...,n$, and $g_{d}$ as given below each node in
figure 12.
$$
\matrix {
 & & & & & & V_{n-1} V_{n(B)}^{-{
1 \over 2}} & &  &&\cr  & & & & & & \bullet & & && \cr
 & & & & & & | & & &&\cr
  \bullet & - &
\ldots &-&\bullet&-&\bullet&-&\bullet&-&\bullet \cr
 {r_{d+1} \over l_{d}}& &  & & V_{n-3}  &  &
V_{n-2} &  & V_{n (B)}^{{1 \over 2}} &&
g_{d}^{-{4 \over 8-n}} }
$$
\bigskip
\centerline{Figure 12. The $E_{n+1}$ Dynkin diagram labelled by the $d
$ dimensional type IIB parameters}
\bigskip

\bigskip
\noindent
{\bf {3. Relations between the fields and parameters in the
different theories}}
\bigskip
The maximal string theory in $d$ dimensions is unique, however, it can be derived by dimensional reduction on a torus from the type IIA theory or the IIB
string theories or even the eleven dimensional M-theory. The parameters of the theory in $d$ dimensions can be thought of arising from the volume of the torus and all its subvolumes, and in the case of the ten dimensional theories also the expectation value of the dilaton. As such for dimensional reduction from M-theory, say, one finds a set of parameters that are related in a one to one way with those found by dimensional reduction from the IIB theory. 
 In the previous section we derived the relations between the parameters and the expectation values of the fields associated with the Cartan subalgebra. In this section we will first find the relations between the fields in the $d$-dimensional theory, as they arise from the different dimensional reductions from eleven dimensions and the IIA and IIB theories. This is straightforward to do using the $E_{11}$ formulation of the theories as the fields appear in the $E_{11}$ group element and, as the group element is the same no matter what dimensional reduction is used, we can simply equate the different group elements to find the desired relations between the fields. 
We will then find how the parameters in the theory in $d$ dimensions are given in terms of the expectation values of the fields occurring in the different possible dimensional reductions.  As a result we can then find the relations between the different descriptions of the parameters in the $d$-dimensional theory arising from the different dimensional reductions.  
\medskip
\noindent
{\bf {3.1. M-theory and IIA}}
\medskip
In the $E_{11}$ non-linear realisation of eleven dimensional
supergravity
dimensionally reduced on an $m=n+1$ torus the group element restricted
to the Cartan subalgebra and written in terms of the physical fields was
given in equation (2.2.9), that is,
$$
g_{M}=e^{\sum_{a=1}^{d} h^{a}{}_{a} K^{a}{}_{a} + e_{1} \rho \sum_
{a=1}^{d} K^{a}{}_{a}}e^{ \sum_{i=d+1}^{11} h^{i}{}_{i} K^{i}{}_{i} +
e_{2} \rho \sum_{i=d+1}^{11} K^{i}{}_{i} }.
\eqno(3.1.1)
$$
On the other hand the $d$ dimensional theory obtained from the ten
dimensional IIA supergravity by dimensionally reducing on an $n$ torus
possesses a  group element,  which when restricted
to the Cartan subalgebra and written in terms of the physical fields,
take the form given in (2.3.8), that is,
$$
g_{IIA}=e^{\sum_{a=1}^{d} \tilde{h}^{a}{}_{a} \tilde{K}^{a}{}_{a} +
\tilde{e}_{1} \tilde{\rho} \sum_{a=1}^{d} \tilde{K}^{a}{}_{a}}e^
{ \sum_{i=d+1}^{10} \tilde{h}^{i}{}_{i} \tilde{K}^{i}{}_{i} + e_{2}
\tilde{\rho} \sum_{i=d+1}^{10} \tilde{K}^{i}{}_{i} } e^{\tilde
{\sigma} \tilde{R}}.
\eqno(3.1.2)
$$
The relationship between the M-theory generators in $d$
dimensions $K^{a}{}_{a}$, $a=1,...,d$, $K^{i}{}_{i}$, $i=d+1,...,11$ and
the type IIA physical generators in $d$ dimensions $\tilde{K}^{a}{}_
{a}$,
$a=1,...,d$, $\tilde{K}^{i}{}_{i}$, $a=d+1,...,10$, $\tilde{R}$ may be found by
equating the Cartan subalgebra generators $H_{a}$, $a=1,...,11$ of each
theory. These expressions were given in equations (2.2.1) and (2.3.2) and using
these one finds that [52]
$$
K^{a}{}_{a} = \tilde{K}^{a}{}_{a}, \quad a=1,...,d,
\quad
K^{i}{}_{i} =\tilde{K}^{i}{}_{i}, \quad i=d+1,...,10,
\quad
K^{11}{}_{11} = {1 \over 8} \sum_{i=1}^{10}\tilde{K}^{i}{}_{i} + {3
\over 2} \tilde{R}.
\eqno(3.1.3)
$$
The theory in $d$ dimensions is unique as is evident from viewing
the $E_{11}$ Dynkin diagrams of figures 2 and 4 for M-theory and type IIA respectively. As such we may equate the M-theory and
type IIA group elements,
$$
g_{M}=g_{IIA}.
\eqno(3.1.4)
$$
\par
One may use equations (3.1.3) to substitute for  either the M-theory, or
type IIA physical generators, in equation (3.1.4) and read off the
relationship between the physical fields of these theories by equating
the coefficients of the generators.  This gives
$$
\tilde{h}^{a}{}_{a} + \tilde{e}_{2} \tilde{\rho} = h^{a}{}_{a} + e_
{2} \rho + {1 \over 8}\left( h^{11}{}_{11} + e_{2} \rho \right),
\quad a=1,...,d,
$$
$$
\tilde{h}^{i}{}_{i} + \tilde{e}_{2} \tilde{\rho} = h^{i}{}_{i} + {1
\over 8} h^{11}{}_{11} + {9 \over 8} e_{2} \rho, \quad i=d+1,...,10,
$$
$$
\tilde{\sigma} = {3 \over 2} \left( h^{11}{}_{11} + e_{2} \rho  \right).
\eqno(3.1.5)
$$
\par
The volume of the type IIA  torus $V_{n (A)}$ is given as a function of
the type IIA physical field $\tilde{\rho}$ and the $E_{n+1}$
Chevalley fields in equation (2.3.12).  In the following equation we
recall
these results, but express the result when
written in terms of physical IIA fields as the physical M-theory
fields using equation (3.1.5) and the $d$ dimensional parameters $V_{j(M)}$, $j=1,...,m$ arising from the dimensional reduction of M-theory on an $m$ torus
$$
V_{n(A)}=e^{\left( {8 \over 8-n} \right) n \tilde{e}_{2} \tilde{\rho}}
=e^{\left( \dot{\varphi}_{10}+ 2\dot{\varphi}_{11}
\right)}=e^{ \sum_{i=d+1}^{10} h^{i}{}_{i} + {9 \over 8-n} n e_{2} \rho}=V_{m-1(M)}.
\eqno(3.1.6)
$$
One can verify that one finds the same result if one expresses the
Chevalley fields in terms of the M-theory physical fields using equation
(2.2.16). Thus we see that the volume of the type IIA torus $V_{n(A)}$
has
the same expression in terms of the $E_{n+1}$ Chevalley fields from both
the type IIA and M-theory perspectives, as must be the case due to the uniqueness of the Chevalley fields.  
\par
A similar story applies to the volume of the M-theory torus which was
given in terms of the M-theory physical fields and
Chevalley fields in equation (2.2.17), that is
$$
V_{m(M)}= e^{ \left( {9 \over 8-n} \right)(n+1) e_{2} \rho}
=e^{3 \dot{\varphi}_{11}}=e^{ \left( {9 \over 8-n} \right) n\tilde{e}_{2} \tilde{\rho}
+  { 3 \over 4} \tilde{\sigma}}=V_{n(A)}^{{3 \over 2}}g_{d}^{{6 \over 8-n }}.
\eqno(3.1.7)
$$
In the final equalities we have converted the volume of the M-theory torus
into the IIA physical fields using equation (3.1.5) and the $d$ dimensional parameters $V_{n(A)}$, $V_{j}$, $j=1,...,n-1$ and $g_{d}$ arising from the dimensional reduction of type IIA on an $n$ torus. One can easily
verify that one finds the same result using equation (2.3.11) to convert
the Chevalley fields into the type IIA physical fields.
\par
We note that the last equation of (3.1.5) implies  the
standard relationship between the IIA coupling $g_{s
(A)}=e^{\tilde{\sigma}}$ in ten dimensions, the eleven dimensional
Planck
length $l_{11}$ and the radius of the compactified eleventh dimension
$r_{11}$ given in equation (2.1.27); indeed one finds the well known
result
$$
g_{s (A)}^{{2 \over 3}}=e^{{2 \over 3} \tilde{\sigma}}
= e^{ \left( h^{11}{}_{11} + e_{2} \rho  \right) } ={r_{11} \over l_
{11}}.
\eqno(3.1.8)
$$
\medskip
\noindent
{ \bf {3.2. IIA and IIB}}
\medskip
In the $E_{11}$ non-linear realisation of type IIA
supergravity
dimensionally reduced on an $n$ torus the group element restricted
to the Cartan subalgebra and written in terms of the physical fields was
given in equation (2.3.8), that is,
$$
g_{IIA}=e^{\sum_{a=1}^{d} \tilde{h}^{a}{}_{a} \tilde{K}^{a}{}_{a} +
\tilde{e}_{1} \tilde{\rho} \sum_{a=1}^{d} \tilde{K}^{a}{}_{a}}e^
{ \sum_{i=d+1}^{10} \tilde{h}^{i}{}_{i} \tilde{K}^{i}{}_{i} + \tilde{e}_{2}
\tilde{\rho} \sum_{i=d+1}^{10} \tilde{K}^{i}{}_{i} } e^{\tilde
{\sigma} \tilde{R}}.
\eqno(3.2.1)$$
On the other hand the $d$ dimensional theory obtained from ten
dimensional type IIB supergravity by dimensionally reducing on an $n$ torus
possesses a  group element, which when restricted
to the Cartan subalgebra and written in terms of the physical fields,
that takes the form given in (2.4.8), that is,
$$
g_{IIB}=e^{\sum_{a=1}^{d} \hat{h}^{a}{}_{a} \hat{K}^{a}{}_{a} + \hat{e}_{1}
\hat{\rho} \sum_{a=1}^{d} \hat{K}^{a}{}_{a}}e^{ \sum_{i=d+1}^{10} \hat
{h}^{i}{}_{i} \hat{K}^{i}{}_{i} + \hat{e}_{2} \hat{\rho} \sum_{i=d+1}^{10}
\hat{K}^{i}{}_{i} }e^{\hat{\sigma} \hat{R}}.
\eqno(3.2.2)
$$
The relationship between the  type IIA generators in $d$
dimensions $\tilde{R}$, $\tilde{K}^{a}{}_{a}$, $a=1,...,d$, $\tilde{K}^{i}{}_{i}$, $i=d+1,...,10$, and
the type IIB physical generators in $d$ dimensions $\hat{R}$, $\hat{K}^{a}{}_
{a}$,
$a=1,...,d$, $\hat{K}^{i}{}_{i}$, $i=d+1,...,10$, $R$ may be found by
equating the Cartan subalgebra generators $H_{a}$, $a=1,...,11$ of each
theory. These expression were given in equations (2.3.2) and (2.4.1) and using
these one finds that [52]
$$
\hat{K}^{a}{}_{a} = \tilde{K}^{a}{}_{a}, \quad a=1,...,d,
\quad
\hat{K}^{i}{}_{i} =\tilde{K}^{i}{}_{i}, \quad i=d+1,...,10,
\quad
\hat{K}^{10}{}_{10}={1 \over 4}\sum_{i=1}^{9} \tilde{K}^{i}{}_{i} -
{3 \over 4} \tilde{K}^{10}{}_{10} - \tilde{R},
$$
$$
\hat{R} = {1 \over 16}\sum_{i=1}^{9}\tilde{K}^{i}{}_{i} - {7 \over
16} \tilde{K}^{10}{}_{10} + {3 \over 4} \tilde{R}
\eqno(3.2.3)
$$
The theory in $d$ dimensions is unique as is evident from viewing
the $E_{11}$ Dynkin diagrams of figures 8 and 11 for the type IIA and type IIB supergravity theories respectively. As such we may equate the type IIA and
type IIB group elements,
$$
g_{IIA}=g_{IIB},
\eqno(3.2.4)
$$
\par
One may use equations (3.2.3) to substitute for either the type IIA, or
type IIB physical generators, in equation (3.2.4) and read off the
relationship between the physical fields of these theories by equating
the coefficients of the generators.  This gives
$$
\tilde{h}^{a}{}_{a} + \tilde{e}_{2} \tilde{\rho} = \hat{h}^{a}{}_{a}
+ \hat{e}_{2} \hat{\rho} + {1 \over 4} \left( \hat{h}^{10}{}_{10} +
\hat{e}_{2} \hat{\rho} \right), \quad a=1,...,d,
$$
$$
\tilde{h}^{i}{}_{i} + \tilde{e}_{2} \tilde{\rho} = \hat{h}^{i}{}_{i}
+ {1 \over 4}\hat{h}^{10}{}_{10} + {5 \over 4} \hat{e}_{2} \hat{\rho}
+ {1 \over 16} \hat{\sigma}, \quad i=d+1,...,9,
$$
$$
\tilde{h}^{10}{}_{10}+ \tilde{e}_{2} \tilde{\rho} = -{3 \over 4}\left
( \hat{h}^{10}{}_{10} + \hat{e}_{2} \hat{\rho}  \right) - {7 \over
16} \hat{\sigma}
$$
$$
\tilde{\sigma} = - \left( \hat{h}^{10}{}_{10} + \hat{e}_{2} \hat
{\rho}  \right)+ {3 \over 4} \hat{\sigma}.
\eqno(3.2.5)
$$
\par
The volume of the type IIA  torus $V_{n (A)}$ is given as a function of
the type IIA physical field $\rho$ and the $E_{n+1}$
Chevalley fields in equation (2.3.12).  In the following equation we
recall
these results, but express the result when
written in terms of physical type IIA fields as the physical type IIB
fields using equation (3.2.5) and the $d$ dimensional parameters $V_{n(B)}$, $V_{j}$, $j=1,...,n-1$ and $g_{d}$ arising from the dimensional reduction of type IIB on an $n$ torus
$$
V_{n(A)}=e^{\left( {8 \over 8-n} \right)n \tilde{e}_{2} \tilde{\rho}}= e^{ \left( \dot{\varphi}_{10}+ 2\dot{\varphi}_
{11} \right)}=e^{ 2 \sum_{i=d
+1}^{9} \hat{h}^{i}{}_{i} + 2\left({5n-8 \over 8-n}\right) n \hat{e}_{2} \hat{\rho}- 
{1 \over 2} \hat{\sigma}}=V_{n-1(B)}^{2} V_{n(B)}^{-1} g_{d}^{-{4 \over 8-n}},
\eqno(3.2.6)
$$
One can verify that one finds the same result if one express the
Chevalley fields in terms of the type IIB theory physical fields using equation
(2.4.11). Thus we see that the volume of the type IIA torus $V_{n(A)}$
has
the same expression in terms of the $E_{n+1}$ Chevalley fields from both
the type IIA and type IIB perspectives, as must be the case through the uniqueness of the Chevalley fields.
\par
A similar story applies to the volume of the type IIB torus which was
given in terms of the type IIB theory physical fields and
Chevalley fields in equation (2.4.12), that is
$$
V_{n(B)}= e^{ \left( {8  \over 8-n} \right) n \hat{e}_{2} \hat{\rho}}= e^{ 2 \dot{\varphi}_{9}} =e^{ 2 \sum_{i=d+1}^
{9} \tilde{h}^{i}{}_{i} + 2\left({5n-8 \over 8-n }\right)\tilde{e}_{2} \tilde{\rho} - {1 \over 2}  \tilde{\sigma}}=V_{n-1(A)}^{2} V_{n(A)}^{-1} g_{d}^{-{4 \over 8-n}}.
\eqno(3.2.7)$$
In the final equalities we have converted volume of the type IIB torus
into the type IIA physical fields using equation (3.1.5) and the $d$ dimensional parameters $V_{n(A)}$, $V_{j}$, $j=1,...,n-1$ and $g_{d}$ arising from the dimensional reduction of type IIA on an $n$ torus. One can easily
verify that one finds the same result using equation (2.3.11) to convert
the Chevalley fields into the IIA physical fields.
\par
One may note that the last two equations of (3.2.5) give the T-duality correspondence
between the radius of the circle upon which the IIA theory is
compactified $\tilde{r}_{10}$ and the radius of the circle upon which
the IIB theory is compactified $\hat{r}_{10}$ and the T-duality
correspondence between the coupling constants $g_{s (A)}$ and $g_{s
(B)}$.  The T-duality correspondence between the radii is given by
$$
\tilde{r}_{10}=e^{\tilde{h}^{10}{}_{10} + \tilde{e}_{2} \tilde{\rho}}
l_{10
(A)}  =e^{\tilde{h}^{10}{}_{10} + \tilde{e}_{2} \tilde{\rho}}l_{s}
e^{{\hat{\sigma} \over 4}}
=e^{-{3 \over 4}\left( \hat{h}^{10}{}_{10} + \hat{e}_{2} \hat{\rho}
\right)-{7 \over 16} \hat{\sigma} }e^{- {1 \over 4} \left( \hat{h}^
{10}{}_{10} + \hat{e}_{2} \hat{\rho} \right) + {3 \over 16}\hat
{\sigma} } l_{s}
$$
$$
=e^{-\left( \hat{h}^{10}{}_{10} + \hat{e}_{2} \hat{\rho}  \right)}e^{-
{1 \over 4} \hat{\sigma}
}l_{s}  ={l_{10 (B)} \over \hat{r}_{10}}e^{-{1 \over 4} \hat{\sigma}
}l_{s}
={l_{s}^{2} \over \hat{r}_{10}}.
\eqno(3.2.8)
$$
While the T-duality correspondence between the coupling constants is
given by
$$
g_{s(A)}=e^{\tilde{\sigma}}
=e^{-\left( \hat{h}^{10}{}_{10} + \hat{e}_{2} \hat{\rho} \right)+ {3
\over 4} \hat{\sigma}
}  ={l_{10 (B)} \over \hat{r}_{10}}e^{{3 \over 4} \hat{\sigma}}
={l_{s} \over \hat{r}_{10}}g_{s(B)}.
\eqno(3.2.9)
$$
These relations are the Buscher rules for T-duality
that relate the string coupling $g_{s(A)}$ and radius $\tilde{r}$ of
the type IIA string to the string coupling $g_{s(B)}$ and radius $\hat
{r}$ of the type IIB string.

\medskip
\noindent
{ \bf {3.3. M-theory and IIB}}
\medskip
In the $E_{11}$ non-linear realisation of eleven dimensional
supergravity
dimensionally reduced on an $m=n+1$ torus the group element restricted
to the Cartan subalgebra and written in terms of the physical fields was
given in equation (2.2.10), that is,
$$
g_{M}=e^{\sum_{a=1}^{d} h^{a}{}_{a} K^{a}{}_{a} + e_{1} \rho \sum_
{a=1}^{d} K^{a}{}_{a}}e^{ \sum_{i=d+1}^{11} h^{i}{}_{i} K^{i}{}_{i} +
e_{2} \rho \sum_{i=d+1}^{11} K^{i}{}_{i} }.
\eqno(3.3.1)
$$
On the other hand the $d$ dimensional theory obtained from the ten
dimensional IIB supergravity by dimensionally reducing on an $n$ torus
possess a  group element,  which when restricted
to the Cartan subalgebra and written in terms of the physical fields,
that take the form given in (2.4.8), that is,
$$
g_{IIB}=e^{\sum_{a=1}^{d} \hat{h}^{a}{}_{a} \hat{K}^{a}{}_{a} + \hat
{e}_{1} \hat{\rho} \sum_{a=1}^{d} \hat{K}^{a}{}_{a}}e^{ \sum_{i=d+1}^
{10} \hat{h}^{i}{}_{i} \hat{K}^{i}{}_{i} + \hat{e}_{2} \hat{\rho}
\sum_{i=d+1}^{10} \hat{K}^{i}{}_{i} } e^{\hat{\sigma} \hat{R}}.
\eqno(3.3.2)
$$
The relationship between the  M-theory  generators in $d$
dimensions $K^{a}{}_{a}$, $a=1,...,d$, $K^{i}{}_{i}$, $i=d+1,...,11$ and
the type IIB physical generators in $d$ dimensions $\hat{K}^{a}{}_
{a}$,
$a=1,...,d$, $\hat{K}^{i}{}_{i}$, $i=d+1,...,10$, $\hat{R}$ may be found by
equating the Cartan subalgebra generators $H_{a}$, $a=1,...,11$ of each
theory. These expression were given in equations (2.2.1) and (2.4.1) and using
these one finds that [52]
$$
\hat{K}^{a}{}_{a} = K^{a}{}_{a}, \quad a=1,...,d,
\quad
\hat{K}^{i}{}_{i} =K^{i}{}_{i}, \quad i=d+1,...,10,
$$
$$
\hat{K}^{10}{}_{10} = {1 \over 3} \sum_{i=1}^{9}K^{i}{}_{i} - {2
\over 3} \left( K^{10}{}_{10} + K^{11}{}_{11}  \right), \quad \hat{R} = -{1 \over 2}\left( K^{10}{}_{10} - K^{11}{}_{11}  \right)
\eqno(3.3.3)
$$
The theory in $d$ dimensions is unique as is evident from viewing
the $E_{11}$ Dynkin diagrams of figures 5 and 11 for the M-theory and type IIB
theories respectively. As such we may equate the M-theory and
type IIB group elements,
$$
g_{M}=g_{IIB},
\eqno(3.3.4)
$$
\par
One may use equations (3.3.3) to substitute for either the M-theory or
type IIB physical generators in equation (3.3.4) and read off the
relationship between the physical fields of these theories by equating
the coefficients of the generators.  This gives
$$
h^{a}{}_{a} + e_{1} \rho = \hat{h}^{a}{}_{a} + \hat{e}_{1} \hat{\rho}
+ {1 \over 3}\left( \hat{h}^{10}{}_{10} + \hat{e}_{2} \hat{\rho}
\right),
\quad a=1,...,d,
$$
$$
h^{i}{}_{i} + e_{2} \rho = \hat{h}^{i}{}_{i} + {4 \over 3} \hat{e}_
{2} \rho + {1 \over 3}\hat{h}^{10}{}_{10}, \quad i=d+1,...,9,
$$
$$
h^{10}{}_{10} + e_{2} \rho = - {2 \over 3} \hat{h}^{10}{}_{10} - {2
\over 3} \hat{e}_{2} \hat{\rho} - {1 \over 2} \hat{\sigma},
$$
$$
h^{11}{}_{11} + e_{2} \rho = - {2 \over 3} \hat{h}^{10}{}_{10} - {2
\over 3} \hat{e}_{2} \hat{\rho} + {1 \over 2} \hat{\sigma}.
\eqno(3.3.5)
$$

\par
The volume of the type IIB  torus $V_{n (B)}$ is given as a function of
the type IIB physical field $\hat{\rho}$ and the $E_{n+1}$
Chevalley fields in equation (2.4.12).  In the following equation we
recall
these results, but in the final equalities we express the result when
written in terms of physical type IIB fields as the physical fields of M-theory using equation (3.3.5) and the $d$ dimensional parameters $V_{j(M)}$, $j=1,...,m,$ arising from the dimensional reduction of M-theory on an $m$ torus
$$
V_{n(B)}=e^{ \left( {8 \over 8-n} \right)n \hat{e}_{2} \hat{\rho}}=e^{2 \dot{\varphi}_{9}}=e^{2 \sum_{i=d+1}^
{9} h^{i}{}_{i} + {12\left( n-2 \right) \over 8-n} e_{2} \rho }=V_{m-2(M)}^{2} V_{m(M)}^{-2}.
\eqno(3.3.6)
$$
One can verify that one finds the same result if one expresses the
Chevalley fields in terms of the M-theory physical fields using equation
(2.2.16). Thus we see that the volume of the type IIB torus $V_{n(B)}$
has
the same expression in terms of the $E_{n+1}$ Chevalley fields from both
the type IIB and M-theory perspectives, as must be the case.
\par
A similar story applies to the volume of the M-theory torus which was
given in terms of the M-theory physical fields and
Chevalley fields in equation (2.2.17), that is
$$
V_{m(M)}= e^{\left( {9 \over 8-n} \right)(n+1) e_{2} \rho}
=e^{3 \dot{\varphi}_{11}}=e^{ 12\left( {n-2 \over  8-n} \right)\hat{e}_
{2} \hat{\rho} - 3 \hat{h}^{10}{}_{10} }=V_{n-1}^{3}V_{n(B)}^{-{3 \over 2}}.
\eqno(3.3.7)
$$
In the final equalities we have converted the volume of the M-theory torus
into the IIB physical fields using equation (3.3.5) and the $d$ dimensional parameters $V_{n(B)}$, $V_{j}$, $j=1,...,n-1$ and $g_{d}$ arising from the dimensional reduction of type IIB on an $n$ torus. One can easily
verify that one finds the same result using equation (2.4.11) to convert
the Chevalley fields into the IIB physical fields.

\bigskip
\noindent
{ \bf {4. Limits in automorphic forms}}
\bigskip

In this chapter we are interested in the behaviour of automorphic forms
as the parameters are taken to  their  limits.
The $E_{n+1}$ automorphic forms appearing as the coefficient
functions of
the higher derivative terms in the type II string effective action in
$d=10-n$ dimensions are built from representations of $E_{n+1}$ whose
states are denoted by $| \psi_{E_{n+1}} \rangle$. 
\par
The $E_{n+1}$ automorphic forms can be written as functions of the state
$| \varphi_{E_{n+1}} \rangle$ which is given by
$$
| \varphi_{E_{n+1}} \rangle = L(g^{-1}_{E_{n+1}})| \psi_{E_{n+1}}
\rangle.
\eqno(4.0.1)$$
Here the group element $g_{E_{n+1}}$ belongs to the coset
$E_{n+1}/H$ where $H$ is the maximal compact
subgroup which is the same as the Cartan involution invariant subgroup.
The symbol $L$ denotes that the group element is in the
representation to
which
$|\psi_{E_{n+1}}\rangle$ belongs.
Using the Iwasawa decomposition, and the local subgroup,  we may write
the coset representative $L(g^{-1}_{E_{n+1}})$ in terms of the Cartan
subalgebra elements $\vec{H}$ and the positive root generators
$E_{\vec{\alpha}}$ of
$E_{n+1}$ and so write the state $| \varphi_{E_{n+1}} \rangle$ in the
form
$$
| \varphi_{E_{n+1}} \rangle= e^{ {1 \over
\sqrt{2}}\sum_{a=1}^{n+1}\dot {\varphi}_a H_a } e^{-\sum_{\vec{\alpha}
 >0}
\chi_{\vec{\alpha}}E_{\vec{\alpha}}}  | \psi_{E_{n+1}} \rangle,
\eqno(4.0.2)$$
We note that we may write the Cartan subalgebra part in
terms of Weyl basis generators rather than Chevalley
generators as $\sum_{a=1}^{n+1}\dot
{\varphi}_a H_a=\dot {\vec{\varphi}}.\vec{H}$; the relation being
determined by the equation
$H_a=\vec\alpha \cdot \vec H$. Further discussion of the theory of
non-linear realisations can be found in appendix B. We are interested in
the behaviour of the automorphic forms in the limits of the parameters, but
as the parameters  are related to the Chevalley fields $\dot \varphi_a$,
the limit can first be carried out on  the state $| \varphi_{E_{n+1}}
\rangle$.
\par
This is most easily explained by giving an example from which the general case is apparent.  
Given a basis of states of the representation we take the states of the representation to be
a sum of the basis states with integer coefficients. For  example for
the case of seven dimensions we have the group $SL(5)$ and we are
interested in the ${\bf 5}$ dimensional representation of $SL(5)$ and in
this case the state
$|\psi_{SL(5)} \rangle$
is given  by
$$
|\psi_{SL(5)} \rangle = m_{1} | \vec{\mu}_{1} \rangle
+  m_{2} | \vec{\mu}_{2} \rangle +...+ m_{5} | \vec{\mu}_{5} \rangle
\eqno(4.0.3)$$
where $| \vec{\mu}_{i} \rangle$, $i=1,...5$ are  the states in the ${\bf
5}$ of $SL(5)$ with weight $\vec{\mu}_{i}$ and $m_{i}$, $i=1,...,5$ are
the corresponding integer coefficients.
\par
The representations that occur in the automorphic forms are those that
the charges of strings belong to. As such the
states $|\psi_{SL(5)} \rangle$ can be thought of as carrying the
string charges  and the presence of the
integer coefficients corresponds  to the well known charge quantisation.
By adopting integers in the states we have in effect taken a
discrete group and it turns out that this corresponds to taking a discretisation of the group $E_{n+1}$ found by using the Chevalley
method [55].
\medskip
\noindent
{ \bf {4.1. The general case}}
\medskip
We will now formulate the constraints placed on an arbitrary higher derivative term in $d$ dimensions by taking the type IIA, type IIB, M-theory and single dimension decompactification limits along with the $j$ dimensional subtorus volume limit and the $d$ dimensional perturbative limit. 

\medskip
\noindent
{ \bf {4.1.1. Perturbative Limit} }
\medskip
String perturbation theory in $d=10-n$ dimensions is an expansion in
powers of the string coupling $g_{d}$.  As we found earlier in
this paper the coupling
$g_{d}$ in
$d=10-n$ dimensions may be expressed in terms of the $E_{n+1}$ Chevalley
fields and is given by
$$
g_{d}=e^{-\left({8-n \over 4} \right)\dot{\varphi}_{10}},
\eqno(4.1.1)
$$
where we have made use of equations (2.1.24) and (2.4.11).  The dependence of the $d$ dimensional string coupling is independent of the perspective of the $d$ dimensional theory, i.e. whether we choose to express the $d$ dimensional theory in terms of the dimensionally reduced fields of type IIA/B string theory or M-theory. When deriving
the parameters in terms of the Chevalley fields earlier in this paper we
deleted node $d$ in the $E_{11}$ Dynkin diagram and we  labeled the
nodes
of the resulting Dynkin diagram of
$E_{n+1}$ by
$11-n,  10-n, \ldots 11$. However, in this
chapter  it is more logical to use the labelling $1, \ldots ,
n+1$. In this relabelling the node $n$ becomes node 10 and so
$\dot{\varphi}_{10}$ is now $\dot{\varphi}_{n}$
We note that  in
$d<10$ dimensions the effective actions of type IIA and IIB string
  and M-theory are equivalent and so there is only one string
coupling in
$d$ dimensions.  Examining equation (4.1.1), we find that the
perturbative limit
$g_{d}
\rightarrow 0$ in
$d$ dimensions is equivalent to $\dot{\varphi}_{n} \rightarrow \infty$.
Taking
$\dot{\varphi}_{n} \rightarrow \infty $ corresponds to deleting node $n$
in the $E_{n+1}$ Dynkin diagram, as shown in figure 13.
$$
\matrix {
& & & & & n+1 & &  &&\cr
& & & &  & \bullet & & && \cr
& & & & & | & & &&\cr
\bullet &-&\ldots -&\bullet&-&\bullet&-&\bullet&-& \otimes \cr
1  & & &n-3  &  & n-2 &  & n-1& &n }
$$
\medskip
\centerline {Figure 13. The $E_{n+1}$ Dynkin diagram with node $n$
deleted}
\medskip
The  algebra remaining after this deletion is
the $GL(1)\times SO(n,n)$ subalgebra of $E_{n+1}$.  Let us denote the
generator of the  $GL(1)$ by $X$ which we may write as $X=\sum_{a=1}^{n+1}
c_a H_a$. Demanding that it commute with $SO(n,n)$ and in particular the
Chevalley generators $E_a, \ a=1,\ldots , n-1, n+1$ implies that
$$
X= H_1+2H_2+\ldots +(n-2)H_{n-2} +{n\over 2} H_{n-1}+ {(n-2)\over 2}
H_{n+1}+ 2H_n.
\eqno(4.1.2)
$$
Using the relation between the Chevalley $H_a$ and Weyl
$\vec H$ description of the generators in the Cartan subalgebra,
given by
$H_a=\vec \alpha_a\cdot \vec H$, and the decomposition of the
$E_{n+1}$ roots given in appendix A, which we recall here with the
appropriate labelling
$\vec{\alpha}_a=(0,\underline{\alpha}_a) , \ a=1,\ldots , n-1, n+1$ and $\alpha_n=(x,
-\underline{\lambda}_{n-1})$and $x^2={ 8-n \over 4}$,   we find that
$$
X= x(1,0)\cdot \vec H\equiv x(\vec H)_{1}.
\eqno(4.1.3)
$$
In deriving this equation we
have used that the roots and fundamental weights of $SO(n,n)$ obey the
equation
$\underline{\alpha}_1+2\underline{\alpha}_2+\ldots +(n-2)\underline{\alpha}_{n_2} +{n\over 2} \underline{\alpha}_{n-1}
+ {(n-2)\over 2}
\underline{\alpha}_{n+1}- 2 \underline{\lambda}_{n-1}=0$.  As explained in appendix B, the group
element that appears in the automorphic form, see equation (4.0.2),  contains
the expression
$e^{-\sum_{a=1}^{n+1} \dot \varphi_a H_a}$, which in terms of the
$GL(1)\times SO(n,n)$ decomposition becomes
$$
\exp(-\sum_{a=1}^{n+1} \dot {\varphi}_a H_a )= \exp (-x\dot {\varphi}_n
(\vec H)_1)\exp (-\left(
\sum_{a=1}^{n-1}{\dot \varphi}_a \underline{\alpha}_{a} - \dot{\varphi}_{n} \underline{\lambda}_{n-1}+ \dot{\varphi}_{n+1} \underline{\alpha}_{n} \right) .
(\underline{H}))
\eqno(4.1.4)
$$
where $\underline{H}$ is a vector consisting of the last $n$ components of $\vec{H}$, the first term is in $GL(1)$ and the second term in $SO(n,n)$. We have
identified the coefficient of the first term  as ${\dot \varphi}_n$ by
examining the first component of the left and right hand sides of this
vector equation when written in terms of the Weyl
generators $\vec H$. Thus the $GL(1)$ factor
corresponds to the  factor
$e^{-x\dot{\varphi}_{n}(\vec{H})_{1}} $ in the $E_{n+1}$ group element and,
from equation (4.1.1), to the powers of the effective coupling
$g_{d}$.  In taking the $g_{d} \rightarrow 0$ limit we must fix the $n$ quantities $ \sum_{a=1}^{n-1}{\dot \varphi}_a \underline{\alpha}_{a} - \dot{\varphi}_{n} \underline{\lambda}_{n-1}+ \dot{\varphi}_{n+1} \underline{\alpha}_{n} $ to preserve the $SO(n,n)$ symmetry.  
\par
The decomposition of  the full
$E_{n+1}$ algebra into representations of  its $GL(1)\times SO(n,n)$
subalgebra can be classified into a level [56-58]. The level is just the
number of times the simple root $\vec{\alpha}_n$ occurs in the corresponding
root when decomposed in terms of simple roots. Clearly,  the level zero
part of the decomposition is just
$GL(1)\times SO(n,n)$, as is clear from figure 13. The decomposition of
the representations of
$E_{n+1}$ into  representations of  $GL(1)\times SO(n,n)$ can
similarly be
classified according to the level, the level in this case is the
number of times the simple root $\vec{\alpha}_n$ occurs in the root string
constructed from the highest weight of the representation [59].

\par
As we discussed above the perturbative limit $g_{d} \rightarrow 0 $
corresponds to deleting node
$n$ in the
$E_{n+1}$ Dynkin diagram. In this limit
an  $E_{n+1}$ automorphic form  has an
expansion in powers of the string  coupling in $d$ dimensions $g_{d}$,
which is controlled by the $GL(1)$ factor. The coefficient functions in
this expansion  are automorphic forms of
$SO(n,n)$ built from the representations of $SO(n,n)$ that occur in the
the decomposition of the representation of $E_{n+1}$, from which the
original representation is built, into those of $SO(n,n)$. These latter
$SO(n,n)$ automorphic forms can be labelled by the
level, discussed above. Using equation (4.1.4) and
(4.0.2)  we can write the state
$|
\varphi_{E_{n+1}}
\rangle $ from which the automorphic form is built  as
$$
| \varphi_{E_{n+1}} \rangle = e^{- x\dot{\varphi}_{n} (\vec H)_{1}} |
\varphi_{SO(n,n)}^{(0)} \rangle +... =e^{-x \dot{\varphi}_{n}
w_{1}}| \varphi_{SO(n,n)}^{(0)}
\rangle + ...
= g_{d}^{{4xw_{1} \over 8-n}}  |
\varphi_{SO(n,n)}^{(0)}
\rangle +...
\eqno(4.1.5)
$$
where $\varphi_{SO(n,n)}^{(0)}$ is the level zero contribution
and so is built from the level zero representation, with
highest weight $w$ in the
decomposition and $w_{1}$ is the first component of $w$. In this equation
$+\ldots$ denotes the states formed from the higher level
representations in the decomposition. Clearly the $g_d$ dependence of
the level
$l$  contributions is given by $g_{d}^{{4xw_{1} - 4lx^{2} \over
8-n}}$
\par
Generic $E_{n+1}$ automorphic forms $\Phi$ constructed from $|
\varphi_{E_{n+1}} \rangle $ are expected to be homogeneous functions
which should satisfy the relation
$$
\Phi_{E_{n+1}} \left( a | \varphi_{E_{n+1}} \rangle  \right)  = a^{c}
\Phi_{E_{n+1}} \left( | \varphi_{E_{n+1}} \rangle  \right),
\eqno(4.1.6)
$$
where $a$ is a real number and $c$ is a scale factor that  depends on
the
particular structure of the automorphic form. Using this  homogeneity
property of $E_{n+1}$ automorphic forms, and equation (4.1.5), one may write
$$
\Phi_{E_{n+1}}\left( | \varphi_{E_{n+1}} \rangle  \right)  =
g_{d}^{{4cxw_{1} \over 8-n}}  \Phi_{E_{n+1}} ( |
\varphi_{SO(n,n)}^{(0)} \rangle ) + \ldots
\eqno(4.1.7)
$$
where  $+\ldots $ are terms that contain higher order
contributions in
$g_{d}$ and level.  In the $g_{d} \rightarrow 0$ limit $\Phi_{E_{n+1}}$ becomes a sum of $SO(n,n)$ automorphic forms with coefficients that are powers of the $d$ dimensional effective coupling $g_{d}$. 
\par
We require that the perturbative terms are consistent with a
perturbative expansion in $g_{d}$.  In string frame this implies that
each term has a $g_d$ dependence that is of the form
$g_{d}^{2g-2}$ where
$g$ is the genus.  String frame in $d$ dimensions is related to Einstein
frame by
$g_{E \mu
\nu}=g_{d}^{-{4 \over d-2}}g_{S \mu \nu}$. Upon rescaling to string
frame
we find
$$
\int d^{d}x \sqrt{-g_{S}}g_{d}^{{4\Delta-2d \over d-2}}\Phi_{E_{n+1}}
{\cal{O}}_{S},
\eqno(4.1.8)
$$
where ${\cal{O}}$  is some polynomial in the $d$ dimensional curvature
$R$, Cartan forms $P$ or field strengths $F$, the subscript $S$ denotes
string frame quantities and $\Delta$ is the number of space time metrics
minus the number of inverse space time
metrics in ${\cal{O}}_{S}$.
Demanding that the perturbative limit of  this generic higher derivative
term exists from a string theory perspective means that in the limit
$g_{d} \rightarrow 0$ equation (4.1.8) agrees with a perturbative
expansion in $g_{d}$,  for this one requires that each term is
multiplied
by a factor of the form $g_{d}^{-2+2n}$, where $n$ is either zero or a
positive integer.  Considering the $E_{n+1}$ automorphic form as a
function of the state $| \varphi_{E_{n+1}}
\rangle $ we see from equations $(4.1.7)$ and $(4.1.8)$ that this condition
is given by
$$ 
\lim_{g_{d} \rightarrow 0}
g_{d}^{{4\Delta-2d
\over 8-n}} \left( g_{d}^{{ 4cx
w_{1}
\over 8-n}} \Phi_{E_{n+1}}(  | \tilde{\varphi}_{SO(n,n)}^{(0)} \rangle)
+\ldots \right)  =  g_{d}^{-2+2n_{0}} \Phi_{SO(n,n)}^{(0)} +\dots
\eqno(4.1.9)
$$
where $n_{0}$ is a non-negative integer.

\medskip
\noindent
{ \bf {4.1.2. Type IIB volume Limit} }
\medskip
Type IIB string theory in $d=10$ dimensions exhibits an $SL(2,{Z})$ symmetry.  So an arbitrary higher derivative term in $d=10-n$ dimensions should, in the large volume limit $V_{n(B)} \rightarrow \infty $ give a sum of $d=10$ higher derivative terms whose coefficient functions are $SL(2,{Z})$ automorphic forms. 
As we found earlier, the volume of the type IIB torus $V_{n(B)}$ may be expressed in terms of the $E_{n+1}$ Chevalley fields and is given by
$$
V_{n(B)}=(2 \pi)^{n} {r_{10}r_{9}...r_{d+1} \over l_{d}^{n}}=e^{{8 \over 8-n} n\beta \rho}=e^{2 \dot{\varphi}_{9}}.
\eqno(4.1.10)
$$
where we have made use of equation (2.4.12).  The dependence of the volume of the type IIB torus $V_{n(B)}$ on the Chevalley field $\dot{\varphi}_{9}$ is independent of the perspective of the $d$ dimensional theory, i.e. whether we choose to express the $d$ dimensional theory in terms of the dimensionally reduced fields of type IIA/B string theory or M-theory.  Relabelling the $E_{n+1}$ part of the Dynkin diagram as in section 4.1.1, node 9 becomes node $n-1$ and so
$\dot{\varphi}_{9}$ is now $\dot{\varphi}_{n-1}$.  Examining equation (4.1.10), we find that the type IIB volume limit
$V_{n(B)}
\rightarrow \infty$ in
$d$ dimensions is equivalent to $\dot{\varphi}_{n-1} \rightarrow \infty$.
Taking
$\dot{\varphi}_{n-1} \rightarrow \infty $ corresponds to deleting node $n-1$
in the $E_{n+1}$ Dynkin diagram, as shown in figure 14.
$$
\matrix {
& & & & & n+1 & &  &&\cr
& & & &  & \bullet & & && \cr
& & & & & | & & &&\cr
\bullet &-&\ldots -&\bullet&-&\bullet&-& \otimes &-& \bullet \cr
1  & & &n-3  &  & n-2 &  & n-1& &n }
$$
\medskip
\centerline {Figure 14. The $E_{n+1}$ Dynkin diagram with node $n-1$ deleted}
\medskip

The  algebra remaining after this deletion is
the $GL(1)\times SL(2) \times SL(n)$ subalgebra of $E_{n+1}$.  Let us denote the
generator of the  $GL(1)$ by $X$ which we may write as $X=\sum_{a=1}^{n+1}
c_a H_a$. Demanding that it commute with $SL(2) \times SL(n)$ and in particular the
Chevalley generators $E_a, \ a=1,\ldots , n-2,n, n+1$ implies that
$$
X= H_1+2H_2+\ldots +(n-2)H_{n-2} +{n\over 2} H_{n-1}+ {(n-2)\over 2}
H_{n+1}+ {n \over 4}H_n.
\eqno(4.1.11)
$$
Using the relation between the Chevalley $H_a$ and Weyl
$\vec H$ description of the generators in the Cartan subalgebra,
given by
$H_a=\vec \alpha_a\cdot \vec H$, and the decomposition of the
$E_{n+1}$ roots given in appendix A, which we recall here with the
appropriate labelling
$\vec{\alpha}_a=(0,\underline{\alpha}_a) , \ a=1,\ldots , n-2$, $\vec{\alpha}_{n+1}=\left(0, \underline{\alpha}_{n-1} \right) $, $\vec{\alpha}_{n}=\left( 0, \beta, \underline{0} \right)$ and $\vec{\alpha}_{n-1}=(x,-\mu,
-\underline{\lambda}_{n-1})$and $x^2={8-n \over 2n}$, we find that
$$
X= x(1,0)\cdot \vec H\equiv x(\vec H)_1
\eqno(4.1.12)
$$
up to an overall scale factor.  In deriving this equation we
have used that the simple root $\beta$ and fundamental weight $\mu$ of $SL(2)$ satisfy $2 \mu - \beta=0$ and the simple roots $\underline{\alpha_{i}}$ and fundamental weight $\underline{\lambda_{n-2}}$ of $SL(n)$ obey the
equation
$\alpha_1+2\alpha_2+\ldots +(n-2)\alpha_{n_2} + {(n-2)\over 2} \alpha_{n-1} - {n \over 2}\lambda_{n-2}=0$.  As explained in appendix B, the group
element that appears in the automorphic form, see equation (4.0.2),  contains
the expression
$e^{-\sum_{a=1}^{n+1} \dot \varphi_a H_a}$ which in terms of the
$GL(1)\times SL(2) \times SL(n)$ decomposition becomes
$$
\exp(-\sum_{a=1}^{n+1} \dot{\varphi}_a H_a )= \exp (-x\dot {\varphi}_{n-1}
(\vec H)_1)\exp(\left(\mu\dot{\varphi}_{n-1}- \beta \dot{\varphi}_{n}\right)\left( \vec{H} \right)_{2} )  
$$ 
$$
\times\exp (-\left(
\sum_{a=1}^{n-2}{\dot \varphi}_a \underline{\alpha}_{a} + \dot{\varphi}_{n+1} \underline{\alpha}_{n-1} - \dot{\varphi}_{n-1} \underline{\lambda}_{n-2} \right) .
\underline{H}),
\eqno(4.1.13)
$$
where $\underline{H}$ is a vector consisting of the last $n-1$ components of $\vec{H}$, the first term is in $GL(1)$, the second term in $SL(2)$ and the third term is in $SL(n)$. We have
identified the coefficient of the first term  as ${\dot \varphi}_{n-1}$ by
examining the first component of the left and right hand sides of this
vector  equation when written in terms of the generator in Weyl
generators $\vec H$. Thus the  $GL(1)$ factor
corresponds to the  factor
$e^{-x\dot{\varphi}_{n-1}(\vec{H})_{1}} $ in the $E_{n+1}$ group element and, from equation (4.1.10), to the powers of the volume of the type IIB torus $V_{n(B)}$.  In taking the $V_{n(B)} \rightarrow \infty$ limit we must fix the quantities $\left(
\sum_{a=1}^{n-2}{\dot \varphi}_a \underline{\alpha}_{a} + \dot{\varphi}_{n+1} \underline{\alpha}_{n-1} - \dot{\varphi}_{n-1} \underline{\lambda}_{n-2} \right)$ and $\mu\dot{\varphi}_{n-1}- \beta \dot{\varphi}_{n}$ to preserve the $SL(2) \times SL(n)$ symmetry.  
\par
The decomposition of  the full
$E_{n+1}$ algebra into representations of its $GL(1)\times SL(2) \times SL(n)$
subalgebra can be classified into a level [56-58]. The level is just the
number of times the simple root $\vec{\alpha}_{n-1}$ occurs in the corresponding
root when decomposed in terms of simple roots. Clearly, the level zero
part of the decomposition is just
$GL(1)\times SL(2) \times SL(n)$, as is clear from figure 14. The decomposition of
the representations of
$E_{n+1}$ into  representations of  $GL(1)\times SL(2) \times SL(n)$ can
similarly be
classified according to the level, the level in this case is the
number of times the simple root $\vec{\alpha}_{n-1}$ occurs in the root string
constructed from the highest weight of the representation [59].

\par
As we discussed above, the large volume limit of the type IIB torus $V_{n(B)} \rightarrow \infty $
corresponds to  deleting node
$n-1$ in the
$E_{n+1}$ Dynkin diagram. In this limit
an  $E_{n+1}$ automorphic form  has an
expansion in powers of the volume of the type IIB torus $V_{n(B)}$,
which is controlled by the $GL(1)$ factor. The  coefficient functions in
this expansion  are automorphic forms of
$SL(2) \times SL(n)$ built from the representations of $SL(2) \times SL(n)$ that occur in the
the decomposition of the representation of $E_{n+1}$, from which the
original representation is built, into those of $SL(2) \times SL(n)$. These latter
$SL(2) \times SL(n)$ automorphic forms can be labelled by the
level, discussed above.  We note that since the type IIB theory in ten dimensions can not depend on the moduli of the $n$ torus it is necessary for the automorphic forms found after taking the $V_{n(B)} \rightarrow \infty$ limit to be constructed from the trivial representation of $SL(n)$.  Using equation (4.1.13) and
(4.0.2)  we can write the state
$|
\varphi_{E_{n+1}}
\rangle $ from which the automorphic form is built  as
$$
| \varphi_{E_{n+1}} \rangle = e^{- x \dot{\varphi}_{n} (\vec H)_{1}} |
\varphi_{SL(2) \times SL(n) }^{(0)} \rangle +...
$$
$$
 =e^{-x \dot{\varphi}_{n}
w_{1}}| \varphi_{SL(2) \times SL(n)}^{(0)}
\rangle + ...
$$
$$
= V_{n(B)}^{-{xw_{1} \over 2}}  |
\varphi_{SL(2) \times SL(n)}^{(0)}
\rangle +...
\eqno(4.1.14)
$$
where $\varphi_{SL(2) \times SL(n)}^{(0)}$ is the level zero contribution
and so is built from the level zero representation, with
highest weight $w$ in the
decomposition and $w_{1}$ is the first component of $w$. In this equation
$+\ldots$ denotes the states formed from the higher level
representations in the decomposition. Clearly the $V_{n(B)}$ dependence of
the  level
$l$  contributions is given by $V_{n(B)}^{-\left({xw_{1}-lx^{2} \over 2}\right)}$
\par
Generic $E_{n+1}$ automorphic forms $\Phi$  constructed from $|
\varphi_{E_{n+1}} \rangle $ are expected to be homogeneous functions
which should satisfy the relation
$$
\Phi_{E_{n+1}} \left( a | \varphi_{E_{n+1}} \rangle  \right)  = a^{c}
\Phi_{E_{n+1}} \left( | \varphi_{E_{n+1}} \rangle  \right),
\eqno(4.1.15)
$$
where $a$ is a real number and $c$ is a scale factor that depends on
the
particular structure of the automorphic form. Using this homogeneity
property of $E_{n+1}$ automorphic forms, and equation (4.1.14), one may write
$$
\Phi_{E_{n+1}}\left( | \varphi_{E_{n+1}} \rangle  \right)  =
V_{n(B)}^{-{cxw_{1} \over 2}} \Phi_{E_{n+1}} ( |
\varphi_{SL(2) \times SL(n)}^{(0)} \rangle ) + \ldots
\eqno(4.1.16)
$$
where  $+\ldots $ are terms that contain higher order
contributions in
$V_{n(B)}$.
\par
We require that the terms remaining in the large volume limit of the type IIB torus match the known coefficient functions of the higher derivative terms in the type IIB effective action in ten dimensions.  By dimensional analysis one sees that an arbitrary $d$ dimensional higher derivative term in Einstein frame takes the form
$$
l_{d}^{k-d}  \int d^{d}x \sqrt{-g}\Phi_{E_{n+1}} {\cal{O}} = l_{10(B)}^{k-\left( 10-n \right)}V_{n(B)}^{-{k-\left(10-n\right) \over 8}} \int d^{d}x \sqrt{-g}\Phi_{E_{n+1}} {\cal{O}}, 
\eqno(4.1.17)
$$
where ${\cal{O}}$ is a $k$ derivative polynomial in the $d$ dimensional curvature
$R$, Cartan forms $P$ or field strengths $F$ and we have used equation (2.1.37) to express the $d$ dimensional Planck length in terms of the ten dimensional type IIB Planck length $l_{10(B)}$ and the volume of the type IIB torus $V_{n(B)}$. From equation (2.1.17) we see that in the large volume limit of the type IIB torus one has 
$$
\lim_{V_{n(B)} \rightarrow \infty}  l_{10(B)}^{n}\int d^{d}x \sqrt{-g}V_{n}^{{8-n \over 8 }}= \int d^{10}x \sqrt{-\hat{g}},  
\eqno(4.1.18)
$$
where the $V_{n(B)}$ factor is that found upon dimensional reduction of type IIB string theory to $d=10-n$ dimensions and the power ${8-n \over 8}$ associated with this factor is a consequence of the definition of $V_{n(B)}$ in equation (2.1.16). 
Therefore any term with $V_{n(B)}$ dependence $V_{n(B)}^{{8-n \over 8 }}$ is preserved in the limit while any term with a lesser power of $V_{n(B)}$ vanishes in the limit. Note that one must be careful when considering non-analytic terms in the action that appear divergent in the $V_{n(B)} \rightarrow \infty$ limit.
Demanding that the large volume limit $V_{n(B)} \rightarrow \infty$ of this generic higher derivative term exists from a string theory perspective means that the limit in equation (4.1.17) exists and that the resulting terms are ten dimensional higher derivative terms with a coefficient function that is a sum of $SL(2)$ automorphic forms. Examining equation (4.1.17) and substituting the $E_{n+1}$ automorphic form as a function of the state $| \varphi_{E_{n+1}}
\rangle $ in (4.1.16) one finds that in the $V_{n(B)} \rightarrow \infty$ limit
$$
\lim_{V_{n(B)} \rightarrow \infty} l_{10}^{k-\left( 10-n \right)}V_{n(B)}^{-{k-\left(10-n\right) \over 8}} \int d^{d}x \sqrt{-g}(V_{n(B)}^{-{cxw_{1} \over 2}} \Phi_{E_{n+1}} ( |
\varphi_{SL(2) \times SL(n)}^{(0)} \rangle ) + \ldots) {\cal{O}}
$$
$$
=\lim_{V_{n(B)} \rightarrow \infty} l_{10}^{k-\left( 10-n \right)}V_{n}^{-{k-\left(10-n\right) \over 8}} \int d^{d}x \sqrt{-g}V_{n}^{{8-n \over 8}} 
$$
$$
\times V_{n}^{-{8-n \over 8}}(V_{n(B)}^{-{cxw_{1} \over 2}} \Phi_{E_{n+1}} ( |
\varphi_{SL(2) \times SL(n)}^{(0)} \rangle ) + \ldots) {\cal{O}} 
$$
$$
= l_{10}^{k-10} \int d^{10}x  \sqrt{-\hat{g}}\lim_{V_{n(B)} \rightarrow \infty} V_{n}^{{2-k \over 8}} \left( 
V_{n(B)}^{-{cxw_{1} \over 2}} \Phi_{E_{n+1}} ( |
\varphi_{SL(2) \times SL(n)}^{(0)} \rangle ) + \ldots \right) \hat{\cal{O}}. 
\eqno(4.1.19)
$$
where we have made use of equation (4.1.18) and denoted the generic $d=10$ type IIB polynomials in the ten dimensional curvature $\hat{R}$, Cartan form $\hat{P}$ and field strengths $\hat{F}$ that arise in the decompactification of the $d$ dimensional polynomial in the curvature $R$, Cartan forms $P$ and field strengths $F$ by $\hat{\cal{O}}$.  The object $\hat{\cal{O}}$ contains fields with dependence on the compactified directions.  Although the automorphic form possesses integer sums associated with the Kaluza-Klein modes of the compactification it is not entirely clear if this is enough to reinstate this coordinate dependence.  In the above equation we have simply restored by hand this dependence in a general coordinate manner.  
\par
Therefore as a function of the state $| \varphi_{E_{n+1}}
\rangle $ the type IIB volume limit condition is given by
$$
\lim_{V_{n(B)} \rightarrow \infty} V_{n(B)}^{{2-k \over 8}} \left( 
V_{n(B)}^{-{cxw_{1} \over 2}} \Phi_{E_{n+1}} ( |
\varphi_{SL(2) \times SL(n)}^{(0)} \rangle ) + \ldots \right) = a_{0} \hat{\Phi}^{(0)}_{SL(2)} + \ldots
\eqno(4.1.20)
$$
where $a_{0}$ is a constant that depends on the $E_{n+1}$ automorphic form.  Any higher derivative term in $d$ dimensions that converges to a higher derivative term that is not compatible with type IIB string theory in $d=10$ dimensions, must be rejected as a possible higher derivative term in $d$ dimensions.  As we noted earlier, any non-vanishing term in the $V_{n(B)} \rightarrow \infty$ limit must have a coefficient function that is constructed from the trivial representation of $SL(n)$ so that the ten dimensional theory does not depend on the moduli of the torus.

\medskip
\noindent
{ \bf {4.1.3. Decompactification of a single dimension limit} }
\medskip
Type II string theory in $d=10-n$ dimensions exhibits an $E(n+1,{Z})$ symmetry.  So, in the decompactification of a single dimension limit ${r_{d+1} \over l_{d}} \rightarrow \infty$ an arbitrary higher derivative term in $d=10-n$ dimensions should become a linear combination of $d+1$ dimensional higher derivative terms with coefficient functions that transform as $E_{n}(Z)$ automorphic forms.  
As we found earlier, the ratio of the radius in the compact $d+1$ direction to the $d$ dimensional Planck length $l_{d}$, in $d=10-n$ dimensions, may be expressed in terms of the $E_{n+1}$ Chevalley fields and is given by
$$
{r_{d+1} \over l_{d}}={l_{10} \over l_{d} }{ r_{d+1} \over l_{10}} = e^{\dot{\varphi}_{d+1}},
\eqno(4.1.21) 
$$
where we have made use of equations (2.1.6), (2.1.15), (2.1.16) and (2.4.11).  The dependence of the ratio of the radius in the compact $d+1$ direction $r_{d+1}$ to the $d$ dimensional Planck length $l_{d}$ on the Chevalley field $\dot{\varphi}_{d+1}$ is independent of the perspective of the $d$ dimensional theory, i.e. whether we choose to express the $d$ dimensional theory in term of the dimensionally reduced fields of type IIA/B string theory or M-theory.
Relabelling the $E_{n+1}$ part of the Dynkin diagram as in section 4.1.1, node d+1 becomes node $1$ and so
$\dot{\varphi}_{d+1}$ is now $\dot{\varphi}_{1}$. Examining equation (4.1.21), we find that the decompactification of a single dimension limit
${r_{d+1} \over l_{d}}
\rightarrow \infty$ in
$d$ dimensions is equivalent to $\dot{\varphi}_{1} \rightarrow \infty$.
Taking
$\dot{\varphi}_{1} \rightarrow \infty $ corresponds to deleting node $1$
in the $E_{n+1}$ Dynkin diagram, as shown in figure 15.
$$
\matrix {
& && & & & & n+1 & &  &&\cr
& && & & &  & \bullet & & && \cr
& && & & & & | & & &&\cr
\otimes &- & \bullet &-&\ldots -&\bullet&-&\bullet&-&\bullet&-&\bullet \cr
1  & & 2 & & &n-3  &  & n-2 &  & n-1& &n }
$$
\medskip
\centerline {Figure 15. The $E_{n+1}$ Dynkin diagram with node 1 deleted}
\medskip

The  algebra remaining after this deletion is
the $GL(1)\times E_{n}$ subalgebra of $E_{n+1}$.  Let us denote the
generator of the  $GL(1)$ by $X$ which we may write as $X=\sum_{a=1}^{n+1}
c_a H_a$. Demanding that it commute with $E_{n}$ and in particular the
Chevalley generators $E_a, \ a=2,\ldots , n+1$ implies that
$$
X=\sum_{a=1}^{n-2}\left( {8 - n + a \over 2} \right)H_{a}+ 2H_{n-1}+ H_{n} + {3 \over 2} H_{n+1}.
\eqno(4.1.22)
$$
Using the relation between the Chevalley $H_a$ and Weyl
$\vec H$ description of the generators in the Cartan subalgebra,
given by
$H_a=\vec \alpha_a\cdot \vec H$, and the decomposition of the
$E_{n+1}$ roots given in appendix A, which we recall here with the
appropriate labelling $\vec{\alpha}_{1}=\left(x, - \underline{\lambda}_{1} \right) $,
$\vec{\alpha}_a=(0,\underline{\alpha}_{a-1}) , \ a=2,\ldots , n+1$,  and $x^2={8-n \over 9-n}$, we find that
$$
X= x(1,0)\cdot \vec H\equiv x(\vec H)_1
\eqno(4.1.23)
$$
up to an overall scale factor.  In deriving this equation we
have used that the simple roots $\underline{\alpha_{i}}$ and fundamental weight $\underline{\lambda_{1}}$ of $E_{n}$ obey the
equation
$\sum_{a=2}^{n-2}\left({8-n+a \over 2}\right)\underline{\alpha}_{a-1}+2\underline{\alpha}_{n-2} + \underline{\alpha}_{n-1} + {3 \over 2} \underline{\alpha}_{n} -\left({9-n \over 2}\right)\underline{\lambda}_{1}=0$.  As explained in appendix B, the group
element that appears in the automorphic form, see equation (4.0.2),  contains
the expression
$e^{-\sum_{a=1}^{n+1} \dot \varphi_a H_a}$ which in terms of the
$GL(1)\times E_{n}$ decomposition becomes
$$
\exp(-\sum_{a=1}^{n+1} \dot{\varphi}_a H_a )= \exp (-x\dot {\varphi}_{1} (\vec{H}_{1})) \exp (-\left(
\sum_{a=2}^{n+1}{\dot \varphi}_a \underline{\alpha}_{a-1} - \dot{\varphi}_{1} \underline{\lambda}_{1} \right) .
\underline{H})
\eqno(4.1.24)
$$
where $\underline{H}$ is a vector consisting of the last $n$ components of $\vec{H}$, the first term is in $GL(1)$ and the second term in $E_{n}$. We have
identified the coefficient of the first term  as ${\dot \varphi}_{1}$ by
examining the first component of the left and right hand sides of this
vector  equation when written in terms of the generator in Weyl
generators $\vec H$. Thus the  $GL(1)$ factor
corresponds to the  factor
$e^{-x\dot{\varphi}_{1}(\vec{H})_{1}} $ in the $E_{n+1}$ group element and,
from  equation (4.1.20),  to the ratio of the radius in the compact $d+1$ direction to the $d$ dimensional Planck length $l_{d}$.  In taking the ${r_{d+1} \over l_{d}} \rightarrow \infty$ limit we must fix the $n$ quantities $\left(
\sum_{a=2}^{n+1}{\dot \varphi}_a \underline{\alpha}_{a-1} - \dot{\varphi}_{1} \underline{\lambda}_{1} \right)$ to preserve the $E_{n}$ symmetry.  
\par
The decomposition of  the full
$E_{n+1}$ algebra into representations of  its $GL(1)\times E_{n}$
subalgebra can be classified into a level [56-58]. The level is just the
number of times the simple root $\vec{\alpha}_{1}$ occurs in the corresponding
root when decomposed in terms of simple roots. Clearly,  the level zero
part of the decomposition is just
$GL(1)\times E_{n}$, as is clear from figure 9. The decomposition of
the representations of
$E_{n+1}$ into  representations of  $GL(1)\times E_{n}$ can
similarly be
classified according to the level, the level in this case is the
number of times the simple root $\vec{\alpha}_{1}$ occurs in the root string
constructed from the highest weight of the representation [59].

\par
As we discussed above, the decompactification of a single dimension limit ${r_{d+1} \over l_{d}} \rightarrow \infty $
corresponds to  deleting node
$1$ in the
$E_{n+1}$ Dynkin diagram. In this limit
an  $E_{n+1}$ automorphic form  has an
expansion in powers of the ratio of the radius in the compact $d+1$ direction to the $d$ dimensional Planck length $l_{d}$,
which is controlled by the $GL(1)$ factor. The  coefficient functions in
this expansion  are automorphic forms of
$E_{n}$ built from the representations of $E_{n}$ that occur in the decomposition of the representation of $E_{n+1}$, from which the
original representation is built, into those of $E_{n}$. These latter
$E_{n}$ automorphic forms can be labelled by the
level, discussed above.  Using equation (4.0.2) and
(4.1.23)  we can write the state
$|
\varphi_{E_{n+1}}
\rangle $ from which the automorphic form is built  as
$$
| \varphi_{E_{n+1}} \rangle = e^{- x \dot{\varphi}_{1} (\vec H)_{1}} |
\varphi_{E_{n} }^{(0)} \rangle +...
$$
$$
 =e^{-x \dot{\varphi}_{1}
w_{1}}| \varphi_{E_{n}}^{(0)}
\rangle + ...
$$
$$
=\left({r_{d+1} \over l_{d}}\right)^{-{xw_{1}}}  |
\varphi_{E_{n}}^{(0)}
\rangle +...
\eqno(4.1.25)
$$
where $\varphi_{E_{n}}^{(0)}$ is the level zero contribution
and so is built from the level zero representation, with
highest weight $w$, in the
decomposition and $w_{1}$ is the first component of $w$. In this equation
$+\ldots$ denotes the states formed from the higher level
representations in the decomposition. Clearly the ${r_{d+1} \over l_{d}}$ dependence of
the  level
$l$  contributions is given by $\left( {r_{d+1} \over l_{d}} \right)^{-{xw_{1}-lx^{2} }}$.
\par
Generic $E_{n+1}$ automorphic forms $\Phi$  constructed from $|
\varphi_{E_{n+1}} \rangle $ are expected to be homogeneous functions
which should satisfy the relation
$$
\Phi_{E_{n+1}} \left( a | \varphi_{E_{n+1}} \rangle  \right)  = a^{c}
\Phi_{E_{n+1}} \left( | \varphi_{E_{n+1}} \rangle  \right),
\eqno(4.1.26)
$$
where $a$ is a real number and $c$ is a scale factor that depends on
the
particular structure of the automorphic form. Using this homogeneity
property of $E_{n+1}$ automorphic forms, and equation (4.1.25), one may write
$$
\Phi_{E_{n+1}}\left( | \varphi_{E_{n+1}} \rangle  \right)  =
{r_{d+1} \over l_{d}}^{-{cxw_{1} }} \Phi_{E_{n+1}} ( |
\varphi_{E_{n}}^{(0)} \rangle ) + \ldots
$$
$$
=\left(l_{d+1}^{-\left({9-n \over 8-n} \right)} r_{d+1}^{ \left({9-n \over 8-n } \right) } \right)^{- c x w_{1}} \Phi_{E_{n+1}} ( |
\varphi_{E_{n}}^{(0)} \rangle ) + \ldots
\eqno(4.1.27)
$$
where  $+\ldots $ are terms that contain higher order
contributions in
${r_{d+1}}$ and we have used equation $(2.1.35)$ to express the state in terms of the $d+1$ dimensional Planck length $l_{d+1}$.
\par
We require that the terms remaining in the decompactification of a single dimension limit match the known coefficient functions of the higher derivative terms in the type II effective action in $d+1$ dimensions. 
By dimensional analysis one sees that an arbitrary $d$ dimensional higher derivative term in Einstein frame takes the form
$$
l_{d}^{k-d}  \int d^{d}x \sqrt{-g}\Phi_{E_{n+1}} {\cal{O}} =\int d^{d}x  \sqrt{-g} l_{d+1}^{{\left( d-1 \right) \left( k-d \right) \over d-2}} r_{d+1}^{-{k-d \over d-2}} \Phi_{E_{n+1}} {\cal{O}}
\eqno(4.1.28)
$$
where we have used equation (2.1.35) to express $l_{d}$ in terms of $r_{d+1}$ and $l_{d+1}$ and ${\cal{O}}$ is a $k$ derivative polynomial in the $d$ dimensional curvature
$R$, Cartan forms $P$ or field strengths $F$. 
 From equation (2.1.11) we see that in the decompactification of a single dimension limit one has 
$$
\lim_{{r_{d+1} \over l_{d+1}} \rightarrow \infty}  l_{d+1} \int d^{d}x \sqrt{-g} {r_{d+1} \over l_{d+1}}= \int d^{d+1}x \sqrt{-\hat{g}}, 
\eqno(4.1.29) 
$$
where the ${r_{d+1} \over l_{d+1}}$ factor is that found upon dimensional reduction of $d+1$ dimensional type II string theory to $d$ dimensions. 
Therefore any term with ${r_{d+1} \over l_{d+1}}$ dependence ${r_{d+1} \over l_{d+1}}$ is preserved in the limit while any term with a lesser power of $r_{d+1} \over l_{d+1}$ vanishes in the limit. Note that one must be careful when considering non-analytic terms in the action that appear divergent in the ${r_{d+1} \over l_{d+1}} \rightarrow \infty$ limit.
Demanding that the decompactification of a single dimension limit ${r_{d+1} \over l_{d+1}}$ of this generic higher derivative term exists from a string theory perspective means that the ${r_{d+1} \over l_{d+1}} \rightarrow \infty$ limit of equation (4.1.27) exists and that the resulting terms are $d+1$ dimensional higher derivative terms with a coefficient function that is a sum of $E_{n}$ automorphic forms (or zero). 
Examining equation (4.1.28) and substituting the $E_{n+1}$ automorphic form as a function of the state $| \varphi_{E_{n+1}}
\rangle $ in (4.1.27) one finds that in the ${r_{d+1} \over l_{d+1}} \rightarrow \infty$ limit an arbitrary $d$ dimensional higher derivative term becomes
$$
\lim_{{r_{d+1} \over l_{d+1}} \rightarrow \infty}  \int d^{d}x  \sqrt{-g} l_{d+1}^{{\left( d-1 \right) \left( k-d \right) \over d-2}} r_{d+1}^{-{k-d \over d-2}} \Phi_{E_{n+1}} {\cal{O}} = \int d^{d+1}x  \sqrt{-\hat{g}}\lim_{{r_{d+1} \over l_{d+1}} \rightarrow \infty}  l_{d+1}^{{\left( d-1 \right) \left( k-d \right) \over d-2}} r_{d+1}^{{2-k \over d-2}} \Phi_{E_{n+1}} {\cal{O}},
$$
$$
= \int d^{d+1}x  \sqrt{-\hat{g}}\lim_{{r_{d+1} \over l_{d+1}} \rightarrow \infty}  l_{d+1}^{{\left( d-1 \right) \left( k-d \right) \over d-2}} r_{d+1}^{{2-k \over d-2}} \left(  \left(l_{d+1}^{-\left({9-n \over 8-n} \right)} r_{d+1}^{ \left({9-n \over 8-n } \right) } \right)^{- c x w_{1}} \Phi_{E_{n+1}} ( |
\varphi_{E_{n}}^{(0)} \rangle ) + \ldots \right)  \hat{\cal{O}},  
\eqno(4.1.30)
$$
where $\hat{\cal{O}}$ labels the different $d+1$ dimensional type II polynomials in the $d+1$ dimensional curvature $\hat{R}$, Cartan form $\hat{P}$ and field strengths $\hat{F}$ that arise in the decompactification of the $d$ dimensional polynomial in the curvature $R$, Cartan forms $P$ and field strengths $F$.  The fields $\hat{\cal{O}}$ are the $d+1$ dimensional analogues of the decompactified type IIB $\hat{\cal{O}}$ fields discussed around equation (4.1.19).
Therefore as a function of the state $| \varphi_{E_{n+1}}
\rangle $ the decompactification of a single dimension condition is given by
$$
\lim_{{r_{d+1} \over l_{d+1}} \rightarrow \infty}  l_{d+1}^{{\left( d-1 \right) \left( k-d \right) \over d-2}} r_{d+1}^{{2-k \over d-2}} \left(  \left(l_{d+1}^{-\left({9-n \over 8-n} \right)} r_{d+1}^{ \left({9-n \over 8-n } \right) } \right)^{- c x w_{1}} \Phi_{E_{n+1}} ( |
\varphi_{E_{n}}^{(0)} \rangle ) + \ldots \right) 
$$
$$
=  a_{0} \hat{\Phi}^{(0)}_{E_{n}} + \ldots
\eqno(4.1.31)
$$
where $a_{0}$ is constant that depends on the $E_{n+1}$ automorphic form.   Any higher derivative term in $d$ dimensions that converges to a higher derivative term that is not compatible with type II string theory in $d+1$ dimensions, must be rejected as a possible higher derivative term in $d$ dimensions.

\medskip
\noindent
{ \bf {4.1.3. M-theory volume Limit} }
\medskip
Type II string theory in $d$ dimensions may be decompactified to eleven dimensional supergravity on an $\left( n+1 \right)$ torus by taking the limit $V_{m(M)} \rightarrow \infty$.  As we found earlier, the volume of the M-theory torus $V_{m(M)}$ may be expressed in terms of the $E_{n+1}$ Chevalley fields and is given by
$$
V_{m(M)}=(2 \pi)^{m} {r_{11}r_{10}r_{9}...r_{d+1} \over l_{d}^{m}}=e^{{9 \over 9-m} m\beta \rho}=e^{3 \dot{\varphi}_{11}},
\eqno(4.1.32)
$$
where we have made use of equation (2.2.17).  The dependence of the volume of the M-theory torus $V_{m(M)}$ on the Chevalley field $\dot{\varphi}_{11}$ is independent of the perspective of the $d$ dimensional theory, i.e. whether we choose to express the $d$ dimensional theory in terms of the dimensionally reduced fields of type IIA/B string theory or M-theory. 
Relabelling the $E_{n+1}$ part of the Dynkin diagram as in section 4.1.1, node 11 becomes node $n+1$ and so
$\dot{\varphi}_{11}$ is now $\dot{\varphi}_{n+1}$. Examining equation (4.1.32), we find that the large volume limit of the M-theory torus
$V_{m(M)}
\rightarrow \infty$ in
$d$ dimensions is equivalent to $\dot{\varphi}_{n+1} \rightarrow \infty$.
Taking
$\dot{\varphi}_{n+1} \rightarrow \infty $ corresponds to deleting node $n+1$
in the $E_{n+1}$ Dynkin diagram, as shown in figure 16.
$$
\matrix {
& & & & & n+1 & &  &&\cr
& & & &  & \otimes & & && \cr
& & & & & | & & &&\cr
\bullet &-&\ldots -&\bullet&-&\bullet&-&\bullet&-& \bullet \cr
1  & & &n-3  &  & n-2 &  & n-1& &n }
$$
\medskip
\centerline {Figure 16. The $E_{n+1}$ Dynkin diagram with node $n+1$ deleted}
\medskip

The  algebra remaining after this deletion is
the $GL(1)\times SL(n+1)$ subalgebra of $E_{n+1}$.  Let us denote the
generator of the  $GL(1)$ by $X$ which we may write as $X=\sum_{a=1}^{n+1}
c_a H_a$. Demanding that it commute with $SL(n+1)$ and in particular the
Chevalley generators $E_a, \ a=1,\ldots , n$ implies that
$$
X= \sum_{a=1}^{n-2}aH_{a}+{2 \over 3}\left(n-2\right)H_{n-1}+\ldots +{n-2 \over 3}H_{n} +{n+1\over 3} H_{n+1}.
\eqno(4.1.33)
$$
Using the relation between the Chevalley $H_a$ and Weyl
$\vec H$ description of the generators in the Cartan subalgebra,
given by
$H_a=\vec \alpha_a\cdot \vec H$, and the decomposition of the
$E_{n+1}$ roots given in appendix A, which we recall here with the
appropriate labelling
$\vec{\alpha}_a=(0,\underline{\alpha}_a) , \ a=1,\ldots , n$, $\vec{\alpha}_{n+1}=\left(x, \underline{\alpha}_{n-2} \right) $ and $x^2={8-n \over n+1}$, we find that
$$
X= x(1,0)\cdot \vec H\equiv x(\vec H)_{1},
\eqno(4.1.34)
$$
up to an overall scale factor.  In deriving this equation we
have used that the simple roots $\underline{\alpha_{i}}$ and fundamental weight $\underline{\lambda_{n-2}}$ of $SL(n+1)$ obey the
equation
$\sum_{a=1}^{n-2} a \underline{\alpha}_{a}+{2 \over 3}\left(n-2 \right)\underline{\alpha}_{n-1}+ {n-2 \over 3} \underline{\alpha}_{n}- {n+1 \over 3} \underline{\lambda}_{n-2}=0$.  As explained in appendix B, the group
element that appears in the automorphic form, see equation (4.0.2),  contains
the expression
$e^{-\sum_{a=1}^{n+1} \dot \varphi_a H_a}$ which in terms of the
$GL(1)\times SL(n+1)$ decomposition becomes
$$
\exp(-\sum_{a=1}^{n+1} \dot{\varphi}_a H_a )= \exp (-x\dot {\varphi}_{n+1}
(\vec H)_1) \exp (-\left(
\sum_{a=1}^{n}{\dot \varphi}_a \underline{\alpha}_{a} - \dot{\varphi}_{n+1} \underline{\lambda}_{n-2} \right) .
\underline{H})
\eqno(4.1.35)
$$
where $\underline{H}$ is a vector consisting of the last $n$ components of $\vec{H}$, the first term is in $GL(1)$ and the second term is in $SL(n+1)$. We have
identified the coefficient of the first term  as ${\dot \varphi}_{n+1}$ by
examining the first component of the left and right hand sides of this
vector  equation when written in terms of the Weyl
generators $\vec H$. Thus the  $GL(1)$ factor
corresponds to the  factor
$e^{-x\dot{\varphi}_{n+1}(\vec{H})_{1}} $ in the $E_{n+1}$ group element and,
from  equation (4.1.32), to the powers of the volume of the M-theory torus $V_{m(M)}$.  In taking the $V_{m(M)} \rightarrow \infty$ limit we must fix the $n$ quantities $\left(
\sum_{a=1}^{n}{\dot \varphi}_a \underline{\alpha}_{a} - \dot{\varphi}_{n+1} \underline{\lambda}_{n-2} \right)$ to preserve the $ SL(n+1)$ symmetry.  
\par
The decomposition of the full
$E_{n+1}$ algebra into representations of  its $GL(1) \times SL(n+1)$
subalgebra can be classified into a level [56-58]. The level is just the
number of times the simple root $\vec{\alpha}_{n+1}$ occurs in the corresponding
root when decomposed in terms of simple roots. Clearly,  the level zero
part of the decomposition is just $GL(1)\times SL(n+1)$, as is clear from figure 16. The decomposition of
the representations of
$E_{n+1}$ into  representations of  $GL(1) \times SL(n+1)$ can
similarly be
classified according to the  level, the level in this case is the
number of times the simple root $\vec{\alpha}_{n+1}$ occurs in the root string
constructed from the highest weight of the representation [59].

\par
As we discussed above, the large volume limit of the M-theory torus $V_{m(M)} \rightarrow \infty $
corresponds to  deleting node
$n+1$ in the
$E_{n+1}$ Dynkin diagram. In this limit
an  $E_{n+1}$ automorphic form  has an
expansion in powers of the volume of the M-theory torus $V_{m(M)}$,
which is controlled by the $GL(1)$ factor. The  coefficient functions in
this expansion  are automorphic forms of
$SL(n+1)$ built from the representations of $SL(n+1)$ that occur in the
the decomposition of the representation of $E_{n+1}$, from which the
original representation is built, into those of $SL(n+1)$. These latter
$SL(n+1)$ automorphic forms can be labelled by the
level, discussed above.  We note that since M-theory can not depend on the moduli of the $m$ torus it is necessary for the automorphic forms found after taking the $V_{m(M)} \rightarrow \infty$ limit to be constructed from the trivial representation of $SL(n+1)$.  Using equation (4.0.2) and
(4.1.35)  we can write the state
$|
\varphi_{E_{n+1}}
\rangle $ from which the automorphic form is built  as
$$
| \varphi_{E_{n+1}} \rangle = e^{- x \dot{\varphi}_{n+1} (\vec H)_{1}} |
\varphi_{SL(n+1) }^{(0)} \rangle +...
$$
$$
 =e^{-x \dot{\varphi}_{n+1}
(\vec \Lambda^{(0)} )_{1}}| \varphi_{SL(n+1)}^{(0)}
\rangle + ...
$$
$$
= V_{m(M)}^{-{xw_{1} \over 3}}  |
\varphi_{SL(n+1)}^{(0)}
\rangle +...
\eqno(4.1.36)
$$
where $\varphi_{ SL(n+1)}^{(0)}$ is the level zero contribution
and so is built from the level zero representation, with
highest weight $w$ in the
decomposition and $w_{1}$ is the first component of $w$. In this equation
$+\ldots$ denotes the states formed from the higher level
representations in the decomposition. Clearly the $V_{m(M)}$ dependence of
the  level
$l$  contributions is given by $V_{m(M)}^{-\left({xw_{1}-lx^{2} \over 3}\right)}$.
\par
Generic $E_{n+1}$ automorphic forms $\Phi$  constructed from $|
\varphi_{E_{n+1}} \rangle $ are expected to be homogeneous functions
which should satisfy the relation
$$
\Phi_{E_{n+1}} \left( a | \varphi_{E_{n+1}} \rangle  \right)  = a^{c}
\Phi_{E_{n+1}} \left( | \varphi_{E_{n+1}} \rangle  \right),
\eqno(4.1.37)
$$
where $a$ is a real number and $c$ is a scale factor that depends on
the
particular structure of the automorphic form. Using this homogeneity
property of $E_{n+1}$ automorphic forms, and equation (4.1.36), one may write
$$
\Phi_{E_{n+1}}\left( | \varphi_{E_{n+1}} \rangle  \right)  =
V_{m(M)}^{-{cxw_{1} \over 3}} \Phi_{E_{n+1}} ( |
\varphi_{SL(n+1)}^{(0)} \rangle ) + \ldots
\eqno(4.1.38)
$$
where  $+\ldots $ are terms that contain higher order
contributions in
$V_{m(M)}$.
\par
We require that the terms remaining in the large volume limit of the M-theory torus match the known coefficient functions of the higher derivative terms in the M-theory effective action in eleven dimensions.  By dimensional analysis one sees that an arbitrary $d$ dimensional higher derivative term in Einstein frame takes the form
$$
l_{d}^{k-d}  \int d^{d}x \sqrt{-g}\Phi_{E_{n+1}} {\cal{O}}=l_{11}^{k-d} V_{m(M)}^{-{k-d \over 9}} \int d^{d}x \sqrt{-g} \Phi_{E_{m}} {\cal{O}},
\eqno(4.1.39)
$$
where ${\cal{O}}$ is a $k$ derivative polynomial in the $d$ dimensional curvature
$R$, Cartan forms $P$ or field strengths $F$.  From equation (2.1.17) we see that in the large volume limit of the M-theory torus one has 
$$
\lim_{V_{m(M)} \rightarrow \infty} l_{11}^{n+1} \int d^{d} x \sqrt{-g} V_{m(M)}^{{9-m \over 9 }} = \int d^{11}x \sqrt{-\hat{g}}, 
\eqno(4.1.40)
$$
therefore any term with $V_{m(M)}$ dependence $V_{m(M)}^{{9-m \over 9 }}$ is preserved in the limit while any term with a lesser power of $V_{m(M)}$ vanishes in the limit.  Note that one must be careful when considering non-analytic terms in the action that appear divergent in the $V_{m(M)} \rightarrow \infty$ limit.
Demanding that the large volume limit $V_{m(M)}$ of this generic higher derivative term exists from an M-theory perspective means that the large volume limit $V_{m(M)} \rightarrow \infty$ of equation (4.1.39) exists and that the resulting terms are eleven dimensional higher derivative terms with coefficient functions that are $SL(n+1)$ automorphic forms that can only be constructed from the trivial representation of $SL(n+1)$ since the M-theory effective action can not depend on the moduli of the torus.  Examining equation (4.1.39) and substituting the $E_{n+1}$ automorphic form as a function of the state $| \varphi_{E_{n+1}}
\rangle $ in (4.1.38) one finds that in the $V_{m(M)} \rightarrow \infty$ limit
$$
\lim_{V_{m(M)} \rightarrow \infty}l_{11}^{k-d} V_{m(M)}^{-{k-d \over 9}} \int d^{d}x \sqrt{-g} (
V_{m(M)}^{-{cxw_{1} \over 3}} \Phi_{E_{n+1}} ( |
\varphi_{SL(n+1)}^{(0)} \rangle ) + \ldots)  {\cal{O}}
$$

$$
=\lim_{V_{m(M)} \rightarrow \infty} l_{11}^{k-d}  \int d^{d}x \sqrt{-g} V_{m(M)}^{{9-m \over 9}} V_{m(M)}^{{ 2-k \over 9}} \Phi_{E_{n+1}} {\cal{O}}
$$
$$
= l_{11}^{k-11} \int d^{11}x  \sqrt{-\hat{g}}\lim_{V_{m(M)} \rightarrow \infty} \left( V_{m(M)}^{{2-k \over 9}} \left( 
V_{m(M)}^{-{cxw_{1} \over 3}} \Phi_{E_{n+1}} ( |
\varphi_{SL(n+1)}^{(0)} \rangle ) + \ldots \right)  \hat{\cal{O}} \right), 
\eqno(4.1.41)
$$
where $\hat{\cal{O}}$ labels the different $d=11$ M-theory polynomials in the eleven dimensional curvature $\hat{R}$, and field strengths $\hat{F}$ that arise in the decompactification of the $d$ dimensional polynomial in the curvature $R$, Cartan forms $P$ and field strengths $F$.  The fields $\hat{\cal{O}}$ are the eleven dimensional M-theory analogues of the decompactified type IIB $\hat{\cal{O}}$ fields discussed around equation (4.1.19).
Therefore, as a function of the state $| \varphi_{E_{n+1}}
\rangle $, the M-theory volume limit condition is given by
$$
\lim_{V_{m(M)} \rightarrow \infty} \left( V_{m(M)}^{{2-k \over 9}} \left( 
V_{m(M)}^{-{cxw_{1} \over 3}} \Phi_{E_{n+1}} ( |
\varphi_{SL(n+1)}^{(0)} \rangle ) + \ldots \right)  \right)=    a_{0}  + \ldots
\eqno(4.1.42)
$$
where $a_{0}$ is a constant arising from the level zero contribution that depends on the $E_{n+1}$ automorphic form and $\ldots$ denote contributions at higher levels.    Any higher derivative term in $d$ dimensions that converges to a higher derivative term that is not compatible with the effective action of M-theory in $d=11$ dimensions, must be rejected as a possible higher derivative term in $d$ dimensions.  As we noted earlier, any non-vanishing term in the $V_{m(M)} \rightarrow \infty$ limit must have a coefficient function that is constructed from the trivial representation of $SL(n+1)$ so that the M-theory effective action in $d=11$ dimensions does not depend on the moduli of the torus.

\medskip
\noindent
{ \bf {4.1.3. Type IIA volume Limit} }
\medskip
Type IIA string theory in $d=10$ dimensions possesses a global $GL(1,{R})$ symmetry.  In this case the scalar sector that parameterises the coset associated with the global $GL(1,{R})$ symmetry is trivial. However, in the large volume limit $V_{n(A)} \rightarrow \infty$, one still requires that the higher derivative terms in the effective action of the type IIA theory in $d=10$ dimensions in string frame be multiplied by a factor of $e^{\left( -2+2g \right) \phi}$ where $g$ is the genus of the $d=10$ type IIA perturbative contribution. 
\par
The volume of the $n$ torus in the type IIA theory may be expressed in terms of the $E_{n+1}$ Chevalley fields and is given by
$$
V_{n(A)}=(2 \pi)^{n} {r_{10}r_{9}...r_{d+1} \over l_{d}^{n}}=e^{{8 \over 8-n }n \beta \rho}=e^{ \dot{\varphi}_{10} + 2 \dot{\varphi}_{11} },
\eqno(4.1.43)
$$
where we have made use of equations (2.3.12).   The dependence of the volume of the type IIA torus $V_{n(A)}$ on the Chevalley fields $\dot{\varphi}_{10}$ and $\dot{\varphi}_{11}$  is independent of the perspective of the $d$ dimensional theory, i.e. whether we choose to express the $d$ dimensional theory in terms of the dimensionally reduced fields of type IIA/B string theory or M-theory.  Similarly, from equations (2.3.11), one finds that the type IIA string coupling in ten dimensions $g_{s(A)}$ may be written 
$$
g_{s(A)}=e^{-{3 \over 2} \dot{\varphi}_{10} + \dot{\varphi}_{11}}.
\eqno(4.1.44)
$$
Relabelling the $E_{n+1}$ part of the Dynkin diagram as in section 4.1.1, nodes 10 and 11 become node $n$ and $n+1$ and so
$\dot{\varphi}_{10}$ is now $\dot{\varphi}_{n}$ and $\dot{\varphi}_{11}$ is now $\dot{\varphi}_{n+1}$. Examining equation (4.1.43), we find that the type IIA volume limit
$V_{n(A)}
\rightarrow \infty$ in
$d$ dimensions is equivalent to $\dot{\varphi}_{n} + 2 \dot{\varphi}_{n+1} \rightarrow \infty$.
Taking
$\dot{\varphi}_{n} + 2 \dot{\varphi}_{n+1} \rightarrow \infty $ corresponds to deleting nodes $n$ and $n+1$
in the $E_{n+1}$ Dynkin diagram, as shown in figure 17.
$$
\matrix {
& & & & & n+1 & &  &&\cr
& & & &  & \otimes & & && \cr
& & & & & | & & &&\cr
\bullet &-&\ldots -&\bullet&-&\bullet&-&\bullet&-& \otimes \cr
1  & & &n-3  &  & n-2 &  & n-1& &n }
$$
\medskip
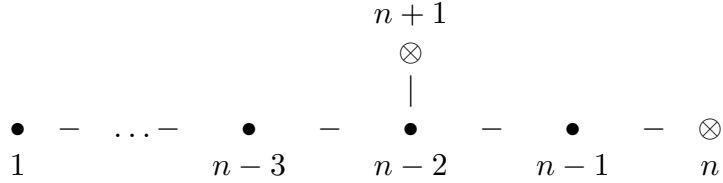
\centerline {Figure 17. The $E_{n+1}$ Dynkin diagram with nodes $n$ and $n+1$  deleted}
\medskip

The  algebra remaining after this deletion is
the $GL(1)\times GL(1) \times SL(n)$ subalgebra of $E_{n+1}$.  Let us denote the
generators of the $GL(1) \times GL(1) $ part of the subalgebra by $X$ which we may write as $X=\sum_{a=1}^{n+1}
c_a H_a$. Demanding that $X$ commutes with $ SL(n)$ and in particular the
Chevalley generators $E_a, \ a=1,\ldots , n-1$ implies that
$$
X= \sum_{a=1}^{n-2}caH_{a}+\left((n-1)c-d\right)H_{n-1} + \left(nc-2d\right) H_{n}+ d H_{n+1}
\eqno(4.1.45)$$
where $c$ and $d$ are real numbers.  
Using the relation between the Chevalley $H_a$ and Weyl
$\vec H$ description of the generators in the Cartan subalgebra,
given by
$H_a=\vec \alpha_a\cdot \vec H$, and the decomposition of the
$E_{n+1}$ roots given in appendix A, which we recall here with the
appropriate labelling
$\vec{\alpha}_a=(0,\underline{\alpha}_a) , \ a=1,\ldots , n-1$, $\vec{\alpha}_{n}=\left(x, - { \underline{\lambda}_{n-2}.\underline{\lambda}_{n-1} \over y}, - \underline{\lambda}_{n-1}  \right) $, $\vec{\alpha}_{n+1}=\left( 0, y, -\underline{\lambda}_{n-2} \right)$ and $x^2={ 8-n \over 4}$, $y^{2}={4 \over n}$ we find that
$$
X= (Ax,By,\underline{0})\cdot \vec H\equiv Ax(\vec H)_{1}+ By(\vec H)_{2},
\eqno(4.1.46)
$$
up to scale factors $A$ and $B$.  In deriving this equation we
have used that the simple roots $\underline{\alpha}_{i}$ and fundamental weights $\underline{\lambda}_{n-1}$, $\underline{\lambda}_{n-2}$ of $SL(n)$ obey the
equations
$\underline{\alpha}_1+2\underline{\alpha}_2+\ldots +(n-2)\underline{\alpha}_{n-2} + (n-1) \underline{\alpha}_{n-1} - n \underline{\lambda}_{n-1}=0$ and $-\underline{\alpha}_{n-1} +2\underline{\lambda}_{n-1} - \underline{\lambda}_{n-2}=0$.  As explained in appendix B, the group
element that appears in the automorphic form, see equation (4.0.2), contains
the expression
$e^{-\sum_{a=1}^{n+1} \dot \varphi_a H_a}$ which in terms of the
$GL(1)\times GL(1) \times SL(n)$ decomposition becomes
$$
\exp(-\sum_{a=1}^{n+1} \dot{\varphi}_a H_a )= \exp (-x\dot {\varphi}_{n}
(\vec H)_1)\exp(-y(\dot{\varphi}_{n+1} - {\underline{\lambda}_{n-2}. \underline{\lambda}_{n-1} \over y^{2}} \dot{\varphi}_{n})(\vec H)_2)
$$
$$
\times \exp (-(\sum_{a=1}^{n-1}{\dot \varphi}_a \underline{\alpha}_{a} - \dot{\varphi}_{n} \underline{\lambda}_{n-1} - \dot{\varphi}_{n+1} \underline{\lambda}_{n-2} ).\underline{H}),
\eqno(4.1.47)
$$
where $\underline{H}$ is a vector consisting of the last $n-1$ components of $\vec{H}$, the first term and second term correspond to the two $GL(1)$ factors and the third term is in $SL(n)$. We have
identified the coefficient of the first term as ${\dot \varphi}_{n}$ and the second term as a linear combination of ${\dot \varphi}_{n}$ and ${\dot \varphi}_{n+1}$  by
examining the first component of the left and right hand sides of this
vector  equation when written in terms of the generator in Weyl
generators $\vec H$. Thus the two $GL(1)$ factors
correspond to $e^{-x\dot{\varphi}_{n}\vec H_{1}} $ and $e^{-y(\dot{\varphi}_{n+1} - {\underline{\lambda}_{n-2}. \underline{\lambda}_{n-1} \over y^{2}} \dot{\varphi}_{n})(\vec H)_2}$ in the $E_{n+1}$ group element and, from  equations (4.1.43) and (4.1.44), to products of the type IIA volume $V_{n(A)}$ and ten dimensional type IIA string coupling $g_{s(A)}$.  In taking the $V_{n(A)} \rightarrow \infty$ limit we must fix the $n-1$ quantities $(\sum_{a=1}^{n-1}{\dot \varphi}_a \underline{\alpha}_{a} - \dot{\varphi}_{n} \underline{\lambda}_{n-1} - \dot{\varphi}_{n+1} \underline{\lambda}_{n-2} )$ to preserve the $SL(n)$ symmetry and $-{3 \over 2} \dot{\varphi}_{10} + \dot{\varphi}_{11}$ to preserve the type IIA string coupling as given in equation (4.1.44).  
\par
The decomposition of  the full
$E_{n+1}$ algebra into representations of  its $GL(1)\times GL(1) \times SL(n)$
subalgebra can be classified into a level [56-58]. The level is indexed by the
number of times the simple root $\vec{\alpha}_n$ and the simple root $\vec{\alpha}_{n+1}$ occur in the corresponding
root when decomposed in terms of simple roots. Clearly,  the level zero
part of the decomposition is just
$GL(1)\times GL(1) \times SL(n)$, as is clear from figure 17. The decomposition of
the representations of
$E_{n+1}$ into  representations of  $GL(1)\times GL(1) \times SL(n)$ can
similarly be
classified according to the  level, the level in this case is the
number of times the simple root $\vec{\alpha}_{n}$ and the simple root $\vec{\alpha}_{n+1}$ occur in the root string
constructed from the highest weight of the representation.

\par
As we discussed above, the large volume limit of the type IIA torus $V_{n(A)} \rightarrow \infty $
corresponds to  deleting nodes
$n$ and $n+1$ in the
$E_{n+1}$ Dynkin diagram. In this limit
an  $E_{n+1}$ automorphic form  has a simultaneous
expansion in powers of the volume of the type IIA volume $V_{n(A)}$ and the type IIA string coupling in ten dimensions $g_{s(A)}$
which are controlled by the $GL(1) \times GL(1)$ part of the $E_{n+1}$ group element. The  coefficient functions in
this expansion  are automorphic forms of
$ SL(n)$ built from the representations of $ SL(n)$ that occur in the
the decomposition of the representation of $E_{n+1}$, from which the
original representation is built, into those of $ SL(n)$. These latter
$ SL(n)$ automorphic forms can be labelled by the
level, discussed above.  We note that since the type IIA theory in ten dimensions can not depend on the moduli of the $n$ torus it is necessary for the automorphic forms found after taking the $V_{n(A)} \rightarrow \infty$ limit to be constructed from the trivial representation of $SL(n)$.  Using equations (4.0.2), (4.1.43) and
(4.1.44)  we can write the state
$|
\varphi_{E_{n+1}}
\rangle $ from which the automorphic form is built  as
$$
| \varphi_{E_{n+1}} \rangle = e^{-x\dot {\varphi}_{n}
(\vec H)_1}e^{-y(\dot{\varphi}_{n+1} - {\underline{\lambda}_{n-2}. \underline{\lambda}_{n-1} \over y^{2}} \dot{\varphi}_{n})(\vec H)_2} |
\varphi_{ SL(n) }^{(0,0)} \rangle +...
$$
$$
=e^{-x\dot {\varphi}_{n}
w_1}e^{-y(\dot{\varphi}_{n+1} - {\underline{\lambda}_{n-2}. \underline{\lambda}_{n-1} \over y^{2}} \dot{\varphi}_{n})w_2} |
\varphi_{ SL(n) }^{(0,0)} \rangle + ...
$$
$$
= V_{n(A)}^{-{xw_{1} \over 4}+({\underline{\lambda}_{n-2}.\underline{\lambda}_{n-1} \over 4y} - {3 \over 8}y)w_{2}} g_{s(A)}^{{xw_{1} \over 2}-({\underline{\lambda}_{n-2}.\underline{\lambda}_{n-1} \over 2y} +  {1 \over 4}y)w_{2}} |
\varphi_{SL(n)}^{(0,0)}
\rangle +...
\eqno(4.1.48)
$$
where $\varphi_{SL(n)}^{(0,0)}$ is the level (0,0) contribution
and so is built from the level $l_{1}=l_{2}=0$ representation, with
highest weight $w$ in the
decomposition and $w_{1}$, $w_{2}$ are the first and second components of $w$ respectively. In this equation
$+\ldots$ denotes the states formed from the higher level
representations in the decomposition. Clearly the $V_{n(A)}$ dependence of
the  level
$(l_{1},l_{2})$ contributions is given by
$$
V_{n(A)}^{-\left({xw_{1}-l_{1}x^{2} \over 4}\right)+ \left({\underline{\lambda}_{n-2}.\underline{\lambda}_{n-1} \over 4y} - {3 \over 8}y\right) \left(w_{2}+l_{1}{\underline{\lambda}_{n-2}.\underline{\lambda}_{n-1} \over y}-l_{2}y  \right)}
\eqno(4.1.49)
$$
while the $g_{s(A)}$ dependence of the level $(l_{1},l_{2})$ contributions is given by 
$$g_{s(A)}^{{xw_{1}-l_{1}x^{2} \over 2}-\left({\underline{\lambda}_{n-2}.\underline{\lambda}_{n-1} \over 2y} + {1 \over 4}y \right) \left(w_{2}+l_{1} {\underline{\lambda}_{n-2}.\underline{\lambda}_{n-1} \over y} - l_{2} y \right) }.
\eqno(4.1.50)
$$
\par
Generic $E_{n+1}$ automorphic forms $\Phi$  constructed from $|
\varphi_{E_{n+1}} \rangle $ are expected to be homogeneous functions
which should satisfy the relation
$$
\Phi_{E_{n+1}} \left( a | \varphi_{E_{n+1}} \rangle  \right)  = a^{c}
\Phi_{E_{n+1}} \left( | \varphi_{E_{n+1}} \rangle  \right),
\eqno(4.1.51)
$$
where $a$ is a real number and $c$ is a scale factor that depends on
the
particular structure of the automorphic form. Using this homogeneity
property of $E_{n+1}$ automorphic forms, and equation (4.1.48), one may write
$$
\Phi_{E_{n+1}}\left( | \varphi_{E_{n+1}} \rangle  \right)  =
V_{n(A)}^{-{cxw_{1} \over 4}+({\underline{\lambda}_{n-2}.\underline{\lambda}_{n-1} \over 4y} - {3 \over 8}y)cw_{2}} g_{s(A)}^{{cxw_{1} \over 2}-({\underline{\lambda}_{n-2}.\underline{\lambda}_{n-1} \over 2y} + {1 \over 4}y)cw_{2}}
$$
$$
\times  \Phi_{E_{n+1}} ( |  SL(n)^{(0,0)} \rangle ) + \ldots
\eqno(4.1.52)
$$
where  $+\ldots $ are terms that contain higher order
contributions in
$V_{n(A)}$ and $g_{s(A)}$ at levels $l_{1}>0$ or $l_{2}>0$.
\par
We require that the terms remaining in the large volume limit of the type IIA torus match the known coefficient functions of the higher derivative terms in the type IIA effective action in ten dimensions.  By dimensional analysis one sees that an arbitrary $d$ dimensional higher derivative term in Einstein frame takes the form
$$
l_{d}^{k-d}  \int d^{d}x \sqrt{-g}\Phi_{E_{n+1}} {\cal{O}}=l_{10}^{k-d}V_{n(A)}^{-{k-d \over 8}} \int d^{d}x \sqrt{-g}\Phi_{E_{n+1}} {\cal{O}}, 
\eqno(4.1.53)
$$
where ${\cal{O}}$ is a $k$ derivative polynomial in the $d$ dimensional curvature
$R$, Cartan forms $P$ or field strengths $F$.  In the large volume limit of the type IIA torus one has 
$$
\lim_{V_{n(A)} \rightarrow \infty}  l_{10}^{n}\int d^{d}x \sqrt{-g}V_{n(A)}^{{8-n \over 8 }}= \int d^{10}x \sqrt{-\hat{g}},  
\eqno(4.1.54)
$$
therefore any term with $V_{n(A)}$ dependence $V_{n(A)}^{{8-n \over 8 }}$ is preserved in the limit while any term with a lesser power of $V_{n(A)}$ vanishes in the limit.  Note that one must be careful when considering non-analytic terms in the action that appear divergent in the $V_{n(A)} \rightarrow \infty$ limit.
Demanding that the large volume limit $V_{n(A)}$ of this generic higher derivative term exists from a type IIA perspective means that the limit $V_{n(A)} \rightarrow \infty$ of equation (4.1.53) exists and that the resulting terms are ten dimensional type IIA higher derivative terms with coefficient functions that are $SL(n)$ automorphic forms that can only be constructed from the trivial representation of $SL(n)$ since the ten dimensional type IIA effective action can not depend on the moduli of the torus. Examining equation (4.1.53) and substituting the $E_{n+1}$ automorphic form as a function of the state $| \varphi_{E_{n+1}}
\rangle $ in (4.1.52) one finds that in the $V_{n(A)} \rightarrow \infty$ limit
$$
\lim_{V_{n(A)} \rightarrow \infty} l_{10}^{k-d}V_{n(A)}^{-{k-d \over 8}} \int d^{d}x \sqrt{-g} (
V_{n(A)}^{-{cxw_{1} \over 4}+c({\underline{\lambda}_{n-2}.\underline{\lambda}_{n-1} \over 4y} - {3 \over 8}y)w_{2}} 
$$
$$
\times g_{s(A)}^{{cxw_{1} \over 2}-c({\underline{\lambda}_{n-2}.\underline{\lambda}_{n-1} \over 2y} + {1 \over 4}y)w_{2}}  \Phi_{E_{n+1}} ( |  SL(n)^{(0,0)} \rangle ) + \ldots ) {\cal{O}} 
$$
$$
=\lim_{V_{n(A)} \rightarrow \infty} l_{10}^{k-d}  \int d^{d}x \sqrt{-g} V_{n(A)}^{{8-n \over 8}} 
$$
$$
\times V_{n(A)}^{-\left({8-n \over 8}\right)}(
V_{n(A)}^{-{cxw_{1} \over 4}+c({\underline{\lambda}_{n-2}.\underline{\lambda}_{n-1} \over 4y} - {3 \over 8}y)w_{2}} g_{s(A)}^{{cxw_{1} \over 2}-c({\underline{\lambda}_{n-2}.\underline{\lambda}_{n-1} \over 2y} + {1 \over 4}y)w_{2}}  \Phi_{E_{n+1}} ( |  SL(n)^{(0,0)} \rangle ) + \ldots )  {\cal{O}}
$$
$$
= l_{10}^{k-10} \int d^{10}x  \sqrt{-\hat{g}}\lim_{V_{n(A)} \rightarrow \infty} ( V_{n(A)}^{{2-k \over 8}} ( V_{n(A)}^{-{cxw_{1} \over 4}+c({\underline{\lambda}_{n-2}.\underline{\lambda}_{n-1} \over 4y} - {3 \over 8}y)w_{2}} 
$$
$$
\times g_{s(A)}^{{cxw_{1} \over 2}-c({\underline{\lambda}_{n-2}.\underline{\lambda}_{n-1} \over 2y} + {1 \over 4}y)w_{2}} \Phi_{E_{n+1}} ( |
\varphi_{SL(n)}^{(0)} \rangle ) + \ldots )  \hat{\cal{O}} ). 
\eqno(4.1.55)
$$
where we have made use of equation (4.1.54) and denoted the different polynomials in the type IIA ten dimensional curvature $\hat{R}$, field strengths $\hat{F}$ and derivatives of the type IIA dilaton that arise in the decompactification of the $d$ dimensional polynomial in the curvature $R$, Cartan forms $P$ and field strengths $F$ by $\hat{\cal{O}}$.  The fields $\hat{\cal{O}}$ are the ten dimensional type IIA analogues of the decompactified type IIB $\hat{\cal{O}}$ fields discussed around equation (4.1.19).
Therefore, as a function of the state $| \varphi_{E_{n+1}}
\rangle $, the type IIA volume limit condition is given by
$$
\lim_{V_{n(A)} \rightarrow \infty} ( V_{n(A)}^{{2-k \over 8}} V_{n(A)}^{-{cxw_{1} \over 4}+c({\underline{\lambda}_{n-2}.\underline{\lambda}_{n-1} \over 4y} - {3 \over 8}y)w_{2}} 
$$
$$
\times g_{s(A)}^{{cxw_{1} \over 2}-c({\underline{\lambda}_{n-2}.\underline{\lambda}_{n-1} \over 2y} + {1 \over 4}y)w_{2}} \Phi_{E_{n+1}} ( |
\varphi_{SL(n)}^{(0,0)} \rangle ) + \ldots )   \hat{\cal{O}}
$$
$$
=    a_{0}  g_{s(A)}^{{cxw_{1} \over 2}-c({\underline{\lambda}_{n-2}.\underline{\lambda}_{n-1} \over 2y} + {1 \over 4}y)w_{2}}  {{\cal{O}}} + \ldots
\eqno(4.1.56)
$$
where $a_{0}$ is a constant arising from the level zero contribution that depends on the $E_{n+1}$ automorphic form and $\ldots$ denote contributions at higher levels.  As we noted earlier, any non-vanishing term in the $V_{n(A)} \rightarrow \infty$ limit must have a coefficient function that is constructed from the trivial representation of $SL(n)$ so that the type IIA effective action in $d=10$ dimensions does not depend on the moduli of the torus.
\par
In addition, the perturbative terms remaining after taking the limit must agree with a perturbative expansion in the ten dimensional type IIA string coupling $g_{s(A)}$.  In string frame this implies that each term has a $g_{s(A)}$ dependence of the form $g_{s(A)}^{-2 + 2g}$, where $g$ is the genus.  String frame in ten dimensions is related to Einstein frame by $g_{E \mu \nu}=e^{-{1 \over 2}\tilde{\sigma}}g_{S \mu \nu}$.  Upon rescaling to Einstein frame in the type IIA ten dimensional theory we find
$$
\int d^{10} x \sqrt{-g_{S}} g_{s(A)}^{ {\Delta-5 \over 2}} {\cal O}_{S},
\eqno(4.1.57)
$$
where ${\cal O}$ is some polynomial in the ten dimensional curvature $R$, fields strengths $F$ or derivatives of the type IIA dilaton, $S$ denotes string frame quantities and $\Delta$ is the number of ten dimensional type IIA space time metrics minus the number of inverse space time metrics.
\par
Considering the $E_{n+1}$ automorphic form as a function of the state $| \varphi_{E_{n+1}}
\rangle $ we see from equations (4.1.56) and (4.1.57) that for the terms remaining in the type IIA volume limit to agree with a perturbative expansion in $g_{s (A)}$ we require
$$
g_{s(A)}^{{\Delta-5
\over 2}} \left(  g_{s(A)}^{{cxw_{1} \over 2}-c({\underline{\lambda}_{n-2}.\underline{\lambda}_{n-1} \over 2y} + {1 \over 4}y)w_{2}} \Phi_{E_{n+1}} ( |
\varphi_{SL(n)}^{(0,0)} \rangle )
+\ldots \right)  \hat{{\cal O}}  =  g_{s(A)}^{-2+2n_{0}}  \hat{{\cal O}} +\dots
\eqno(4.1.58)
$$
where $n_{0}$ is a non-negative integer.

\medskip
\noindent
{ \bf {4.1.6. Large volume Limit of a $j$ dimensional subtorus} }
\medskip
Type II string theory in $k<10$ dimensions exhibits an $E_{11-k}(Z)$ symmetry.  So in the large volume limit of a $j$ dimensional subtorus $V_{j} \rightarrow \infty $ an arbitrary higher derivative term in $d=10-n$ dimensions should give a sum of $d+j$ dimensional higher derivative terms whose coefficient functions are $E_{11-(d+j)}(Z)$ automorphic forms. 
As we found earlier, the volume of a $j$ dimensional subtorus $V_{j}$ may be expressed in terms of the $E_{n+1}$ Chevalley fields and is given by
$$
V_{j}=(2 \pi)^{j} {r_{d+1} r_{d+2} ... r_{d
+j} \over \left( l_{d} \right)^{j}}=e^{\dot{\varphi}_{d+j}}.
\eqno(4.1.59)
$$
where we have made use of equations (2.2.19), (2.3.15) and (2.4.15).  The dependence of the volume of the j dimensional subtorus $V_{j}$ on the Chevalley field $\dot{\varphi}_{d+j}$ is independent of the perspective of the $d$ dimensional theory, i.e. whether we choose to express the $d$ dimensional theory in terms of the dimensionally reduced fields of type IIA/B string theory or M-theory. 
Relabelling the $E_{n+1}$ part of the Dynkin diagram as in section 4.1.1, node $d+j$ becomes node $j$ and so
$\dot{\varphi}_{d+j}$ is now $\dot{\varphi}_{j}$. Examining equation (4.1.59), we find that the large volume limit of the $j$ dimensional subtorus
$V_{j}
\rightarrow \infty$ in
$d$ dimensions is equivalent to $\dot{\varphi}_{j} \rightarrow \infty$.
Taking
$\dot{\varphi}_{j} \rightarrow \infty $ corresponds to deleting node $j$
in the $E_{n+1}$ Dynkin diagram, as shown in figure 18.
$$
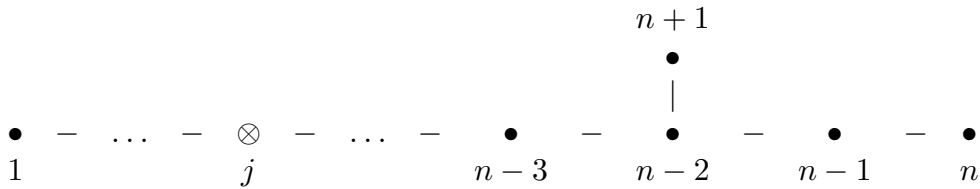
\matrix {
& & &&&&&& & & n+1 & &  &&\cr
& & & &&&&&&  & \bullet & & && \cr
& & & &&&& &&& | & & &&\cr
\bullet &-&\ldots&-&\otimes&-& \ldots & -&\bullet&-&\bullet&-& \bullet &-& \bullet \cr
1  & & & &j&&&&n-3  &  & n-2 &  & n-1& &n }
$$
\medskip
\centerline {Figure 18. The $E_{n+1}$ Dynkin diagram with node $j$ deleted}
\medskip

The  algebra remaining after this deletion is
the $GL(1)\times SL(j) \times E_{n+1-j}$ subalgebra of $E_{n+1}$.  Let us denote the
generator of the  $GL(1)$ by $X$ which we may write as $X=\sum_{a=1}^{n+1}
c_a H_a$. Demanding that it commute with $SL(j) \times E_{n+1-j}$ and in particular the
Chevalley generators $E_a, \ a=1,\ldots j-1 ,j+1, \ldots, n+1$ implies that
$$
X= \sum_{a=1}^{j} aH_{a} +\left({j-n+9 \over j-n+8}\right)jH_{j+1}+ \left({j-n+10 \over j-n+8}\right)j H_{j+2}+...+ \left({6 \over j-n+8}\right)jH_{n-2} 
$$
$$
+ \left({4 \over j-n+8}\right)jH_{n-1}+\left({2 \over j-n+8}\right)jH_{n}+ \left({3 \over j-n+8}\right)jH_{n+1}.
\eqno(4.1.60)
$$
Using the relation between the Chevalley $H_a$ and Weyl
$\vec H$ description of the generators in the Cartan subalgebra,
given by
$H_a=\vec \alpha_a\cdot \vec H$, and the decomposition of the
$E_{n+1}$ roots given in appendix A, which we recall here with the
appropriate labelling
$\vec{\alpha}_a=(0,\underline{\alpha}_a) , \ a=1,\ldots , j-1$, $\vec{\alpha}_{j}=\left(x, - \underline{\lambda}_{j-1}, - \hat{\lambda}_{1} \right) $, $\vec{\alpha}_{b}=\left( 0, \underline{0}, \hat{\alpha}_{k-j} \right)$, $b=j+1,...,n+1$ and $x^2={(n+1)(8-n+j)-9j \over j(n+1-j)(8-n+j)}$, we find that
$$
X= x(1,0)\cdot \vec H\equiv x(\vec H)_{1},
\eqno(4.1.61)
$$
up to an overall scale factor.  In deriving this equation we
have used that the simple roots $\underline{\alpha_{i}}$ and fundamental weight $\underline{\lambda_{j-1}}$ of $SL(j)$ obey the
equation
$\sum_{i=1}^{j-1} i \underline{\alpha}_{i} - j \underline{\lambda}_{j-1}   =0$ and the simple roots $\hat{\alpha}_{i}$ and fundamental weights $\hat{\lambda}_{i}$ of $E_{n+1-j}$ obey the equation $(j-n+9) \hat{\alpha}_{1}+(j-n+10)\hat{\alpha}_{2} +...+ 6 \hat{\alpha}_{n-2-j}+ 4 \hat{\alpha}_{n-1-j} + 2 \hat{\alpha}_{n-j}  + 3 \hat{\alpha}_{n+1-j}  -  j(j-n+8)\hat{\lambda}_{j-1}   =0$.  As explained in appendix B, the group
element that appears in the automorphic form, see equation (4.0.2), contains
the expression
$e^{-\sum_{a=1}^{n+1} \dot \varphi_a H_a}$ which in terms of the
$GL(1)\times SL(j) \times E(n+1-j)$ decomposition becomes
$$
\exp(-\sum_{a=1}^{n+1} \dot{\varphi}_a H_a )= \exp (-x\dot {\varphi}_{j}
(\vec H)_1)\exp(\left(\sum_{i=1}^{j-1}\dot{\varphi}_{i} \underline{\alpha}_{i} - \dot{\varphi}_{j} \underline{\lambda}_{j-1} \right)\left({\underline{H}} \right) )  
$$ 
$$
\times\exp (-\left(
\sum_{a=j+1}^{n+1}{\dot \varphi}_{a} \hat{\alpha}_{a-j} - \dot{\varphi}_{j} \hat{\lambda}_{1} \right),
\underline{H})
\eqno(4.1.62)
$$
where $\underline{H}$ is a vector consisting of the $j-1$ components of the $SL(j)$ part of $\vec{H}$ and $\underline{H}$ is a vector consisting of the $n+1-j$ components of the $E_{n+1-j}$ part of $\vec{H}$, the first term is in $GL(1)$, the second term in $SL(j)$ and the third term is in $E_{n+1-j}$. We have
identified the coefficient of the first term  as ${\dot \varphi}_{j}$ by
examining the first component of the left and right hand sides of this
vector  equation when written in terms of the Weyl
generators $\vec H$. Thus the  $GL(1)$ factor
corresponds to the  factor
$e^{-x\dot{\varphi}_{j}(\vec{H})_{1}} $ in the $E_{n+1}$ group element and, from equation (4.1.59), to the powers of the $j$ dimensional subtorus volume $V_{j}$.  In taking the $V_{j} \rightarrow \infty$ limit we must fix the $j-1$ quantities $\sum_{i=1}^{j-1}\dot{\varphi}_{i} \underline{\alpha}_{i} - \dot{\varphi}_{j} \underline{\lambda}_{j-1}$ and $\sum_{a=j+1}^{n+1}{\dot \varphi}_{a} \hat{\alpha}_{a-j} - \dot{\varphi}_{j} \hat{\lambda}_{1}$ to preserve the $SL(j) \times E_{n+1-j}$ symmetry.  
\par
The decomposition of  the full
$E_{n+1}$ algebra into representations of its $GL(1)\times SL(j) \times E_{n+1-j}$
subalgebra can be classified into a level [56-58]. The level is just the
number of times the simple root $\vec{\alpha}_{j}$ occurs in the corresponding
root when decomposed in terms of simple roots. Clearly,  the level zero
part of the decomposition is just
$GL(1)\times SL(j) \times E_{n+1-j}$, as is clear from figure 18. The decomposition of
the representations of
$E_{n+1}$ into  representations of  $GL(1)\times SL(j) \times E_{n+1-j}$ can
similarly be
classified according to the level, the level in this case is the
number of times the simple root $\vec{\alpha}_{j}$ occurs in the root string
constructed from the highest weight of the representation [59].

\par
As we discussed above, the large volume limit of the $j$ dimensional subtorus $V_{j} \rightarrow \infty $
corresponds to deleting node
$j$ in the
$E_{n+1}$ Dynkin diagram. In this limit
an  $E_{n+1}$ automorphic form  has an
expansion in powers of the volume of the $j$ dimensional subtorus $V_{j}$,
which is controlled by the $GL(1)$ factor. The  coefficient functions in
this expansion  are automorphic forms of
$SL(j) \times E_{n+1-j}$ built from the representations of $SL(j) \times E_{n+1-j}$ that occur in the
the decomposition of the representation of $E_{n+1}$, from which the
original representation is built, into those of $SL(j) \times E_{n+1-j}$. These latter
$SL(j) \times E_{n+1-j}$ automorphic forms can be labelled by the
level, discussed above.  We note that since the type II theory in $d+j$ dimensions can not depend on the moduli of the $j$ torus it is necessary for the automorphic forms found after taking the $V_{j} \rightarrow \infty$ limit to be constructed from the trivial representation of $SL(j)$.  Using equation (4.1.62) and
(4.0.2)  we can write the state
$|
\varphi_{E_{n+1}}
\rangle $ from which the automorphic form is built  as
$$
| \varphi_{E_{n+1}} \rangle = e^{- x \dot{\varphi}_{j} (\vec H)_{1}} |
\varphi_{SL(j) \times E_{n+1-j} }^{(0)} \rangle +...
$$
$$
 =e^{-x \dot{\varphi}_{j}
w_{1}}| \varphi_{SL(j) \times E_{n+1-j}}^{(0)}
\rangle + ...
$$
$$
= V_{j}^{-{xw_{1}}}  |
\varphi_{SL(j) \times E_{n+1-j}}^{(0)}
\rangle +...
\eqno(4.1.63)
$$
where $\varphi_{SL(j) \times E_{n+1-j}}^{(0)}$ is the level zero contribution
and so is built from the level zero representation, with
highest weight $w$ in the
decomposition and $w_{1}$ is the first component of $w$. In this equation
$+\ldots$ denotes the states formed from the higher level
representations in the decomposition. Clearly the $V_{j}$ dependence of
the  level
$l$  contributions is given by $V_{j}^{-\left({xw_{1}-lx^{2} }\right)}$
\par
Generic $E_{n+1}$ automorphic forms $\Phi$  constructed from $|
\varphi_{E_{n+1}} \rangle $ are expected to be homogeneous functions
which should satisfy the relation
$$
\Phi_{E_{n+1}} \left( a | \varphi_{E_{n+1}} \rangle  \right)  = a^{c}
\Phi_{E_{n+1}} \left( | \varphi_{E_{n+1}} \rangle  \right),
\eqno(4.1.64)
$$
where $a$ is a real number and $c$ is a scale factor that depends on
the
particular structure of the automorphic form. Using this homogeneity
property of $E_{n+1}$ automorphic forms, and equation (4.1.63), one may write
$$
\Phi_{E_{n+1}}\left( | \varphi_{E_{n+1}} \rangle  \right)  =
V_{j}^{-{cxw_{1} }} \Phi_{E_{n+1}} ( |
\varphi_{SL(j) \times E_{n+1-j}}^{(0)} \rangle ) + \ldots
\eqno(4.1.65)
$$
where  $+\ldots $ are terms that contain higher order
contributions in
$V_{j}$ at higher levels.
\par
We require that the terms remaining in the large volume limit of the $j$ dimensional subtorus match the known coefficient functions of the higher derivative terms in the type II string effective action in $d+j$ dimensions.  By dimensional analysis one sees that an arbitrary $d$ dimensional higher derivative term in Einstein frame takes the form
$$
l_{d}^{k-d}  \int d^{d}x \sqrt{-g}\Phi_{E_{n+1}} {\cal{O}} = l_{d+j}^{k-d}V_{j}^{-{k-\left(10-n\right) \over 8+j-n}} \int d^{d}x \sqrt{-g}\Phi_{E_{n+1}} {\cal{O}} ,
\eqno(4.1.66)
$$
where ${\cal{O}}$ is a $k$ derivative polynomial in the $d$ dimensional curvature
$R$, Cartan forms $P$ or field strengths $F$. From iterating equation (2.1.11) we see that in the large volume limit of the $j$ dimensional subtorus one has 
$$
\lim_{V_{j} \rightarrow \infty}  l_{d+j}^{j}\int d^{d}x \sqrt{-g}V_{j}^{{d-2 \over d-2+j}}= \int d^{d+j}x \sqrt{-\hat{g}},  
\eqno(4.1.67)
$$
therefore any term with $V_{j}$ dependence $V_{j}^{{d-2 \over d-2+j}}$ is preserved in the limit while any term with a lesser power of $V_{j}$ vanishes in the limit. Note that one must be careful when considering non-analytic terms in the action that appear divergent in the $V_{j} \rightarrow \infty$ limit.
Demanding that the large volume limit $V_{j} \rightarrow \infty$ of this generic higher derivative term exists from a string theory perspective means that the $V_{j} \rightarrow \infty$ limit of equation (4.1.66) exists and that the resulting terms are $d+j$ dimensional higher derivative terms with a coefficient function that is a sum of $SL(j) \times E_{n+1-j}$ automorphic forms.
Examining equation (4.1.64) and substituting the $E_{n+1}$ automorphic form as a function of the state $| \varphi_{E_{n+1}}
\rangle $ in (4.1.65) one finds that in the $V_{j} \rightarrow \infty$ limit
$$
\lim_{V_{j} \rightarrow \infty} l_{d+j}^{k-d}V_{j}^{-\left({k-\left(10-n\right) \over 8+j-n} \right)} \int d^{d}x \sqrt{-g} (V_{j}^{-{cxw_{1} }} \Phi_{E_{n+1}} ( |
\varphi_{SL(j) \times E_{n+1-j}}^{(0)} \rangle ) + \ldots ) {\cal{O}}
$$
$$
=\lim_{V_{j} \rightarrow \infty} l_{d+j}^{k-d}V_{j}^{-\left({k-\left(10-n\right) \over 8+j-n} \right)} \int d^{d}x \sqrt{-g}V_{j}^{{d-2 \over d-2+j}} V_{j}^{-\left( {d-2 \over d-2+j}\right)}\Phi_{E_{n+1}} {\cal{O}} 
$$
$$
= l_{d+j}^{k-(d+j)} \int d^{d+j}x  \sqrt{-\hat{g}}\lim_{V_{j} \rightarrow \infty} \left( V_{j}^{{2-k \over 8+j-n}} \left( 
V_{j}^{-{cxw_{1} }} \Phi_{E_{n+1}} ( |
\varphi_{SL(j) \times E_{n+1-j}}^{(0)} \rangle ) + \ldots \right) \hat{\cal{O}} \right). 
\eqno(4.1.68)
$$
where $\hat{\cal{O}}$ denotes the different $d+j$ dimensional type II string theory polynomials in the $d+j$ dimensional curvature $\hat{R}$, Cartan forms $\hat{P}$ and field strengths $\hat{F}$ that arise in the decompactification of the $d$ dimensional polynomial in the curvature $R$, Cartan forms $P$ and field strengths $F$.
The fields $\hat{\cal{O}}$ are the $d+j$ dimensional analogues of the decompactified type IIB $\hat{\cal{O}}$ fields discussed around equation (4.1.19).
Therefore as a function of the state $| \varphi_{E_{n+1}}
\rangle $ the large volume limit of the $j$ dimensional subtorus condition is given by
$$
\lim_{V_{j} \rightarrow \infty} \left(V_{j}^{{2-k \over 8+j-n}} \left( 
V_{j}^{-{cxw_{1} }}  \Phi_{E_{n+1}} ( |
\varphi_{SL(j) \times E_{n+1-j}}^{(0)} \rangle ) + \ldots \right)  \hat{\cal{O}}\right)=  a_{0} \hat{\Phi}^{(0)}_{E_{n+1-j}} \hat{{\cal{O}}} + \ldots
\eqno(4.1.69)
$$
where $a_{0}$ is a constant that depends on the $E_{n+1}$ automorphic form.  Any higher derivative term in $d$ dimensions that converges to a higher derivative term that is not compatible with type II string theory in $d+j$ dimensions, must be rejected as a possible higher derivative term in $d$ dimensions.  As we noted earlier, any non-vanishing term in the $V_{j} \rightarrow \infty$ limit must have a coefficient function that is constructed from the trivial representation of $SL(j)$ so that the $d+j$ dimensional theory does not depend on the moduli of the $j$ dimensional subtorus.

\medskip
\noindent
{ \bf {4.2 The Eisenstein-like automorphic form constructed from the {\bf 5} of $SL(5)$ with highest weight $\vec{\Lambda}_{n+1}$} }
\medskip
The automorphic form that appears as the coefficient function of the $R^{4}$ higher derivative terms in $d=7$ dimensions is the unconstrained Eisenstein-like automorphic form constructed from the {\bf 5} of $SL(5)$.  Through taking the limits, discussed in the previous section, we will find conditions under which this automorphic form could exist as a coefficient function for a higher derivative term in the $d=7$ dimensional effective action of type II string theory. 
\par
$$
\matrix {
 \bullet & 3&  \cr
 | & &  \cr
 \bullet &2 & \cr
 | & & \cr
\bullet&-&\bullet \cr
 1 &  & 4}
$$
\medskip
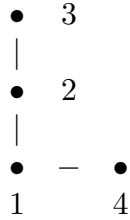
\centerline {Figure 19. The $SL(5)$ Dynkin diagram }
\medskip
The representation of $SL(5)$ with highest weight $\vec{\Lambda}^{3}$ is the {\bf 5} of $SL(5)$.  The five weights in the root string of this representation are
$$
\vec{\Lambda}^{3}_{0000}, \quad \vec{\Lambda}^{3}_{0010}, \quad \vec{\Lambda}^{3}_{0110}, \quad \vec{\Lambda}^{3}_{1110}, \quad \vec{\Lambda}^{3}_{1111},
\eqno(4.2.1)
$$
where we have adopted notation such that
$$
\vec{\Lambda}^{k}_{a_{1} a_{2} a_{3} a_{4}}= \vec{\Lambda}^{k} - \sum_{i=1}^{4}a_{i} \vec{\alpha}_{i}.
\eqno(4.2.2)
$$
\par
The $SL(5)$ lattice state $| \psi \rangle$ transforming under the {\bf 5} of $SL(5)$ may be written in terms of these weights as
$$
| \psi \rangle = m_{1} | \vec{\Lambda}^{3}_{0000} \rangle + m_{2} | \vec{\Lambda}^{3}_{0010} \rangle + m_{3} | \vec{\Lambda}^{3}_{0110} \rangle + m_{4} | \vec{\Lambda}^{3}_{1110} \rangle + m_{5} | \vec{\Lambda}^{3}_{1111} \rangle.
\eqno(4.2.3)
$$
where $m_{i} \in {\bf Z}$, $i=1,2,...,5$. The $SL(5)$ group element is given by 
$$
L(g^{-1})=e^{{1 \over \sqrt{2}} \vec{\phi}. \vec{H}}e^{-\sum_{\vec{\alpha} > 0} \chi_{\vec{\alpha}} E_{\vec{\alpha}}},
\eqno(4.2.4)
$$
where the $SL(5)$ algebra is in Weyl basis.  Comparing the Cartan subalgebra part of the group element $L(g^{-1})$ with the Cartan subalgebra part of the internal part of the $E_{11}$ group element $e^{-\vec{\varphi}.\vec{H}}$, where $\vec{\varphi}=\left(\tilde{\varphi}_{d+1},\tilde{\varphi}_{d+2},...,\tilde{\varphi}_{11} \right)$ and $\tilde{\varphi}_{i}$ are the $E_{n+1}$ Cartan subalgebra fields in the $E_{11}$ group element in Weyl basis, one finds
$$
\phi_{i}= - \sqrt{2} \tilde{\varphi}_{d+i}.
\eqno(4.2.5)
$$
Therefore the $SL(5)$ group element may be written 
$$
L(g^{-1})=e^{- \vec{\varphi}. \vec{H}}e^{-\sum_{\vec{\alpha} > 0} \chi_{\vec{\alpha}} E_{\vec{\alpha}}}.
\eqno(4.2.5)
$$
The non-linearly realised lattice state $| \varphi \rangle$ is defined by
$$
| \varphi \rangle = L(g^{-1})| \psi \rangle
$$
$$
= e^{- \vec{\varphi}. \vec{H}}e^{-\sum_{\vec{\alpha} > 0} \chi_{\vec{\alpha}} E_{\vec{\alpha}}} \left( m_{1} | \vec{\Lambda}^{3}_{0000} \rangle + m_{2} | \vec{\Lambda}^{3}_{0010} \rangle \right) 
$$
$$
+ e^{- \vec{\varphi}. \vec{H}}e^{-\sum_{\vec{\alpha} > 0} \chi_{\vec{\alpha}} E_{\vec{\alpha}}} \left(m_{3} | \vec{\Lambda}^{3}_{0110} \rangle + m_{4} | \vec{\Lambda}^{3}_{1110} \rangle + m_{5} | \vec{\Lambda}^{3}_{1011} \rangle \right).
\eqno(4.2.6)
$$
\par
The invariant unconstrained Eisentein-like automorphic form constructed from $| \varphi \rangle$ is defined by
$$
\Phi(| \varphi \rangle )= \sum_{\Lambda} {1 \over u^{s}}
\eqno(4.2.7)
$$
where $\Lambda$ is the lattice spanned by $m_{1},m_{2},...m_{5}$ and
$$
u=\langle \varphi_{D \tau} | \varphi \rangle. 
\eqno(4.2.8)
$$
From the above equations we see that the scale factor associated with $\Phi$ is $c=-2s$.
\medskip
\noindent
{ \bf {4.2.1. Perturbative Limit}}
\medskip
The perturbative limit is found by deleting node 3 in the $SL(5)$ Dynkin diagram given in figure 19.  Deleting node 3 decomposes $SL(5)$ into a $GL(1) \times SO(3,3)$ subalgebra.  The simple roots and fundamental weights of $SL(5)$ under this decomposition are given in appendix A.1 with $n=3$.  The highest weight $w$ of the automorphic form in this case is $w=\vec{\Lambda}_{3}=({1 \over x}, \tilde{0})$. From equation (4.1.9) with $c=-2s$, $d=7$, $n=3$ one finds that the $d=7$ dimensional perturbative limit condition is given by
$$
\lim_{g_{d} \rightarrow 0} g_{d}^{{4\Delta-14
\over 5}} \left( g_{d}^{{ -8sx
w_{1}
\over 5}} \Phi_{E_{n+1}}(  | \tilde{\varphi}_{SO(n,n)} \rangle)
+\ldots \right)  =  g_{d}^{-2+2n_{0}} \Phi_{SO(n,n)}^{(0)} +\dots.
\eqno(4.2.9)
$$
where $\Delta$ is the number of inverse spacetime metrics minus the number of spacetime metrics in the higher derivative term for which $\Phi_{SL(5)}$ appears as the coefficient function and $n_{0}$ is a non-negative integer. 
Since $w_{1}x=1$, this condition is equivalent to
$$
g_{d}^{{4\Delta-14-8s
\over 5}}=g_{d}^{-2+2n_{0}}.
\eqno(4.2.10)
$$ 
In other words, one finds that for $\Phi$ to appear as the coefficient function of an arbitrary higher derivative term in the $d=7$ dimensional effective action one requires
$$
n_{0}=4 \Delta - 4 - 8s.
\eqno(4.2.11)
$$
We note that the well known $R^{4}$ higher derivative term with $\Delta=4$ has the coefficient function $\Phi$ with $s={3 \over 2}$ [30,31] and contains a perturbative contribution at tree level $n_{0}=0$.  Similarly, the $\partial^{4}R^{4}$ higher derivative term with $\Delta=6$ has the coefficient function $\Phi$ with $s={5 \over 2}$ [30,31] and contains a perturbative contribution at tree level $n_{0}=0$.

\medskip
\noindent
{ \bf {4.2.2. Type IIB volume limit}}
\medskip
The type IIB volume limit is found by deleting node 2 in the $SL(5)$ Dynkin diagram given in figure 19.  Deleting node 2 decomposes $SL(5)$ into a $GL(1) \times SL(2) \times SL(3)$ subalgebra.  The simple roots and fundamental weights of $SL(5)$ under this decomposition are given in appendix A.3 with $n=3$.  The highest weight $w$ of the automorphic form in this case is $w=\vec{\Lambda}_{3}=({1 \over 2x}, \mu, \underline{0})$. From equation (4.1.20) with $c=-2s$, $d=7$, $n=3$ one finds that the large volume limit of the type IIB torus condition is given by
$$
\lim_{V_{n(B)} \rightarrow \infty} V_{n(B)}^{{2-k \over 8}} \left( 
V_{n(B)}^{sxw_{1}} \Phi_{E_{n+1}} ( |
\varphi_{SL(2) \times SL(n)}^{(0)} \rangle ) + \ldots \right) = a_{0} \hat{\Phi}^{(0)}_{SL(2)} + \ldots
\eqno(4.2.12)
$$
Since the $SL(3)$ weight of the level zero contribution is $\underline{0}$ we find that the level zero part of the automorphic form does not depend on the moduli of the torus and therefore may be preserved in the type IIB volume limit.  Since $w_{1}x={1 \over 2}$, for the level zero part of $\Phi$ to be preserved in the large volume limit of the type IIB torus we require
$$
{2-k \over 8}+ {s \over 2}=0.
\eqno(4.2.13)
$$ 
\par
The dependence of $s$ on the number derivatives $k$ of the higher derivative term agrees with the prediction of [28] as expected. The $R^{4}$ higher derivative term has $k=8$ and coefficient function $\Phi$ with $s={3 \over 2}$ [30,31].  Thus, in the large volume limit of the type IIB torus the level zero part of $\Phi$ is preserved and is constructed from the representation of $SL(2)$ with highest weight $\mu$.  Similarly, the $\partial^{4}R^{4}$ higher derivative term has $k=12$ and coefficient function $\Phi$ with $s={5 \over 2}$ [30,31].  Again, in the large volume limit of the type IIB torus the level zero part of $\Phi$ is preserved and is constructed from the representation of $SL(2)$ with highest weight $\mu$.

\medskip
\noindent
{ \bf {4.2.3. M-theory volume limit}}
\medskip
The M-theory volume limit is found by deleting node 4 in the $SL(5)$ Dynkin diagram given in figure 19.  Deleting node 4 decomposes $SL(5)$ into a $GL(1) \times SL(4)$ subalgebra.  The simple roots and fundamental weights of $SL(5)$ under this decomposition are given in appendix A.2 with $n=3$.  The highest weight $w$ of the automorphic form in this case is $w=\vec{\Lambda}_{3}=({\underline{\lambda}_{3}. \underline{\lambda}_{1} \over x}, \underline{\lambda}_{3})$. From equation (4.1.42) with $c=-2s$, $d=7$, $n=3$ one finds that the large volume limit of the M-theory torus condition is given by
$$
\lim_{V_{m(M)} \rightarrow \infty} \left( V_{m(M)}^{{2-k \over 9}} \left( 
V_{m(M)}^{{2sxw_{1} \over 3}} \Phi_{E_{n+1}} ( |
\varphi_{SL(n+1)}^{(0)} \rangle ) + \ldots \right) \right)=    a_{0} + \ldots,
\eqno(4.2.14)
$$
where $a_{0}$ is a constant.
Since the $SL(4)$ weight of the level zero contribution is $\underline{\lambda}_{3}$ we observe that the level zero part of the automorphic form depends on the moduli of the torus and therefore can not be preserved in the M-theory volume limit.  Since $w_{1}x=\underline{\lambda}_{3}. \underline{\lambda}_{1}={1 \over 4}$, for the level zero part of $\Phi$ to be vanish in the large volume limit of the M-theory torus we require
$$
{2-k \over 9}+ {s \over 6}=0.
\eqno(4.2.15)
$$ 
The $R^{4}$ higher derivative term has $k=8$ and coefficient function $\Phi$ with $s={3 \over 2}$ [30,31].  Thus, in the large volume limit of the M-theory torus the level zero part of $\Phi$ vanishes as expected.  Similarly, the $\partial^{4}R^{4}$ higher derivative term has $k=12$ and coefficient function $\Phi$ with $s={5 \over 2}$ [30,31].  Again, in the large volume limit of the M-theory torus the level zero part of $\Phi$ vanishes.  Further analysis of the automorphic form $\Phi$ with $s={3 \over 2}$ demonstrates that in the $V_{m(M)} \rightarrow \infty$ limit the level one contribution of $\Phi$ provides an automorphic form constructed from the trivial representation of $SL(4)$.  For $s={3 \over 2}$ this level one contribution converges to the coefficient of the eleven dimensional $R^{4}$ higher derivative term in the $V_{m(M)} \rightarrow \infty$ limit.

\medskip
\noindent
{ \bf {4.2.4. Type IIA volume limit}}
\medskip
The type IIA volume limit is found by deleting nodes 3 and 4 in the $SL(5)$ Dynkin diagram given in figure 19.  Deleting nodes 3 and 4 decomposes $SL(5)$ into a $GL(1) \times GL(1) \times SL(3)$ subalgebra.  The simple roots and fundamental weights of $SL(5)$ under this decomposition are given in appendix A.5 with $n=3$.  The highest weight $w$ of the automorphic form in this case is $w=\vec{\Lambda}_{3}=({1 \over x}, 0, \underline{0})$. From equation (4.1.56) with $c=-2s$, $d=7$, $n=3$ one finds that the large volume limit of the type IIA torus condition is given by
$$
\lim_{V_{n(A)} \rightarrow \infty} ( V_{n(A)}^{{2 -k \over 8}}V_{n(A)}^{{sxw_{1} \over 2}} g_{s(A)}^{{-sxw_{1}}} \Phi_{E_{n+1}} ( |
\varphi_{SL(n)}^{(0,0)} \rangle ) + \ldots ) 
$$
$$
= a_{0} g_{s(A)}^{{-sxw_{1}}} + \ldots
\eqno(4.2.16)
$$
where $a_{0}$ is a constant.
Since the $SL(3)$ weight of the level zero contribution is $\underline{0}$ we observe that the level zero part of the automorphic form does not depend on the moduli of the torus and therefore may be preserved in the type IIA volume limit.  Since $w_{1}x=1$, for the level zero part of $\Phi$ to be preserved in the large volume limit of the type IIA torus we require
$$
{2-k \over 8}+ {s \over 2}=0.
\eqno(4.2.17)
$$ 
\par
The dependence of $s$ on the number derivatives $k$ of the higher derivative term agrees with the prediction of [29] as expected. The $R^{4}$ higher derivative term has $k=8$ and coefficient function $\Phi$ with $s={3 \over 2}$ [30,31].  Thus, in the large volume limit of the type IIA torus the level zero part of $\Phi$ is preserved and contains a type IIA string coupling factor $g_{s}^{-{3 \over 2}}$.  Similarly, the $\partial^{4}R^{4}$ higher derivative term has $k=12$ and coefficient function $\Phi$ with $s={5 \over 2}$ [30,31].  Again, in the large volume limit of the type IIB torus the level zero part of $\Phi$ is preserved and contains a type IIA string coupling factor $g_{s}^{-{5 \over 2}}$. 
\par
From equation (4.1.58), the type IIA perturbative condition is
$$
g_{s(A)}^{{\Delta-5
\over 2}} \left(  g_{s(A)}^{{-s}}
+\ldots \right)  =  g_{s(A)}^{-2+2n_{0}} +\dots
\eqno(4.2.18)
$$
where $\Delta$ is the number of inverse $d=10$ spacetime metrics minus the number of $d=10$ spacetime metrics in the decompactified type IIA higher derivative term and $n_{0}$ is a non-negative integer.   Therefore if the level zero part of $\Phi$ is preserved in the large volume limit of the type IIA torus, i.e. equation (4.2.16) holds, then for $\Phi$ to appear as the coefficient function of an arbitrary higher derivative term in the $d=7$ dimensional effective action one requires
$$
n_{0}={\Delta -1 - 2s \over 4}.
\eqno(4.2.19)
$$
We note the that the well known $R^{4}$ higher derivative term in the $d=7$ effective action has the coefficient function $\Phi$ with $s={3 \over 2}$ [30,31] gives rise to a $d=10$ type IIA $R^{4}$ term with $\Delta=4$ containing a perturbative contribution at tree level $n_{0}=0$.  Similarly, the $\partial^{4}R^{4}$ higher derivative term in the $d=7$ effective action contains the coefficient function $\Phi$ with $s={3 \over 2}$ [30,31] gives rise to a $d=10$ type IIA $\partial^{4} R^{4}$ term with $\Delta=6$ from a perturbative contribution at tree level $n_{0}=0$.

\medskip
\noindent
{ \bf {4.2.5. Decompactification of a single dimension limit}}
\medskip
The decompactification of a single dimension limit is found by deleting node 1 in the $SL(5)$ Dynkin diagram given in figure 19.  Deleting node 1 decomposes $SL(5)$ into a $GL(1) \times SL(2) \times SL(3)$ subalgebra.  The simple roots and fundamental weights of $SL(5)$ under this decomposition are given in appendix A.4 with $n=3$.  The highest weight $w$ of the automorphic form in this case is $w=\vec{\Lambda}_{3}=({\hat{\lambda}_{2}. \hat{\lambda}_{1} \over x},\hat{\lambda}_{2} )$. From equation (4.1.31) with $c=-2s$, $d=7$, $n=3$ one finds that the $d=7$ decompactification of a single dimension limit condition is given by
$$
\lim_{{r_{d+1} \over l_{d+1}} \rightarrow \infty}  l_{d+1}^{{6\left( k-7 \right) \over 5}} r_{d+1}^{{2-k \over 5}} \left(  \left(l_{d+1}^{-\left({6 \over 5} \right)} r_{d+1}^{ \left({6 \over 5 } \right) } \right)^{2s x w_{1}} \Phi_{E_{n+1}} ( |
\varphi_{E_{n}}^{(0)} \rangle ) + \ldots \right)  
$$
$$
=  a_{0} \hat{\Phi}^{(0)}_{E_{n}}  + \ldots
\eqno(4.2.20)
$$
where $a_{0}$ is a constant. 
Since $w_{1}x=\hat{\lambda}_{2}. \hat{\lambda}_{1}={1 \over 3}$, for the level zero part of $\Phi$ to be preserved in the decompactification of a single dimension limit we require
$$
{2-k}+ {4s}=0.
\eqno(4.2.21)
$$ 
\par
The $R^{4}$ higher derivative term has $k=8$ and coefficient function $\Phi$ with $s={3 \over 2}$ [30,31].  Thus, in the decompactification of a single dimension limit the level zero part of $\Phi$ is preserved and is constructed from the representation of $E_{3}=SL(2) \times SL(3)$ with highest weight $\hat{\lambda}_{2}$ which is equivalent to the ${\bf 3}$ of $SL(3)$ with highest weight $\underline{\lambda}_{2}$.  Similarly, the $\partial^{4}R^{4}$ higher derivative term has $k=12$ and coefficient function $\Phi$ with $s={5 \over 2}$ [30,31].  Again, in the decompactification of a single dimension limit the level zero part of $\Phi$ is preserved and is constructed from the representation of $E_{3}=SL(2) \times SL(3)$ with highest weight $\hat{\lambda}_{2}$ which is equivalent to the ${\bf 3}$ of $SL(3)$ with highest weight $\underline{\lambda}_{2}$.

\medskip
\noindent
{\bf Conclusion}
\medskip
The maximal string  theory in $d=11-n$ dimensions has $n$ parameters.
These  parameters are in one
to one relation with the volumes of the tori and subtori that arise in
the dimensional reduction from eleven dimensions and from the view
point of the IIA  and IIB theories also the expectation value of
the dilaton field. The volumes of the  tori,  and subtori, are
encoded in
the vacuum expectation values of the  diagonal
components of the metric. Indeed,  the parameters
can be thought of as  the expectation values of the scalars in the
part of the scalar coset in the $d$ dimensional theory that  belong
to the Cartan subalgebra of $E_{n}$. In this paper
we have found the precise relationship between these parameters and the
just mentioned scalar fields. In doing so we have also found the
correspondence between the nodes of the
$E_n$ Dynkin diagram and the parameters.
Thus the results in this paper provide a precise way of implementing
the possible limits of the parameters in terms of   the scalar fields
that
belong to the Cartan subalgebra.
\par
As explained in the introduction there has been a recent interest in
the higher derivative terms in the string effective action
and the automorphic forms of $E_{n}$ that they contain.
However, the considerations of these  papers have, with two  exceptions,
been confined to terms in the effective action that have less than 14
space-time derivatives. However, if  one knew all the automorphic forms
that arise then one would know all possible string corrections. Thus it
is desirable to develop some systematic  understanding of the
automorphic
forms that do arise in the string effective action. One method of
investigating if conjectured automorphic forms are acceptable is to
study
their limits as the parameters are varied. Particularly instructive has
been the  study of the limit of small string coupling as the result must
be consistent with the form of perturbation theory that string theory
predicts. In this paper we use  the earlier results to investigate
the behaviour of generic automorphic forms in the possible limits of the
parameters.  The consequences one might draw from these calculations
are left to a future paper.

\medskip
\noindent
{\bf Acknowledgements}
\medskip
We wish to thank Jorge Russo for discussions on the importance and
significance of investigating the many possible limits of the
parameters beyond those that have been previously studied. Indeed the idea to study
this problem had also independently occurred to him.  The authors would also like to thank 
Neil Lambert for useful discussions and comments.   Peter West thanks
the STFC for support through grant ST/J002798/1 awarded to the theory group at King's.

\bigskip
\noindent
{\bf {Appendix A: Decompositions of the $E_{n+1}$ algebra}}
\bigskip
The $d$ dimensional parameters of the effective actions of type IIA/B string theory and M-theory compactified on an $n$ torus are associated with specific nodes of the $E_{n+1}$ Dynkin diagram.  In taking particular limits in these parameters the resulting theory is expected to possess a symmetry given by the deletion of the node $E_{n+1}$ Dynkin diagram node relevant to that parameter.  
\medskip
\noindent
{\bf {A.1. Perturbative Limit}}
\medskip
To investigate the properties of our automorphic form in the $g_{d} \rightarrow 0$ limit it will be expedient to decompose the $E_{n+1}$ algebra into a $GL(1) \times SO(n,n)$ subalgebra.  To do this we delete node $n$ of the Dynkin diagram, the simple roots $\vec{\alpha}$ of $E_{n+1}$ then decompose as
$$
\vec{\alpha}_{i}=\left( 0 , \tilde{\alpha}_{i} \right), \ \ \ i=1,...,n-1, 
\eqno(A.1)
$$
$$
\vec{\alpha}_{n}=\left( x , -\tilde{\lambda}_{n-1} \right), \ \ \ 
\eqno(A.2)
$$
$$
\vec{\alpha}_{n+1}=\left( 0,  \tilde{\alpha}_{n}   \right), 
\eqno(A.3)
$$
where the tilde denotes $SO(n,n)$ simple roots and fundamental weights.  The variable $x$ is fixed by the condition on the length of the simple roots, $\vec{\alpha}_{n+1}^{2}=2=x^{2}+\tilde{\lambda}_{n-1}^{2}$, this leads to 
$$
x^{2}={8-n \over 4}.
\eqno(A.4)
$$
The corresponding fundamental weights are
$$
\vec{\Lambda}^{i}=\left({\tilde{\lambda}_{i}.\tilde{\lambda}_{n-1} \over x}, \tilde{\lambda}_{i} \right), \ \ \ i=1,...,n-1, 
\eqno(A.5)
$$
$$
\vec{\Lambda}^{n}=\left( {1 \over x}, \tilde{0} \right),  
\eqno(A.6)
$$
$$
\vec{\Lambda}^{n+1}=\left({\tilde{\lambda}_{n}.\tilde{\lambda}_{n-1} \over x}, \tilde{\lambda}_{n} \right).
\eqno(A.7)
$$
\medskip
\noindent
{\bf {A.2 M-theory Limit}}
\medskip
To investigate the properties of our automorphic form in the $V_{m(M)} \rightarrow \infty$ limit we decompose the $E_{n+1}$ algebra into a $GL(1) \times SL(n+1)$ subalgebra.  To do this we delete node $n+1$ of the Dynkin diagram, the simple roots $\vec{\alpha}$ of $E_{n+1}$ then decompose as
$$
\vec{\alpha}_{i}=\left( 0 , \underline{\alpha}_{i} \right), \ \ \ i=1,...,n,
\eqno(A.8)
$$
$$
\vec{\alpha}_{n+1}=\left(x, -\underline{\lambda}_{n-2} \right), 
\eqno(A.9)
$$
where the underline denotes $SL(n+1)$ simple roots and fundamental weights.  The variable $x$ is fixed by the condition on the length of the simple roots, $\vec{\alpha}_{n+1}^{2}=2=x^{2}+\underline{\lambda}_{n-2}^{2}$, this leads to $x^{2}={8-n \over n+1}$.  The corresponding fundamental weights are
$$
\vec{\Lambda}^{i}=\left({\underline{\lambda}_{i}.\underline{\lambda}_{n-2} \over x}, \underline{\lambda}_{i} \right), \ \ \ i=1,...,n, 
\eqno(A.10)
$$
$$
\vec{\Lambda}^{n+1}=\left({1 \over x}, \underline{0} \right). 
\eqno(A.11)
$$
\medskip
\noindent
{ \bf {A.3 IIB Volume Limit}}
\medskip
To investigate the properties of our automorphic form in the $V_{n(B)} \rightarrow \infty$ limit we decompose the $E_{n+1}$ algebra into a $GL(1) \times SL(2) \times SL(n)$ subalgebra.  To do this we delete node $n-1$ of the Dynkin diagram, the simple roots $\vec{\alpha}$ of $E_{n+1}$ then decompose as
$$
\vec{\alpha}_{i}=\left( 0 , 0, \underline{\alpha}_{i} \right), \ \ \ i=1,...,n-2,
\eqno(A.12)
$$
$$
\vec{\alpha}_{n-1}=\left( x , -\mu_{1}, - \underline{\lambda}_{n-2} \right), 
\eqno(A.13)
$$
$$
\vec{\alpha}_{n}=\left( 0, \beta_{1},0   \right), 
\eqno(A.14)
$$
$$
\vec{\alpha}_{n+1}=\left( 0 , 0, \underline{\alpha}_{n-1} \right), 
\eqno(A.15)
$$
where the underline denotes $SL(n)$ simple roots and fundamental weights and $\mu_{1}$, $\beta_{1}$ are the fundamental weight and simple root of $SL(2)$ respectively.  The variable $x$ is fixed by the condition on the length of the simple roots, $\vec{\alpha}_{n}^{2}=2=x^{2}+\underline{\lambda}_{n-2}^{2}+ \mu_{1}^{2}$, this leads to $x^{2}={8-n \over 2n}$.  The corresponding fundamental weights are
$$
\vec{\Lambda}^{i}=\left({\underline{\lambda}_{i}.\underline{\lambda}_{n-2} \over x}, 0, \underline{\lambda}_{i} \right), \ \ \ i=1,...,n-2, 
\eqno(A.16)
$$
$$
\vec{\Lambda}^{n-1}=\left({1 \over x}, \underline{0} \right), 
\eqno(A.17)
$$
$$
\vec{\Lambda}^{n}=\left( {1 \over 2x}, \mu_{1}, \underline{0} \right),
\eqno(A.18)
$$
$$
\vec{\Lambda}^{n+1}=\left({\underline{\lambda}_{n-1}.\underline{\lambda}_{n-2} \over x}, 0, \underline{\lambda}_{n-1} \right). 
\eqno(A.19)
$$
\medskip
\noindent
{\bf {A.4 Decompactification of a Single Dimension Limit}}
\medskip
To investigate the properties of our automorphic form in the ${r_{d+1} \over l_{d}} \rightarrow \infty$ limit we decompose the $E_{n+1}$ algebra into a $GL(1) \times E_{n}$ subalgebra.  To do this we delete node $1$ of the Dynkin diagram, the simple roots $\vec{\alpha}$ of $E_{n+1}$ then decompose as
$$
\vec{\alpha}_{1}=\left( x , -\hat{\lambda}_{1} \right),  
\eqno(A.20)
$$
$$
\vec{\alpha}_{i}=\left(0, \hat{\alpha}_{i-1} \right), \ \ \ i=2,...,n+1, 
\eqno(A.21) 
$$
where the hat denotes $E_{n}$ simple roots and fundamental weights.  The variable $x$ is fixed by the condition on the length of the simple roots, $\vec{\alpha}_{1}^{2}=2=x^{2}+\hat{\lambda}_{1}^{2}$.  The corresponding fundamental weights are
$$
\vec{\Lambda}^{1}=\left({1 \over x}, \underline{0} \right), 
\eqno(A.22)
$$
$$
\vec{\Lambda}^{i}=\left({\hat{\lambda}_{i-1}.\hat{\lambda}_{1} \over x}, \hat{\lambda}_{i-1} \right), \ \ \ i=2,...,n+1. 
\eqno(A.23)
$$
We now proceed to calculate the inner products of the $E_{n}$ fundamental weights.  To do this we decompose the $E_{n}$ algebra into a $GL(1) \times SL(n)$ subalgebra by deleting node $n+1$, one finds
$$
\hat{\alpha}_{i}=\left( 0 , \underline{\alpha}_{i} \right), \ \ \ i=1,...,n-1, 
\eqno(A.24)
$$
$$
\hat{\alpha}_{n}= \left(y, -\underline{\lambda}_{n-3} \right), 
\eqno(A.25)
$$
with fundamental weights
$$
\hat{\lambda}_{i}=\left( {\underline{\lambda}_{i}. \underline{\lambda}_{n-3} \over y}, \underline{\lambda}_{i} \right), \ \ \ i=1,...,n-1, 
\eqno(A.26)
$$
$$
\hat{\lambda}_{n}=\left( {1 \over y} , \underline{0}  \right). 
\eqno(A.27)
$$
The variable $y$ is fixed by the condition $\hat{\alpha}_{n-2}^{2}=2$, this gives $y^{2}={9-n \over n}$.  We then have
$$
\hat{\lambda}_{1}.\hat{\lambda}_{1}=\left({ 3 \over ny}, \underline{\lambda}_{1} \right).\left({3 \over ny}, \underline{\lambda}_{1} \right)  
$$
$$
={9 \over n^{2} y^{2} }+{n-1 \over n} 
$$
$$
={10-n \over 9-n},
\eqno(A.28)
$$
where we have made use of the expression $\underline{\lambda}_{i}.\underline{\lambda}_{j}={i\left( n-j \right) \over n}$ for $i \leq j$.
We may now substitute this back into $\vec{\alpha}_{1}.\vec{\alpha}_{1}$ to fix the variable $x$,
$$
x^{2}=2-\hat{\lambda}_{1}.\hat{\lambda}_{1} 
$$
$$
={8-n \over 9-n}.
\eqno(A.29)
$$
\medskip
\noindent
{\bf {A.5 IIA Volume Limit}}
\medskip
The decomposition of representations of $E_{n+1}$ into those of $ GL(1) \times GL(1) \times SL(n)$ is given by deleting nodes $n$ and $n+1$ of the Dynkin diagram appropriate to the type IIA theory. In this section we will
find how the roots and weights of $E_{n+1}$ decompose in terms of those of
$GL(1)\times GL(1) \times SL(n)$.
\par
Let us carry out the decomposition by  first
deleting  node
$n$ to find the roots and fundamental weights of $D_{n}$ and then
delete node
$n+1$ to find the algebra $SL(n)$. Using the methods given
in reference [56], the simple roots of
$E_{n+1}$ can be expressed as
$$
\vec \alpha_{i} = \left(0, \tilde{\alpha}_{i} \right),
\quad i
= 1,...,n-1, n+1 \quad
\vec \alpha_{n}=\left(x,-  \tilde{\lambda}_{n-1} \right).
\eqno(A.30)
$$
Here
$\tilde{\alpha}_{i}, i =1,...,n$ are the roots of $D_
{n}$
and
$\tilde{\lambda}_i$ are its fundamental weights which are given
by
$$
\vec \Lambda_{i} = \left({\tilde{\lambda}_i\cdot \tilde{\lambda}_{n-1} \over x},
\tilde{\lambda}_i \right),
\quad
i = 1,...,n-1, n+1 \quad
\vec \Lambda_{n}=\left({1\over x}, \tilde{0} \right).
\quad\
\eqno(A.31)
$$
The variable
$x$ is fixed by demanding that $\vec{\alpha}_{n}^2=2=x^2+
\tilde{\lambda}_{n-1}^2$.
\par
We now delete node $n$ to find the $A_{n-1}$ algebra. The roots of
$E_{n+1}$ are found from the above roots by substituting the
corresponding
decomposition of the $D_{n}$ roots and weights into those of $A_
{n-1}$.
The roots of $D_{n}$ in terms of those of $A_{n-1}$ are given by
$\tilde{\alpha}_i = \left(0, \underline{\alpha}_{i}\right),\
i=1,...,n -1$ and
$\tilde{\alpha}_{n} = \left(y,- \underline
\lambda_{n-2}\right)$ while the fundamental weights are given by
$\tilde{\lambda}_i = \left({\lambda _{n-2}\cdot
\lambda_{i}\over y},
\underline{\lambda}_{i}\right)\ i=1,...,n -1$ and
$\tilde{\lambda}_{n+1} = \left({1\over y}, \underline{0} \right)$. Requiring
${\tilde \alpha}_{n+1}^2=2$ gives $y^2={4\over n}$
We then find that the roots of $E_{n+1}$ are given by
$$
\vec \alpha_{i} = \left(0, 0,\underline \alpha_{i} \right), \quad
i =1,...,n-1, 
\vec \alpha_{n} = \left(x , -{\lambda _{n-2}\cdot \lambda_{n-1}\over
y}, -\underline    \lambda_{n-1}
\right), 
\vec \alpha_{n+1} =\left(0,y, -\underline \lambda_{n-2} \right).
\eqno(A.32)
$$
The fundamental weights of $E_{n+1}$ are found in the same way to be
$$
\vec{\Lambda}_{i} = \left( {c_i\over x},{\lambda _{n-2}\cdot
\lambda_{i}\over y},
\underline \lambda_{i} \right),  \quad i =1,...,n-1,
\eqno(A.33)
$$
$$
\vec{\Lambda}_{n} = \left( {1\over x},0 , \underline 0  \right),
\eqno(A.34)
$$
$$
\vec{\Lambda}_{n+1} = \left( {n-2\over 4 x},{1\over y} , \underline 0
\right),
\eqno(A.35)
$$
where $ c_i={i\over 2}, \ i=1,\ldots ,n-2$ and  $ c_{n-1}={n \over 4}$.
As $\tilde{\lambda}_{n-1}^2={n\over 4}$ we find that $x^2={8-n\over 4}$.

\medskip
\noindent
{\bf {A.6 $j$ dimensional subtorus limit}}
\medskip
To investigate the properties of our automorphic form in the $V_{j} \rightarrow \infty$ limit we decompose the $E_{n+1}$ algebra into a $GL(1) \times SL(j) \times E_{n+1-j}$ subalgebra.  To do this we delete node $j$ of the Dynkin diagram, the simple roots $\vec{\alpha}_{i}$ of $E_{n+1}$ then decompose as
$$
\vec{\alpha}_{i}=\left(0, \underline{\alpha}_{i}, \hat{0} \right), \ \ \ i=1,...,j-1,
\eqno(A.36)
$$
$$
\vec{\alpha}_{j}=\left( x , -\underline{\lambda}_{j-1}, - \hat{\lambda}_{1} \right),  
\eqno(A.37)
$$
$$
\vec{\alpha}_{k}=\left(0, \hat{\alpha}_{k-j} \right), \ \ \ k=j+1,...,n+1, 
\eqno(A.38) 
$$
where the underline and the hat denote $SL(j)$ and $E_{n+1-j}$ quantities and $\alpha$, $\lambda$ are the respective simple root and fundamental weights of the corresponding algebra.  The corresponding fundamental weights are
$$
\vec{\Lambda}^{i}=\left(  {\underline{\lambda}_{i}.\underline{\lambda}_{j-1} \over x}, \underline{\lambda}_{i}, \hat{0} \right), \ \ \ i=1,...,j-1,
\eqno(A.39)
$$
$$
\vec{\Lambda}^{j}=\left(  {1 \over x}, \underline{0}, \hat{0} \right),
\eqno(A.40)
$$
$$
\vec{\Lambda}^{k}=\left({\hat{\lambda}_{k-j}.\hat{\lambda}_{1} \over x}, \underline{0}, \hat{\lambda}_{k-j} \right), \ \ \ k=j+1,...,n+1. 
\eqno(A.41)
$$
The variable $x$ is fixed by the condition on the length of the simple roots, $\vec{\alpha}_{j}^{2}=2=x^{2}+\hat{\lambda}_{1}.\hat{\lambda}_{1}+\underline{\lambda}_{j-1}.+\underline{\lambda}_{j-1}$.  After some work one finds
$$
x^{2}={(n+1)(8-n+j) - 9j \over j(n+1-j)(8-n+j)}.
\eqno(A.42)
$$

\bigskip
\noindent
{\bf {Appendix B: Automorphic forms and non-linearly realised lattice states}}
\medskip
Generic higher derivative corrections in the effective action of Type II string theory, compactified on an $n$-torus to $d=10-n$ dimensions, are polynomials in the curvature $R$, Cartan forms $P$ and degree $k$ field strengths $F_{k}$ multiplied by an automorphic form $\Phi_{E_{n+1}}$ transforming under the $E_{n+1}$ U-duality group.  One may construct an $E_{n+1}$ automorphic form $\Phi_{E_{n+1}}$ from the function $| \varphi_{E_{n+1}} \rangle$, which is defined by,
$$
| \varphi_{E_{n+1}} \rangle = L(g^{-1}) | \psi \rangle
\eqno(B.1)
$$
where $L(g^{-1})$ is a representation of the coset element $g\in E_{n+1}/K$, $K$ being the maximal compact sub-group of $E_{n+1}$ and $ | \psi \rangle$ is a linear representation of $E_{n+1}({Z})$.  Using the Iwasawa decomposition and fixing the local group element $h \in K$ to be the identity, we may write 
$$
L(g^{-1})= e^{{1 \over \sqrt{2}}\vec{\phi}.\vec{H}}e^{-\sum_{\vec{\alpha}>0} \chi_{\vec{\alpha}}E_{\vec{\alpha}}},
\eqno(B.2)
$$
where $H$ are the generators in the Cartan sub-algebra of $E_{n+1}$, in Weyl basis, and $E_{\vec{\alpha}}$ are the positive root generators, while $\chi_{\vec{\alpha}}$ are the axions and $\vec{\phi}$ is a vector whose components are linear combinations of the physical fields of type IIA/B string theory or M-theory, namely, the type IIA/B dilaton, the $n$-torus volume modulus $\rho$ and the remaining $n$ or $n-1$ moduli $\underline {\phi}$.  Instead of writing the coset element $g\in E_{n+1}/K$ in terms of the type IIA, type IIB or M-theory physical fields we may write it as a function of the $E_{n+1}$ Chevalley fields $\dot{\varphi}_{i}$, $i=d+1,...,11$ parameterising the $E_{n+1}$ symmetry.  The fields in the $E_{n+1}$ part of the $E_{11}$ group element $\dot{\varphi}_{i}$ are equal to those in the physical field parameterisation of the group element $L\left( g^{-1}  \right)$ used to construct the automorphic form, up to a numerical factor.  We find, through comparing the normalisations of the fields associated with the Cartan subalgebra in the $E_{n+1}$ part of the $E_{11}$ group element $e^{\vec{\varphi} .\vec{H}}$ and those in the automorphic form group element $e^{-{1 \over \sqrt{2}} \vec{\phi}. \vec{H}}$ that $\phi_{i}=-\sqrt{2}\tilde{\varphi}_{i}$, where $\tilde{\varphi}_{i}$ are the $E_{n+1}$ fields in Weyl basis.  So the coset element $g\in E_{n+1}/K$ as a function of the $E_{n+1}$ fields $\tilde{\varphi}_{i}$ is
$$
L(g^{-1})= e^{-\vec{\varphi}. \vec{H}}e^{-\sum_{\vec{\alpha}>0} \chi_{\vec{\alpha}}E_{\vec{\alpha}}},
\eqno(B.3)
$$  
where $\vec{\varphi}=\left( \tilde{\varphi}_{1}, \tilde{\varphi}_{2},...,\tilde{\varphi}_{n+1} \right)$.
In Weyl basis, where the commutator of the Cartan subalgebra elements $H_{i}$ with the positive root generators is $[H_{i}, E_{\vec{\alpha}}]=\vec{\alpha}_{i}E_{\vec{\alpha}}$, the action of the Cartan subalgebra $\vec{H}$ on $| \psi_{\vec \Lambda_{k}} \rangle $ is
$$
| \varphi_{\Lambda_{k}} \rangle = e^{-\vec{\varphi}.\vec{H}} | \psi_{\Lambda_{k}} \rangle 
$$
$$
= e^{-\vec{\varphi}. \left[\vec \Lambda_{k} \right] }| \psi_{\Lambda_{k}} \rangle.
\eqno(B.4)
$$
where $[\vec \Lambda_{k} ]$ is the set of weights in the representation of $E_{n+1}$ with highest weight $\vec \Lambda_{k}$. 
The Weyl basis of $E_{n+1}$ fields $\tilde{\varphi}_{i}$ are related to the $E_{n+1}$ Chevalley basis fields $\dot{\varphi}_{a}$ by $\tilde{\varphi}_{i}=\dot{\varphi}_{a}\alpha^{a}_{i}$, where $\alpha^{a}_{i}$ is the i'th component of the a'th simple root.   
We will denote the automorphic form that is a function of the above non-linearly realised $\vec{\Lambda}_{k}$ representation of $E_{n+1}$ by  $\Phi_{E_{n+1}} \left( | \varphi_{\Lambda_{k}} \rangle \right)$. 
\par
The large volume limit of the type IIA/B and M-theory torus, along with the perturbative limit are associated with a single node of the $E_{n+1}$ Dynkin diagram.  To evaluate an $E_{n+1}$ automorphic form in these limits it is expedient to delete the relevant node giving a decomposition of the $E_{n+1}$ algebra in terms of a $GL(1)$ factor, that corresponds to the parameter of interest, and a rank $n$ subalgebra.
Splitting the set of positive roots $\vec{\alpha} >0$ into the set of positive roots $\vec{\alpha}^{\ast}$ which contain the simple root $\vec{\alpha}_{l}$, where $l$ is the deleted node, and the remaining roots $\underline{\alpha}$ that do not contain $\vec{\alpha}_{l}$, we have 
$$
e^{-\sum_{\vec{\alpha} > 0} \chi_{\vec{\alpha}} E_{\vec{\alpha}}}= e^{-\sum_{\underline{\alpha} > 0} \chi_{\underline{\alpha}} E_{\underline{\alpha}} -\sum_{\vec{\alpha}^{*} > 0} \chi_{\vec{\alpha}^{*}} E_{\vec{\alpha}^{*}} }.
\eqno(B.5)
$$
By the Baker-Campbell-Hausdorff lemma we may write this as
$$
e^{-\sum_{\vec{\alpha} > 0} \chi_{\vec{\alpha}} E_{\vec{\alpha}}}= e^{-\sum_{\underline{\alpha} > 0} \chi_{\underline{\alpha}} E_{\underline{\alpha}}}e^{ -\sum_{\vec{\alpha}^{*} > 0} \chi_{\vec{\alpha}^{*}} E_{\vec{\alpha}^{*}} } e^{-{1 \over 2}\left[\sum_{\underline{\alpha} > 0} \chi_{\underline{\alpha}} E_{\underline{\alpha}}, \sum_{\vec{\alpha}^{*} > 0} \chi_{\vec{\alpha}^{*}} E_{\vec{\alpha}^{*}} \right]}\ldots,
\eqno(B.6)
$$
where $\ldots$ denotes the higher order commutators of $\sum_{\underline{\alpha} > 0} \chi_{\underline{\alpha}} E_{\underline{\alpha}}$ and $\sum_{\vec{\alpha}^{*} > 0} \chi_{\vec{\alpha}^{*}} E_{\vec{\alpha}^{*}}$.
The action of $e^{-\sum_{\vec{\alpha} > 0} \chi_{\vec{\alpha}} E_{\vec{\alpha}}}$ on a set of states $| \mu_{i} \rangle$ of weight $\mu_{i}$ can be expressed as
$$
e^{-\sum_{\vec{\alpha} > 0} \chi_{\vec{\alpha}} E_{\vec{\alpha}}} | \mu_{i} \rangle =  e^{-\sum_{\underline{\alpha} > 0} \chi_{\underline{\alpha}} E_{\underline{\alpha}}}e^{ -\sum_{\vec{\alpha}^{*} > 0} \chi_{\vec{\alpha}^{*}} E_{\vec{\alpha}^{*}} } e^{-{1 \over 2}\left[\sum_{\underline{\alpha} > 0} \chi_{\underline{\alpha}} E_{\underline{\alpha}}, \sum_{\vec{\alpha}^{*} > 0} \chi_{\vec{\alpha}^{*}} E_{\vec{\alpha}^{*}} \right]} \ldots | \mu_{i} \rangle
$$
$$
=e^{-\sum_{\underline{\alpha} > 0} \chi_{\underline{\alpha}} E_{\underline{\alpha}}} | \mu_{i} \rangle - e^{-\sum_{\underline{\alpha} > 0} \chi_{\underline{\alpha}} E_{\underline{\alpha}}} \sum_{j} \tilde{\chi}_{ij}  | \mu_{j} \rangle
\eqno(B.7)
$$
where $\tilde{\chi}_{ij}$ is a polynomial in the fields $\chi_{\vec{\alpha}^{\ast}}$, $\chi_{\underline{\alpha}}$.  Note that $| \mu_{j} \rangle$ is at a lower level in $\vec{\alpha}_{l}$ than $| \mu_{i} \rangle$.  
\par
Defining $ | \psi_{A}^{p,j} \rangle $ to be the state carrying the linear representation $j$ of the subalgebra at level $p$.   The non-linearly realised lattice state may be written 
$$
| \varphi_{\Lambda_{k}} \rangle= e^{-x\dot{\varphi}_{d+l}(\vec{H})_{1}-\underline{\varphi}.\underline{H} }e^{-\sum_{\underline{\alpha} > 0} \chi_{\underline{\alpha}} E_{\underline{\alpha}}}
$$
$$
\times e^{ -\sum_{\vec{\alpha}^{*} > 0} \chi_{\vec{\alpha}^{*}} E_{\vec{\alpha}^{*}} } e^{-{1 \over 2}\left[\sum_{\underline{\alpha} > 0} \chi_{\underline{\alpha}} E_{\underline{\alpha}}, \sum_{\vec{\alpha}^{*} > 0} \chi_{\vec{\alpha}^{*}} E_{\vec{\alpha}^{*}} \right]}\ldots \sum_{p,j}  | \psi_{A}^{p,j} \rangle 
$$
$$
=e^{-x\dot{\varphi}_{d+l}(\vec{H})_{1}-\underline{\varphi}.\underline{H} }e^{-\sum_{\underline{\alpha} > 0} \chi_{\underline{\alpha}} E_{\underline{\alpha}}} \sum_{p,j}  | \tilde{\psi}_{A}^{p,j} \rangle 
$$
$$
=e^{-x\dot{\varphi}_{d+l}(\Lambda_{k})_{1}} \sum_{p,j} e^{px^{2}\dot{\varphi}_{d+l}} | \tilde{\varphi}_{A}^{p,j} \rangle, 
\eqno(B.8)
$$
where $| \tilde{\psi}_{A}^{p,j} \rangle$ and $| \tilde{\varphi}_{A}^{p,j} \rangle$ are the shifted, by the action of all positive root generators associated with the deleted node, linearly realised and non-linearly realised lattice states respectively.
\par
One may then use the homogeneity property of a generic $E_{n+1}$ automorphic form to write
$$
\Phi_{E_{n+1}}(| \varphi_{\Lambda_{k}} \rangle)= e^{-cx\dot{\varphi}_{d+l} (\Lambda_{k})_{1} }\Phi_{E_{n+1}}( \sum_{p,j} e^{px^{2}\dot{\varphi}_{d+l}} | \tilde{\varphi}_{A}^{p,j} \rangle)
\eqno(B.9)
$$
where $c$ is a constant that depends on the structure of the particular automorphic form under consideration.

\medskip
\noindent
{\bf References}
\medskip

\item{[1]}
I.~C.~G.~Campbell and P.~C.~West,
{\it N=2 D=10 Nonchiral Supergravity and Its Spontaneous
Compactification},
Nucl.\ Phys. B {\bf 243} (1984) 112.

\item{[2]}
F.~Giani and M.~Pernici,
{\it N=2 Supergravity in Ten-Dimensions},
Phys.\ Rev. D {\bf 30} (1984) 325.

\item{[3]}
M.~Huq and M.~A.~Namazie,
{\it Kaluza-Klein Supergravity in Ten-Dimensions},
Class.\ Quant.\ Grav.\ {\bf 2}, 293 (1985)
[Erratum-ibid.\ {\bf 2}, 597 (1985)].
\medskip
\item{[4]}
J.~H.~Schwarz and P.~C.~West,
{\it Symmetries and Transformations of Chiral N=2 D=10 Supergravity},
Phys.\ Lett. B {\bf 126}, 301 (1983).

\item{[5]}
P.~S.~Howe and P.~C.~West,
{\it The Complete N=2, D=10 Supergravity},
Nucl.\ Phys. B {\bf 238}, 181 (1984).

\item{[6]}
J.~H.~Schwarz,
{\it Covariant Field Equations of Chiral N=2 D=10 Supergravity},
Nucl.\ Phys. B {\bf 226}, 269 (1983).

\item{[7]}
E.~Cremmer, B.~Julia and J.~Scherk,
{\it Supergravity Theory in Eleven-Dimensions},
Phys.\ Lett. B {\bf 76}, 409 (1978).

\item{[8]}
E.~Cremmer and B.~Julia,
{\it The N=8 Supergravity Theory. 1. The Lagrangian},
Phys.\ Lett. B {\bf 80}, 48 (1978).

\item{[9]}
N.~Marcus and J.~H.~Schwarz,
{\it Three-Dimensional Supergravity Theories},
Nucl.\ Phys. B {\bf 228}, 145 (1983).

\item{[10]}
B.~Julia and H.~Nicolai,
{\it Conformal internal symmetry of 2-d sigma models coupled to
gravity and a
dilaton},
Nucl.\ Phys. B {\bf 482}, 431 (1996)
[arXiv:hep-th/9608082].

\item{[11]}
B.~Julia, in {\it Vertex Operators and Mathematical
Physics}, Publications of the Mathematical
Sciences Research Institute no3. Springer Verlag (1984); in
{\it Superspace and
Supergravity}, ed. S. W. Hawking and M. Rocek, Cambridge
University Press (1981)

\item{[12]}
C.~Teitelboim,
{\it Monopoles of Higher Rank},
Phys.\ Lett. B {\bf 167}, 69 (1986).

\item{[13]}
R.~I.~Nepomechie,
{\it Magnetic Monopoles from Antisymmetric Tensor Gauge Fields},
Phys.\ Rev. D {\bf 31}, 1921 (1985).

\item{[14]}
A.~Font, L.~E.~Ibanez, D.~Lust and F.~Quevedo,
{\it Strong - weak coupling duality and nonperturbative effects in
string
theory},
Phys.\ Lett. B {\bf 249}, 35 (1990).
S.J. Rey, {\it The Confining Phase Of Superstrings And Axionic
Strings}, Phys. Rev. {\bf D43} (1991) 526.

\item{[15]}
A.~Sen,
{\it Electric magnetic duality in string theory},
Nucl.\ Phys. B {\bf 404}, 109 (1993)
[arXiv:hep-th/9207053].
A. Sen, {\it Quantization of dyon charge and electric magnetic
duality in
string theory}, Phys. Lett. {\bf 303B} (1993) 22; {\it Strong - weak
coupling duality in four-dimensional string theory}, Int. J. Mod. Phys.
{\bf A9} (1994) 3707. J. Schwarz and A. Sen, {\it Duality symmetric
actions}, Nucl. Phys. {\bf B411} (1994) 35.

\item{[16]}
C.~M.~Hull and P.~K.~Townsend,
{\it Unity of superstring dualities}
Nucl.\ Phys. B {\bf 438}, 109 (1995)
[arXiv:hep-th/9410167].

\item{[17]}
M.~B.~Green and M.~Gutperle,
{\it Effects of D instantons},
Nucl.\ Phys. B {\bf 498}, 195 (1997)
[arXiv:hep-th/9701093].

\item{[18]}
M.~B.~Green, M.~Gutperle and P.~Vanhove,
{\it One loop in eleven-dimensions},
Phys.\ Lett. B {\bf 409}, 177 (1997)
[arXiv:hep-th/9706175].

\item{[19]}
M.~B.~Green and S.~Sethi,
{\it Supersymmetry constraints on type IIB supergravity},
Phys.\ Rev. D {\bf 59}, 046006 (1999)
[arXiv:hep-th/9808061].

\item{[20]}
M.~B.~Green, H.~h.~Kwon and P.~Vanhove,
{\it Two loops in eleven-dimensions},
Phys.\ Rev. D {\bf 61}, 104010 (2000)
[arXiv:hep-th/9910055].

\item{[21]}
M.~B.~Green and P.~Vanhove,
{\it Duality and higher derivative terms in M theory},
JHEP {\bf 0601}, 093 (2006)
[arXiv:hep-th/0510027].

\item{[22]}
M.~B.~Green, J.~G.~Russo and P.~Vanhove,
{\it Modular properties of two-loop maximal supergravity and
connections with
string theory},
JHEP {\bf 0807}, 126 (2008)
[arXiv:0807.0389 [hep-th]].

\item{[23]}
J.~G.~Russo,
{\it Construction of SL(2,Z) invariant amplitudes in type IIB
superstring
theory},
Nucl.\ Phys. B {\bf 535}, 116 (1998)
[arXiv:hep-th/9802090].

\item{[24]}
A.~Basu,
{\it The D**10 R**4 term in type IIB string theory},
Phys.\ Lett. B {\bf 648}, 378 (2007)
[arXiv:hep-th/0610335].

\item{[25]} C. Pope, 
{\it Lectures on Kaluza-Klein Theory}, http://faculty.physics.tamu.edu/pope/ihplec.pdf

\item{[26]}
N.~Lambert and P.~C.~West,
{\it Enhanced Coset Symmetries and Higher Derivative Corrections},
Phys.\ Rev. D {\bf 74}, 065002 (2006)
[arXiv:hep-th/0603255].

\item{[27]}
N.~Lambert and P.~C.~West,
{\it Duality Groups, Automorphic Forms and Higher Derivative
Corrections},
Phys.\ Rev. D {\bf 75}, 066002 (2007)
[arXiv:hep-th/0611318].

\item{[28]}
F.~Gubay, N.~Lambert, P.~West,
{\it Constraints on Automorphic Forms of Higher Derivative Terms
from Compactification},
JHEP {\bf 1008}, 028 (2010).
[arXiv:1002.1068 [hep-th]]

\item{[29]}
  F.~Gubay and P.~West,
  {\it Higher derivative type II string effective actions, automorphic forms and
  E11},
  arXiv:1111.0464 [hep-th].

\item{[30]}
N.~Lambert and P.~West,
{\it Perturbation Theory From Automorphic Forms},
JHEP {\bf 1005}, 098 (2010)
[arXiv:1001.3284 [hep-th]].

\item{[31]}
M.~B.~Green, J.~G.~Russo and P.~Vanhove,
{\it Automorphic properties of low energy string amplitudes in
various
dimensions},
Phys.\ Rev. D {\bf 81}, 086008 (2010)
[arXiv:1001.2535 [hep-th]].

\item{[32]}
M.~B.~Green, J.~G.~Russo and P.~Vanhove,
{\it String theory dualities and supergravity divergences},
JHEP {\bf 1006}, 075 (2010)
[arXiv:1002.3805 [hep-th]].

\item{[33]}
M.~B.~Green, S.~D.~Miller, J.~G.~Russo and P.~Vanhove,
{\it Eisenstein series for higher-rank groups and string theory
amplitudes},
arXiv:1004.0163 [hep-th].

\item{[34]}
N.~Berkovits and C.~Vafa,
{\it Type IIB R**4 H**(4g-4) conjectures},
Nucl.\ Phys. B {\bf 533}, 181 (1998)
[arXiv:hep-th/9803145].

\item{[35]}
M.~B.~Green, J.~G.~Russo and P.~Vanhove,
{\it Non-renormalisation conditions in type II string theory and
maximal
supergravity},
JHEP {\bf 0702}, 099 (2007) [arXiv:hep-th/0610299].

\item{[36]}
N.~Berkovits,
{\it New higher-derivative R**4 theorems},
Phys.\ Rev.\ Lett. {\bf 98}, 211601 (2007)
[arXiv:hep-th/0609006].

\item{[37]}
  A.~Basu,
  {\it The $D^4 R^4$ term in type IIB string theory on $T^2$ and U-duality},
  Phys.\ Rev.\  D {\bf 77}, 106003 (2008)
  [arXiv:0708.2950 [hep-th]].
  
\item{[38]}
  A.~Basu,
  {\it The $D^6 R^4$ term in type IIB string theory on $T^2$ and U-duality},
  Phys.\ Rev.\  D {\bf 77}, 106004 (2008)
  [arXiv:0712.1252 [hep-th]].
  
\item{[39]}
  A.~Basu,
  {\it Supersymmetry constraints on the $R^4$ multiplet in type IIB on $T^2$},
  Class.\ Quant.\ Grav.\  {\bf 28}, 225018 (2011)
  [arXiv:1107.3353 [hep-th]].
  
\item{[40]}
E.~Kiritsis and B.~Pioline,
{\it On R**4 threshold corrections in IIb string theory and (p, q)
string
instantons},
Nucl.\ Phys. B {\bf 508}, 509 (1997)
[arXiv:hep-th/9707018].
  
\item{[41]}
N.~A.~Obers and B.~Pioline,
{\it Eisenstein series and string thresholds},
Commun.\ Math.\ Phys. {\bf 209}, 275 (2000)
[arXiv:hep-th/9903113]; N.~A.~Obers and B.~Pioline,
{\it Eisenstein series in string theory},
Class.\ Quant.\ Grav. {\bf 17}, 1215 (2000)
[arXiv:hep-th/9910115].

\item{[42]}
B.~Pioline,
{\it R**4 couplings and automorphic unipotent representations},
JHEP {\bf 1003}, 116 (2010)
[arXiv:1001.3647 [hep-th]].

\item{[43]}
P.~C.~West,
{\it E(11) and M theory},
Class.\ Quant.\ Grav. {\bf 18}, 4443 (2001)
[arXiv:hep-th/0104081].

\item{[44]}
I.~Schnakenburg and P.~C.~West,
{\it Kac-Moody symmetries of 2B supergravity},
Phys.\ Lett. B {\bf 517}, 421 (2001)
[arXiv:hep-th/0107181].

\item{[45]}
P.~C.~West,
{\it Hidden superconformal symmetry in M theory},
JHEP {\bf 0008}, 007 (2000)
[arXiv:hep-th/0005270].

\item{[46]}
P.~C.~West,
{\it The IIA, IIB and eleven-dimensional theories and their common
E(11) origin},
Nucl.\ Phys. {\bf B693}, 76-102 (2004).
[hep-th/0402140].

\item{[47]}
F.~Riccioni and P.~C.~West,
{\it The E(11) origin of all maximal supergravities},
JHEP {\bf 0707}, 063 (2007)
[arXiv:0705.0752 [hep-th]].

\item{[48]}
  F.~Riccioni, D.~Steele, P.~West,
  {\it The E(11) origin of all maximal supergravities: The Hierarchy of field-strengths},
  JHEP {\bf 0909}, 095 (2009).
  [arXiv:0906.1177 [hep-th]]

\item{[49]}
	E. Witten, 
	{\it Some Properties of O(32) Superstrings},
	Phys. lett. {\bf  149B}, 351 (1984).
	
\item{[50]}		
	E. Fradkin and A. Tseytlin, 
	{\it Effective Field Theory from Quantized Strings}, 
	Phys. lett. {\bf  158B} 316 (1985).

\item{[51]}
  I.~Schnakenburg and P.~C.~West,
  {\it Massive IIA supergravity as a non-linear realisation},
  Phys.\ Lett.\  B {\bf 540}, 137 (2002)
  [arXiv:hep-th/0204207].

\item{[52]}
  P.~C.~West,
  {\it Some simple predictions from $E_{11}$ symmetry},
  Phys.\ Lett.\  {\bf 603B}, 63 (2004).
  [hep-th/0407088]

\item{[53]}
  P.~C.~West,
  {\it Introduction to strings and branes},
	{\it  Cambridge, UK: Cambridge University Press (2012) 672p}.

\item{[54]}
  P.~C.~West,
  {\it E(11), SL(32) and central charges},
  Phys.\ Lett.\  {\bf B575}, 333-342 (2003).
  [hep-th/0307098].

\item{[55]}
  S.~Mizoguchi and G.~Schroder,
  {\it On discrete U-duality in M-theory},
  Class.\ Quant.\ Grav.\  {\bf 17}, 835 (2000)
  [arXiv:hep-th/9909150].
  
\item{[56]}
  M.~R.~Gaberdiel, D.~I.~Olive, P.~C.~West,
  {\it A Class of Lorentzian Kac-Moody algebras},
  Nucl.\ Phys.\  {\bf B645}, 403-437 (2002).
  [hep-th/0205068].

\item{[57]}
  P.~C.~West,
  {\it Very extended E(8) and A(8) at low levels, gravity and supergravity},
  Class.\ Quant.\ Grav.\  {\bf 20}, 2393 (2003)
  [arXiv:hep-th/0212291].

\item{[58]}
  T.~Damour, M.~Henneaux and H.~Nicolai,
  {\it E10 and a 'small tension expansion' of M Theory},
  Phys.\ Rev.\ Lett.\  {\bf 89}, 221601 (2002)
  [arXiv:hep-th/0207267].
  
\item{[59]}
  A.~Kleinschmidt, P.~C.~West,
  {\it Representations of G+++ and the role of space-time},
  JHEP {\bf 0402}, 033 (2004).
  [hep-th/0312247].

\end